\documentclass{aa}  

\usepackage{graphicx}
\usepackage{txfonts}

\usepackage{booktabs}

\usepackage{multicol}
\usepackage{subfig}
\usepackage{hyperref}
\usepackage{gensymb}
\usepackage{xcolor}

\begin{document} 

\title{The Nephele ecosystem: Stars,  globular clusters, and stellar streams associated with the progenitor galaxy of  $\omega$~Centauri}
  \titlerunning{The Nephele ecosystem: Stars, GCs, and stellar streams associated with the progenitor galaxy of $\omega$~Cen}

   \author{G.~Pagnini
          \inst{1, 2}
          \and
          P.~Di Matteo\inst{1}
          \and
          M.~Haywood\inst{1}
          \and
          P. ~Bianchini\inst{2}
          \and
          S. Ferrone\inst{1, 3, 4}
          \and
          A.~Mastrobuono-Battisti\inst{5, 6, 7}
           \and
           O. Agertz\inst{8}
           \and
           S. Khoperskov\inst{9}
           \and
           F. Renaud\inst{2, 10}
           \and
           N. Ryde\inst{8}
          }

   \institute{LIRA, Observatoire de Paris, Université PSL, Sorbonne Université, Université Paris Cité, CY Cergy Paris Université, CNRS, 92190 Meudon, France\\
              \email{pagnini@astro.unistra.fr}
              \and Université de Strasbourg, CNRS, Observatoire Astronomique de Strasbourg, F-67000 Strasbourg, France
              \and Dipartimento di Fisica, Università di Roma “La Sapienza”, Piazza Aldo Moro
              \and Institute for Complex Systems CNR, Piazzale Aldo Moro 2, 00185 Rome, Italy
        \and Dipartimento di Fisica e Astronomia “Galileo Galilei”, Università di Padova, Vicolo dell’Osservatorio 3, 35122 Padova, Italy
        \and Dipartimento di Tecnica e Gestione dei Sistemi Industriali, Università degli Studi di Padova, Stradella S. Nicola 3, I-36100 \\Vicenza, Italy
        \and Istituto Nazionale di Astrofisica – Osservatorio Astronomico di Padova, Vicolo dell’Osservatorio 5, Padova, I-35122, Italy 
           \and Division of Astrophysics, Department of Physics, Lund University, Box 118, SE-221 00 Lund, Sweden
           \and  Leibniz Institut für Astrophysik Potsdam (AIP), An der Sternwarte 16, D-14482, Potsdam, Germany
           \and University of Strasbourg Institute for Advanced Study, 5 all\'ee du G\'en\'eral Rouvillois, F-67083 Strasbourg, France}

   \date{Received xxx; accepted xxx}

  \abstract
  {Globular clusters (GCs) and their associated stellar streams are key tracers of the hierarchical assembly history of the Milky Way. $\omega$~Centauri, the most massive and chemically complex GC in the Galaxy, is widely believed to be the remnant nucleus of an accreted dwarf galaxy. Identifying its associated debris and that of chemically similar clusters can provide important constraints on the nature of this progenitor system.}{We aim to identify stars in the Galactic field that are chemically and kinematically associated with $\omega$~Cen and with a group of GCs hypothesised to share a common origin. This group, recently proposed to form a coherent system named Nephele, may represent the remnants of a single, massive accretion event.}{We analysed APOGEE DR17 data to select field stars with high-quality chemical abundances. We applied a Gaussian mixture model (GMM) in an 8D chemical abundance space to identify stars compatible with $\omega$~Cen chemistry. We then computed the orbital energy and angular momentum of these stars and applied a second GMM, calibrated on simulations from the e-TidalGCs project, to determine the kinematic compatibility with the predicted streams of $\omega$~Cen and the associated Nephele GCs.}{We identify 470 stars chemically compatible with $\omega$~Cen, of which 58 are also Al-rich, consistent with second-generation stars found in GCs. Of these, six stars show kinematics consistent with the predicted $\omega$~Cen stream, and additional stars are linked to the tidal streams of NGC 6205, NGC 6254, NGC 6273, NGC 6656, and NGC 6809. These findings suggest the presence of extended stellar streams that have not been previously detected. We also find overlap in chemical and kinematic properties between Nephele stars and the Gaia Sausage–Enceladus population.}{Our results suggest the presence of stellar debris associated with $\omega$~Cen and its candidate family of GCs. The combined chemical and kinematic analysis supports the scenario in which these systems originated in a common progenitor, which has now been disrupted. While uncertainties remain—particularly due to disc contamination and limited sky coverage—this work illustrates the potential of chemical and dynamical methods to trace the remnants of past accretion events in the inner Galaxy.}


   \maketitle
%

\section{Introduction}
The current system of Galactic globular clusters (GCs) today amounts to about 170 objects \citep[with \textit{Gaia} parameters from][]{vasiliev21}, mostly redistributed in the halo of our Galaxy, but also in the bulge and thick disc. With typical masses and sizes\footnote{As an estimate of the size, we made use the half-mass radius of the cluster.} of $1.3\times10^5$M$_\odot$ and 5~pc, respectively \citep{baumgardt18}, this system probably represents what is left of a population that must have been much more numerous in the past, and even more massive, and that has been reduced because of a variety of physical processes, including tidal effects \citep{gnedin97, murali97a, murali97b, fall01}. Tidal effects are in fact experienced by all finite-sized systems orbiting the Galaxy, due to the different gravitational attraction exerted by the latter in different regions of the system. This differential acceleration causes stars to escape mainly through the L1 and L2 Lagrange points, where gravitational and centrifugal forces are in equilibrium. Consequently, stars preferentially leak out through these gateways, giving rise to tidal tails that extend ahead of and behind the cluster along its orbit in the Galaxy \citep[see, for example,][]{capuzzo05, montuori07}. This is how tidal tails, or stellar streams, are formed. Over time, these structures grow in size (stars lost in earlier times tend to move further and further away from their parent system), and after billions of years of evolution they can reach extensions of tens of kiloparsecs in the Galaxy. This phenomenon has been reproduced in detail in numerous numerical works, which have highlighted how tidal tails and their properties depend on the characteristics of the progenitor (e.g. its mass) but also on its orbit in the Galaxy and the properties of its visible and dark mass distribution \citep[among others, see ][]{keenan75, oh92a, grillmair98, combes99, johnston02, kupper10, mastrobuono12, bovy14, erkal2015, pearson17, carlberg18, carlberg20}.

Observational evidence of the existence of tidal tails around GCs dates back some 30 years, when the first works \citep[see, for example, ][]{grillmair95, leon00} started highlighting the existence of tidal features around some Galactic GCs. The discovery of extended tidal tails around the halo cluster Palomar 5 \citep{odenkirchen01, odenkirchen03}  dates from this period and has arguably remained probably the most spectacular result of the existence of streams around such objects. Because of its characteristics, Palomar~5 is considered to be a cluster in the process of dissolution, and one that will probably not survive the next passage through the Galactic disc \citep[][]{dehnen2004}.
Many other discoveries of streams around GCs followed \citep[see, for example, ][]{belokurov06, chun10, sollima11, sollima12} and were boosted by the release of the \emph{Gaia} catalogues \citep{gaia16, gaia18, gaiaEDR3, gaiadr3}, which made it possible to discover many new stellar streams, some of which associated with known GCs, and to confirm and extend the study of already known ones \citep[see, for example,][]{navarrete17, malhan18a, malhan18b, ibata19a, ibata19b, ibata21, piatti18, piatti21b, piatti21a, thomas20, boldrini21}. 
Despite the \emph{Gaia} data being available for the whole sky, however, the streams discovered so far are mostly confined to absolute latitudes $|b| \gtrsim 15$~deg where the field density is significantly lower than in the bulge, disc, and innermost halo regions, with those regions being particularly inconvenient for the search of streams also due to dust obscuration.  However, a large number of GC streams are also predicted there, even if their morphology is expected to be more complex than that of streams emanating from halo clusters \citep{ferrone2023, pearson2024, grondin2024}.\\

Among the first GCs around which signs of tidal tails were sought was the $\omega$~Centauri cluster. This cluster has a very peculiar set of characteristics: its mass is ten times greater than the average mass of Galactic GCs \citep{baumgardt18}; its stars show a very wide spread in metallicity \citep[see, for example,][]{johnson2010, pancino11, marino11, nitschai23}, accompanied by a possible spread in age \citep{villanova07, villanova14, clontz2024}; and its kinematics is peculiar  \citep{meylan86, merritt97, norris97, vandeven2006, pancino07, bianchini13, kamann2018, sanna20} with part of its stars in counter-rotation \citep{pechetti24}. All the properties have long led to the suspicion that it is the remnant of a dwarf galaxy that was accreted by the Milky Way in the past \citep[][for early works]{lee99, majewski00, carraro00}. Of this galaxy, $\omega$~Cen would have constituted the nuclear regions, which would have survived the destruction \citep{bekki03, tsuchiya03, tsuchiya04}, appearing to us in the guise of a GC, however atypical. The search for tidal tails around $\omega$~Cen began with the work of \citet{leon00}, who pointed out the presence of such structures outside the cluster tidal radius, with an inclination roughly perpendicular to the Galactic plane. These results were partly questioned by \citet{law03}, due to the strong differential reddening in the region where $\omega$~Cen tails were found by \citet{leon00}.  The search for streams around this cluster was subsequently continued with the works of \citet{dacosta08} and \citet{marconi14}, until the discovery of the Fimbulthul stream by \citet{ibata19b} in \emph{Gaia}~DR2 data. This stream was subsequently shown to be related to $\omega$~Centauri \citep{ibata19a}, of which it would constitute part of the trailing tail. \citet{simpson20} explored GALAH data \citep{desilva2015galah} to find two additional members of Fimbulthul, based in particular on their peculiar chemical abundances. Chemical abundances have also been used by \citet{lind15} and \citet{fernandez15} to link some individual stars found in the Galactic field to the cluster.\\

In addition to the stars lost by $\omega$~Cen GC along its current orbit in the Galaxy, more recently, using spectroscopic data and \emph{Gaia} astrometry, the search for traces of the accretion of $\omega$~Cen's progenitor in the Milky Way has begun, in particular using the chemical abundances of stars and GCs as chemical markers of their origin.
Following this approach, in \citet{pagnini2025} we used the APOGEE~DR17 data \citep{abdurro22} and in particular the catalogue of GCs published by \citet{schiavon2023} to search for the population of GCs that once belonged to the progenitor of $\omega$~Cen, and which would later be accreted in the Galaxy. For this, we searched, among Galactic GCs,  those chemically compatible with $\omega$~Centauri, i.e. clusters that overlap with $\omega$~Centauri in part of its chemical abundance space.  If $\omega$~Cen is indeed the nucleus of an accreted galaxy, we assumed that its chemical abundances should be representative of at least a part of its progenitor, as it is the case for M~54 and the nuclear star cluster of the Milky Way, whose chemical patterns overlap with those of their respective host galaxies, i.e. Sagittarius dwarf \citep[as shown by][]{pagnini2025} and the Milky Way \citep[as shown by][]{nandakumar2025, ryde2025}. Following this approach, we were able to highlight the strong link that binds the GCs NGC~6656, NGC~6752, NGC~6254, NGC~6809, NGC~6273 and  NGC~6205 to $\omega$~Cen and interpreted this link as evidence that all these clusters formed in the same progenitor galaxy, which has been long searched for and which we named Nephele. This work was followed by that of \citet{anguiano2025}, who --  using APOGEE~DR17 data as well --  suggested that even among field stars there are some with chemistry compatible with $\omega$~Cen, and that these field stars are also redistributed over a wide range of energies and angular momenta, in agreement with \citet{pagnini2025}'s findings for GCs.\\

In this paper we continue our analysis of the APOGEE data, in particular by exploiting the unique opportunity that this survey offers to compare in a ‘self-consistent’ manner the chemical abundances of stars from different clusters with those of stars in the field, to search, among the latter, stars that once belonged to Nephele, either to its field population or to its GCs. 
By using the same methodology adopted in \citet{pagnini2025}, we exploit the APOGEE~DR17 catalogue and find 470 stars which have a high probability to be chemically compatible with $\omega$~Cen and the Nephele accretion event, 58 of which are Al-rich stars, as typical of second generation GC stars \citep[see, for example, ][]{gratton01, pancino17}. By comparing the kinematic properties of these stars to the numerical models of Galactic GCs' streams of \citet{ferrone2023}, we conclude that, of these 470 stars, 6 have kinematics compatible with the stream of $\omega$~Cen  GC itself, 9 with NGC~6656, 1 with NGC~6254, 2 with NGC~6809, 1 with NGC~6273, and 1 with NGC~6205. Instead, no kinematically compatible stars were found for NGC~6752. 

This paper is structured as follows: in Sect.~\ref{obsdata}, we discuss the observational sample analysed in this work, in Sect.~\ref{methods} we explain the method used to identify field stars belonging to $\omega$~Cen stream and other Nephele's GCs streams, and in Sect.~\ref{results} we present the results, first focusing on the selection of stars chemically compatible with $\omega$~Cen. We then operate further selections on the kinematics of this sample, to restrict the search of stellar streams. In Sect.~\ref{discussion}, we discuss the novelties of our results, also in the context of other recent studies on the subject, and finally, in Sect.~\ref{conclusions} we derive our conclusions.

Before continuing, a word on the nomenclature adopted in this work. Every time we refer in the following to  the $\omega$~Cen stream we refer to stars lost by the $\omega$~Cen GC, along its current orbit in the Galaxy. The same applies when we refer to other GC streams. Because, however, as recalled above, there are strong hints that  this cluster is the left-over of a more massive galaxy, whose accretion into the Milky Way possibly started several Gyr ago, we refer to this galaxy, the  $\omega$~Cen progenitor, as Nephele, as already done in \citet{pagnini2025}\footnote{This may seem like a mere semantic difference, but the possibilities described above (i.e. stars in the Nephele field, stars in the $\omega$~Cen stream, or in the streams of one of the six clusters that \citet{pagnini2025} have associated with Nephele) can be associated with physical processes acting on different timescales: the stars in the Nephele field may have been lost even at very early times, when the interaction between Nephele and the Milky Way began, while those associated with the $\omega$~Cen stream or one of the other clusters, and which are still dynamically coherent with them, may have been lost from the clusters themselves at more recent times.}. Beyond its stars, Nephele -- as stated before -- would have brought into the Galaxy also its system of GCs. It is this whole  ecosystem - made of  Nephele  stars, its GCs, and the streams associated to these latter —  that we want to investigate in this paper.


\section{Observational data}\label{obsdata}

To select the sample of field stars, we exploit the APOGEE~DR17 catalogue \citep[][]{abdurro22} applying the following selection criteria :
\begin{enumerate}
\item a signal-to-noise ratio $\tt{SNREV} > 70$;
\item temperatures in the range $\rm 3500\,K < T_{eff} < 5500\,K$ and surface gravities  $\rm logg < 3.6$; 
\item \tt{APOGEE STARFLAG} and \tt{APOGEE STARBAD} $= 0$.
\end{enumerate}
From this sample of field stars, we further removed stars which have a high probability to belong to GCs, i.e. we removed all stars in the APOGEE~DR17 catalogue with \texttt{VB\_PROB} $\geq 0.9$ according to \citet{vasiliev21}. After these selections, our final sample of field stars comprises 197265 stars. As in \citet{pagnini2025}, for the GCs data of
the Nephele’s family, we make use of data from the APOGEE Value Added Catalogue (VAC) of Galactic GC stars \citep[see][]{schiavon2023} using the membership probabilities from \citet{vasiliev21}. These probabilities are obtained from a mixture--model analysis of \emph{Gaia} EDR3 positions, parallaxes and proper motions combined with the structural parameters of each cluster (see their Sect.~2).
In this work we adopt stars with {\tt VB\_PROB} $\ge 0.9$ as high confidence members, following the practice of \citet{vasiliev21}, who use stars with membership probabilities $\gtrsim 90$ per cent as high-confidence cluster members in their error analysis (their Sect.~3).
Among these stars, only those satisfying the previous criteria and stars which have, according to \citet{vasiliev21}, a high probability of being members of Nephele's GCs (\texttt{VB\_PROB} $\geq 0.9$), have been used for this study.

Besides detailed chemical abundances for 20 species, and radial (i.e. line-of-sight) velocities, the APOGEE catalogue provides also proper motions from \emph{Gaia}~EDR3 release \citep{gaiaEDR3}. As for the distances of individual stars, we make use of those provided by the ASTRONN\footnote{\url{https://www.sdss4.org/dr17/data_access/value-added-catalogs/?vac_id=the-astronn-catalog-of-abundances,-distances,-and-ages-for-apogee-dr17-stars}} catalogue where distances have been determined as described in  \citet{leung2019a}.
We used a right-handed Galactocentric frame that leads to a three-dimensional velocity of the Sun equal to $\rm [U_{\odot}, V_{\odot}, W_{\odot}] = [11.1, 248.0, 8.5] \,km\, s^{-1}$ \citep[][]{horta2023}. We assumed the distance between the Sun and the Galactic centre to be $\rm R_{\odot} = 8.178 \,kpc$ \citep[][]{gravity2019}, and the vertical height of the Sun above the midplane to be $\rm Z_{\odot} = 0.02 \,kpc$ \citep[][]{bennett2019}.

\section{Methods}
\label{methods}
Our analysis is divided into two main steps: first, we search for Galactic field stars that are chemically compatible with $\omega$~Centauri, to identify stars that are likely to belong to Nephele. Then we refine this sample by looking for those that are also kinematically compatible with the streams of Nephele's clusters, making use of the numerical simulations presented in \citet{ferrone2023}.

\subsection{Searching for Nephele stars in the field: Chemical filter}\label{GMMChem}
\begin{figure*}\centering
\includegraphics[clip=true, trim = 20mm 87mm 20mm 85mm, width=0.33\linewidth]{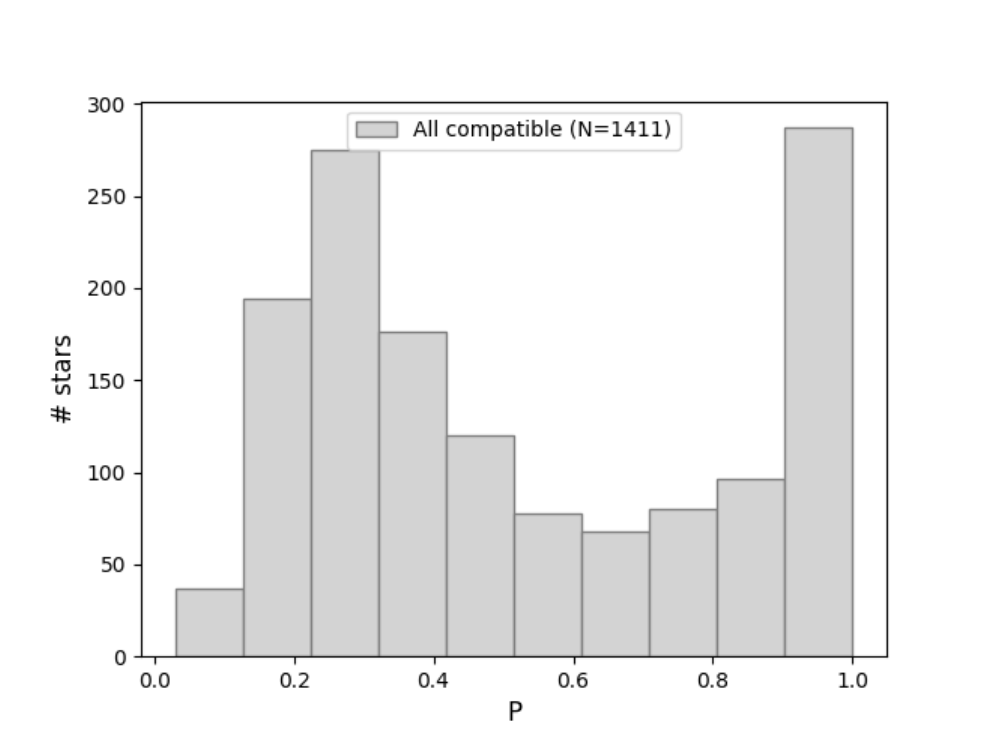}
\includegraphics[clip=true, trim = 3mm 0mm 0mm 0mm, width=0.31\linewidth]{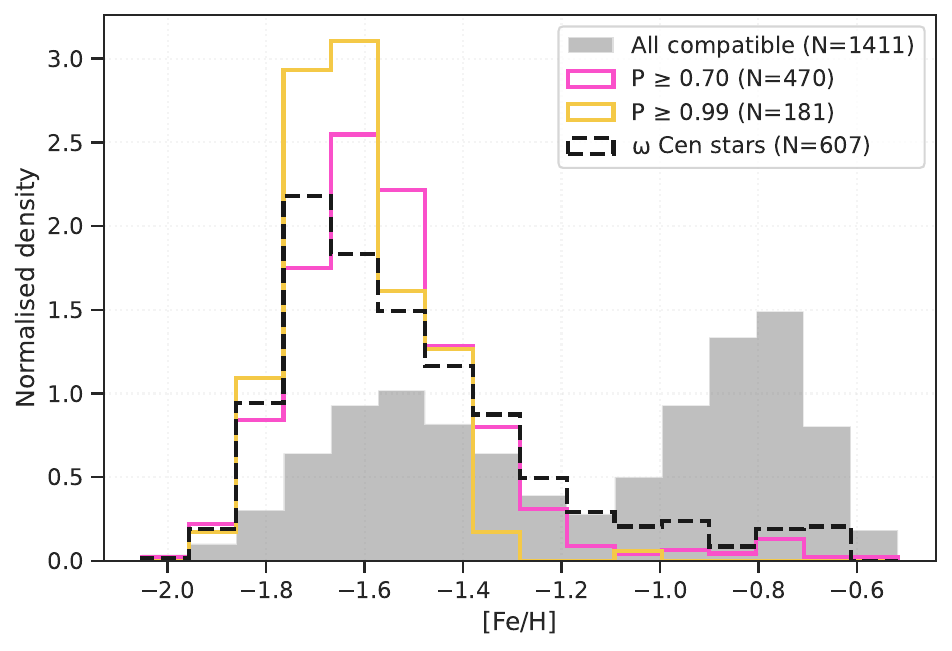}
\includegraphics[clip=true, trim = 4mm 5mm 0mm 4mm, width=0.32\linewidth]{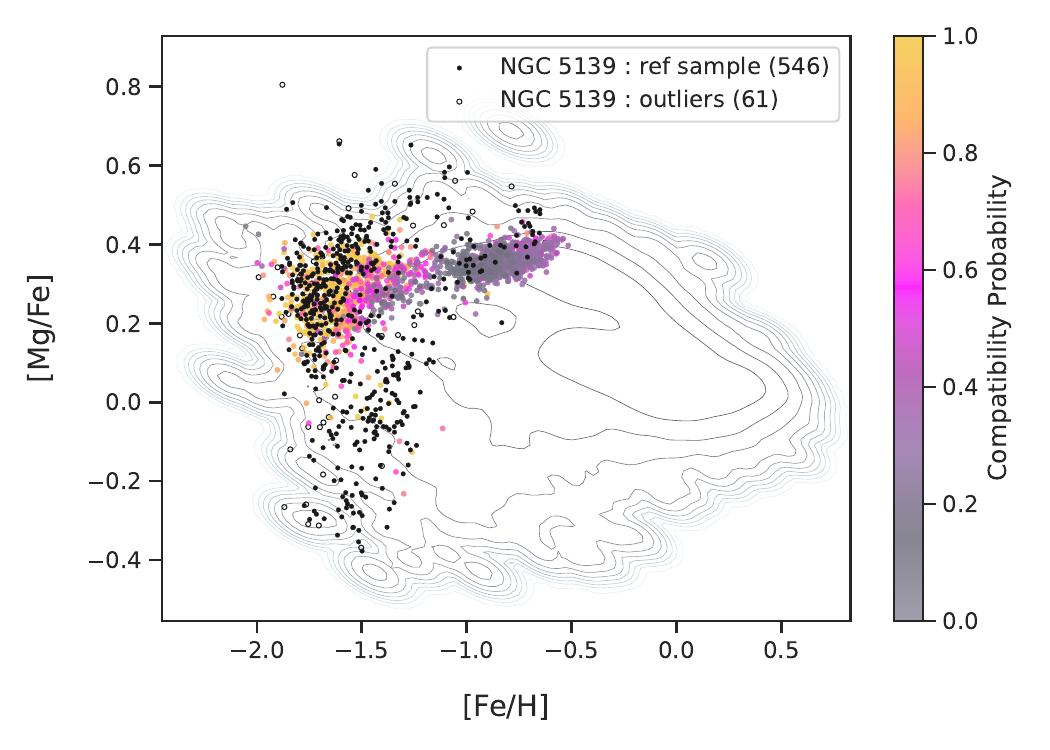}
  \caption{\textbf{Left panel:} Probability distribution of all stars in our sample which do not have a null probability of being chemically compatible with $\omega$~Cen, according to the GMMChem model. \textbf{Middle panel:} [Fe/H] distribution of all stars with a probability $P > 0$ (grey), $P \geq 0.7$ (magenta), and $P \geq 0.99$ (yellow)  to be chemically compatible with $\omega$~Cen (black). \textbf{Right panel:} Distribution in the [Mg/Fe]-[Fe/H] plane of $\omega$~Centauri stars (black points), and of stars chemically compatible to it, coloured-coded by the probability of their compatibility $P$. Outliers, defined as stars with a probability density below the 10th percentile threshold, are shown as empty black circles for $\omega$~Cen and as contours for field stars. Stars $0 < P < 0.7$ mostly distribute in the region occupied by $\alpha$-enhanced thick disc stars; these are interpreted as contaminants from the disc.}\label{fig:prob}
 \end{figure*}

To assess the chemical compatibility of field stars with $\omega$~Cen, we make use of the same Gaussian Mixture Model approach\footnote{In particular we make use of the Gaussian Mixture class available in sklearn, as part of the clustering methods of the scikit-learn python package, see \url{https://scikit-learn.org/stable/modules/generated/sklearn.mixture.GaussianMixture.html\#sklearn.mixture.GaussianMixture}} used in \citet{pagnini2025}, here referred as GMMChem. We consider the 8D abundance space tested in that work and defined by [Fe/H], $\alpha-$elements as  [Mg/Fe], [Si/Fe] and [Ca/Fe], light+odd-Z elements as  [C/Fe], [Al/Fe] and [K/Fe], and an iron-peak element as [Mn/Fe], and we keep only stars which have all the abundance flags of these elements equal to zero. With this choice, the number of field stars used for the analysis described below reduces to 176965. As in \citet{pagnini2025}, the distribution of $\omega$~Cen in the 8D abundance space is fitted with an increasing number of gaussian components. By minimising the Bayesian information criterion (BIC), we retrieve that the number of components that best reproduces the 8D distribution of $\omega$~Cen is five (see  Figure~1 in \citet{pagnini2025} for $\omega$~Cen distribution in the [Mg/Fe]-[Fe/H] plane). We then apply this model to the sample of field stars defining chemically compatible stars as those whose log-likelihood exceeds or is equal to a threshold, which we established as the 10th percentile of the log-likelihood distribution of the training dataset, namely $\omega$~Cen. The subset of $\omega$~Cen stars whose log-likelihood exceeds the defined threshold is designated as the ``reference sample'', whereas stars with a probability density below this threshold are classified as outliers. Similar results were obtained also varying this threshold between the 5th and the 20th percentile. To account for the uncertainties in the abundances of both field stars and $\omega$~Cen, we applied a bootstrap resampling of the data. For each star, we estimated a probability of compatibility with $\omega$~Cen by recording the fraction of bootstrap realisations in which the star satisfied the compatibility criteria, i.e. $\rm P = N_{compatible}/N_{boot}$, where $\rm N_{boot}= 100$ is the number of bootstrap realisations. The list of 1411 stars in the APOGEE~DR17 catalogue which appear to have a probability $P > 0$ to be chemically compatible with $\omega$~Cen, according to the GMMChem model, is given in Table~\ref{table:Nephele_ID}, in which we report also the associated probability $P$. The distribution of probabilities is shown in Fig.~\ref{fig:prob}, left panel, from which we can see that this distribution shows a first peak at P$\sim 0.25$, followed by a minimum at $P \sim 0.7$, after which the distribution rises up to $P=1$. Stars having a moderate probability to be chemically associated to $\omega$~Cen are, on average, more metal-rich than $\omega$~Cen itself (see Fig.~\ref{fig:prob}, middle panel) and are mostly redistributed in the $\alpha$-enhanced thick disc (Fig.~\ref{fig:prob}, right panel). We interpret these stars as contaminants from the disc, which -- especially at these high metallicities -- shows chemical patterns which are close to those of the metal-rich population of $\omega$~Cen and thus can sometimes be classified by the GMMChem model as chemically compatible with it. To remove this contamination, which is significant in terms of number of stars, for the following of this work, we retain as candidate members only a set of 470 stars with probabilities $P \geq 0.7$, which we refer to as Nephele candidate stars in the field. Already with this selection, most of the metal-rich contamination is removed (see Fig.~\ref{fig:prob}). We also adopt a second, more restrictive definition, selecting only stars with $P \geq 0.99$. This leaves us with 181 stars, which constitute the Nephele golden sample. Throughout the following we refer to stars with $P \ge 0.7$ as \emph{Nephele} candidates, while stars with $0 < P < 0.7$ are explicitly treated as contaminants and are not used in our quantitative estimates.

\begin{table}\centering
\caption{APOGEE ID of candidate Nephele stars and their probability of compatibility  $P$.}
\begin{tabular}{llcc}
\toprule
APOGEE ID & P & Al-rich & Nephele candidate\\
\midrule
2M00081703-5132067 & 0.26 & N & N \\
2M00093859-5202193 & 0.62 & N & N \\
2M00103225-2109408 & 0.94 & Y & Y\\
2M00115922+0111203 & 0.21 & N & N \\
2M00134023+8432153 & 0.70 & N & Y \\
2M00255788+0241423 & 0.39 & N & N \\
2M00361137+0111448 & 0.30 & N & N\\
2M00402516+4954415 & 0.39 & N & N\\
2M00421277-6729561 & 0.32 & N & N\\
2M00451197+1420100 & 0.43 & N & N\\
..... & ......& ....&....\\
\bottomrule
\end{tabular}
 \tablefoot{Only stars with a Nephele compatibility probability $P \geq 0.7$ (Sect.~\ref{GMMChem}) are considered as Nephele candidates. The column \texttt{Nephele candidate} flags these stars, and the column \texttt{Al-rich} indicates which stars belong to the Al-rich population, according to the abundance criterion defined by Eq.~\ref{eq:alrich}. (This table is available in full online, with Nephele and Al–rich
given as boolean flags matching the Y/N entries.). \label{table:Nephele_ID}}
\end{table}

\subsection{Searching for stellar streams of Nephele globular clusters: Kinematic filter}
\label{GMMKin}
Among the 470 candidate members of Nephele, some may belong to the stellar streams of GCs which \citet{pagnini2025} have associated to this accretion event, namely NGC~6656, NGC~6752, NGC~6254, NGC~6809, NGC~6273 and  NGC~6205. We recall indeed that all these clusters share part of the chemical patterns of $\omega$~Centauri, and that -- for this reason -- at this stage are indistinguishable from Nephele field stars. To determine whether, among the stars found in GMMChem, there are any that may have been lost from the clusters mentioned above, including $\omega$~Cen itself, we proceed as follows.   
We take advantage of the fact that APOGEE~DR17 and its Value Added Catalogue AstroNN provide full 6D information for stars: together with their positions in the sky, their distances to the Sun, proper motions and radial velocities are also available. We exploit this additional information, by computing their orbital energy (E). and angular momentum ($\rm L_z$) and comparing these properties with the predictions provided by the e-TidalGCs project\footnote{\url{https://etidal-project.obspm.fr}}, a library of simulations of Galactic GCs streams \citep[see][]{ferrone2023}. 

These simulations have modelled so far the formation and evolution of extra-tidal features around 159 Galactic GCs -- including all those of interest for the present study -- by making use of a test-particle methodology \citep[see][ for details]{ferrone2023}. In these models, particles experience the gravitational field from the GC and the Galaxy but do not generate any gravitational field themselves. For each GC, the errors in its distance, proper motions, and radial velocities are also taken into account by assuming Gaussian distributions of these quantities. This results in the generation of 50 random realisations of these parameters, and thus, for each cluster, in a given Galactic potential, we have in total 50+1 simulations. These simulations allow to predict the kinematic properties of the corresponding stream, taking into account also the uncertainties on the current clusters positions and velocities. In \citet{ferrone2023}, three different Galactic potentials have been used for orbit integration, two of which axisymmetric and one containing a rotating stellar bar. For the present study, we make use of the simulations of stellar streams integrated in one of the two axisymmetric potentials (called PII in \citet{ferrone2023} -- see also \citet{pouliasis17} for the original work describing this Galactic mass distribution\footnote{This model is consistent with several key observational constraints, including the stellar density in the solar neighbourhood, the scale lengths and heights of the thin and thick discs, the Galactic rotation curve from \citet{reid14}, and the vertical gravitational force as a function of Galactocentric distance \citep[see][]{pouliasis17}. However, being axisymmetric by construction, the model does not fully capture the non-axisymmetric mass distribution observed in the innermost few kiloparsecs of the Galaxy.}), in which the Milky Way is modelled by three components: two discs (a thin and a thick), both described by  \citet{miyamoto1975three} density distributions, and a spherical dark matter halo with the same functional form as adopted in the \citet{allen91} model. 

Guided by the e-TidalGCs simulations, we start by searching for stars - among those that are chemically compatible with $\omega$~Cen - which are also compatible with the kinematic properties of the streams of the Nephele's GCs. To enable a direct comparison with the observed stars, we compute orbital energy and angular momentum (E, L$_z$) from the simulated streams using the publicly available \texttt{galpy} \citep[][]{bovy2015} under the \citet{mcmillan2017} Galactic potential, the same model adopted for the APOGEE stars. In this way, both the simulations and the data are placed consistently in the same E-L$_z$ space.

We then applied a Gaussian Mixture approach, stacking all the 51 realisations of each simulated stellar stream, considering the 2-dimensional E-L$_z$ space defined above, which is fitted with an increasing number of gaussian components until the optimal number of components is retrieved by minimising the Bayesian information criterion. The resulting set of GMM models for Nephele's GCs in this 2D-kinematic space is hereafter referred to as GMMKin. We then apply these models to the sample of field stars, selected to be compatible with $\omega$~Cen with the GMMChem model, defining kinematically compatible stars as those whose log-likelihood exceeds or is equal to a threshold, which we established as the 10th percentile of the log-likelihood distribution of the training simulated dataset\footnote{We recall that energies are computed from the simulations' outputs using the \citet{mcmillan2017} Galactic potential — the same used for the data — to ensure consistency. Within the e-TidalGCs framework we have also repeated the GMMKin selection using the alternative axisymmetric Galactic potential from \citet{pouliasis17} (referred to as PI in \citealt{ferrone2023}). The total number of stream candidates remains 20, and the number of stars associated with each GC changes by at most one, so our GC–stream associations in the E–L$_z$ plane are robust within the family of e-TidalGCs models. In addition, \citet[][their Sect.~3.3.3 and Figs.~10–12]{ferrone2023} compare the e-TidalGCs-predicted streams with several observed stellar streams using the \texttt{galstreams} library \citep{mateu2023} and find good agreement over the range covered by current data, supporting the use of these simulations as our kinematic prior. Our stream identifications should nevertheless be regarded as conditional on the adopted orbital models and can be revisited as alternative suites of dynamical simulations become available.}
.

\section{Results}
\label{results}

\subsection{Nephele candidate stars }
 
\begin{figure*}\centering
\includegraphics[clip=true, trim = 3mm 15mm 0mm 3mm, width=0.33\linewidth]{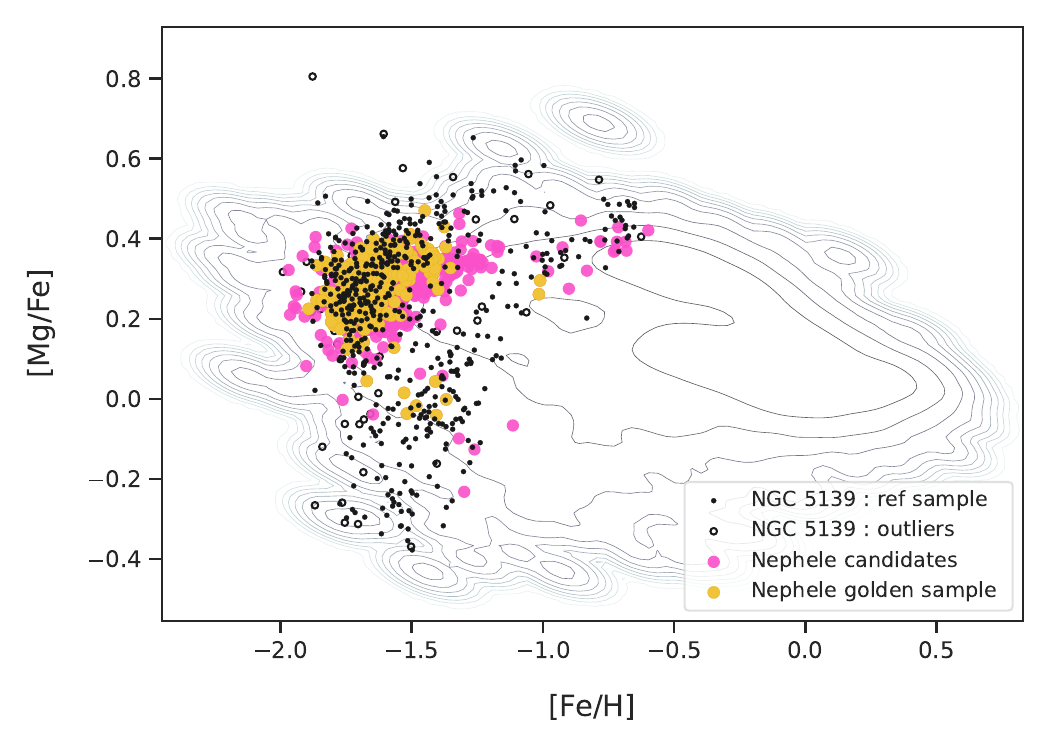}
\includegraphics[clip=true, trim = 3mm 15mm 0mm 3mm, width=0.33\linewidth]{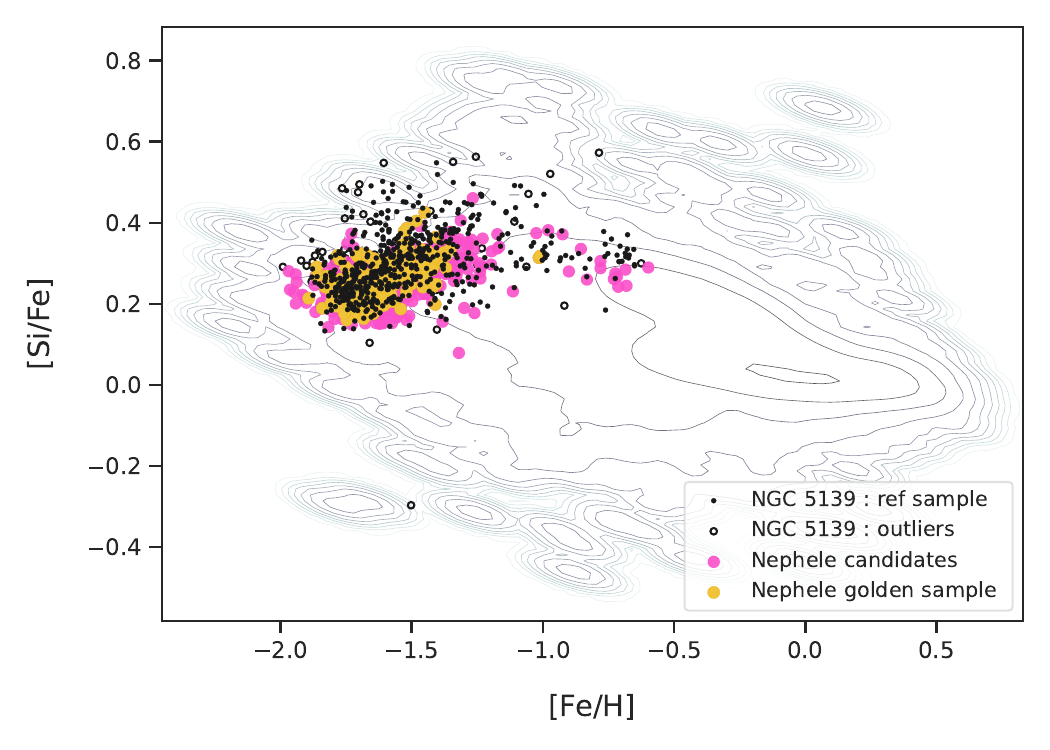}
\includegraphics[clip=true, trim = 3mm 15mm 0mm 3mm, width=0.33\linewidth]{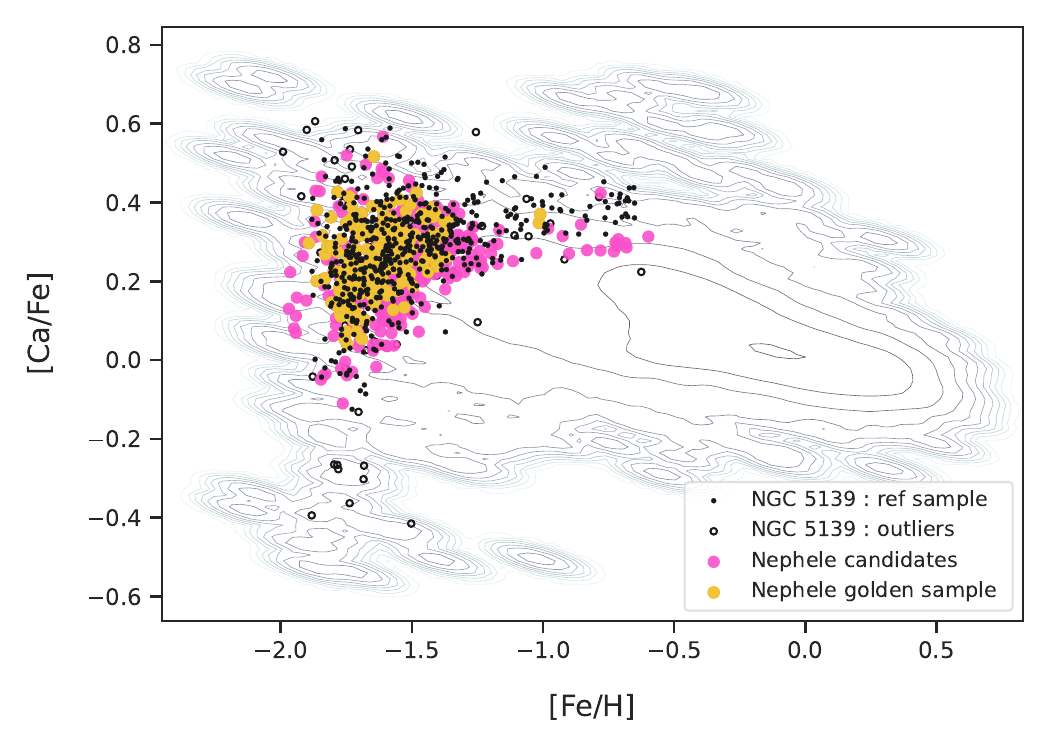}\par
\includegraphics[clip=true, trim = 3mm 15mm 2mm 0mm, width=0.33\linewidth]{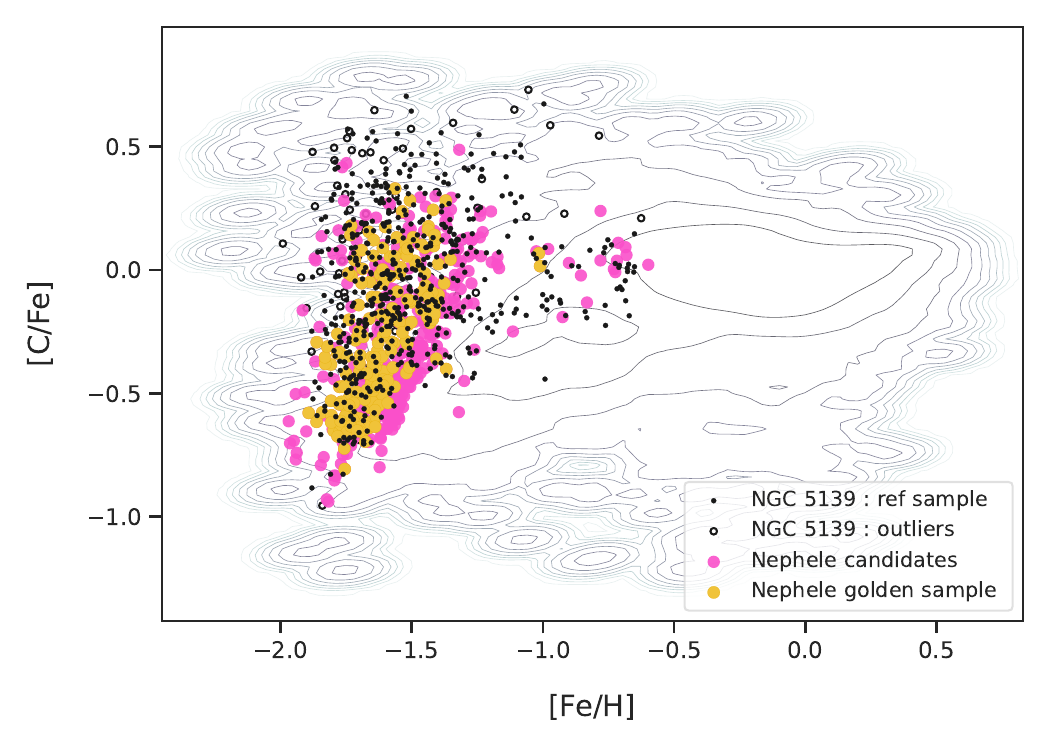}
\includegraphics[clip=true, trim = 2mm 15mm 0mm 1mm, width=0.33\linewidth]{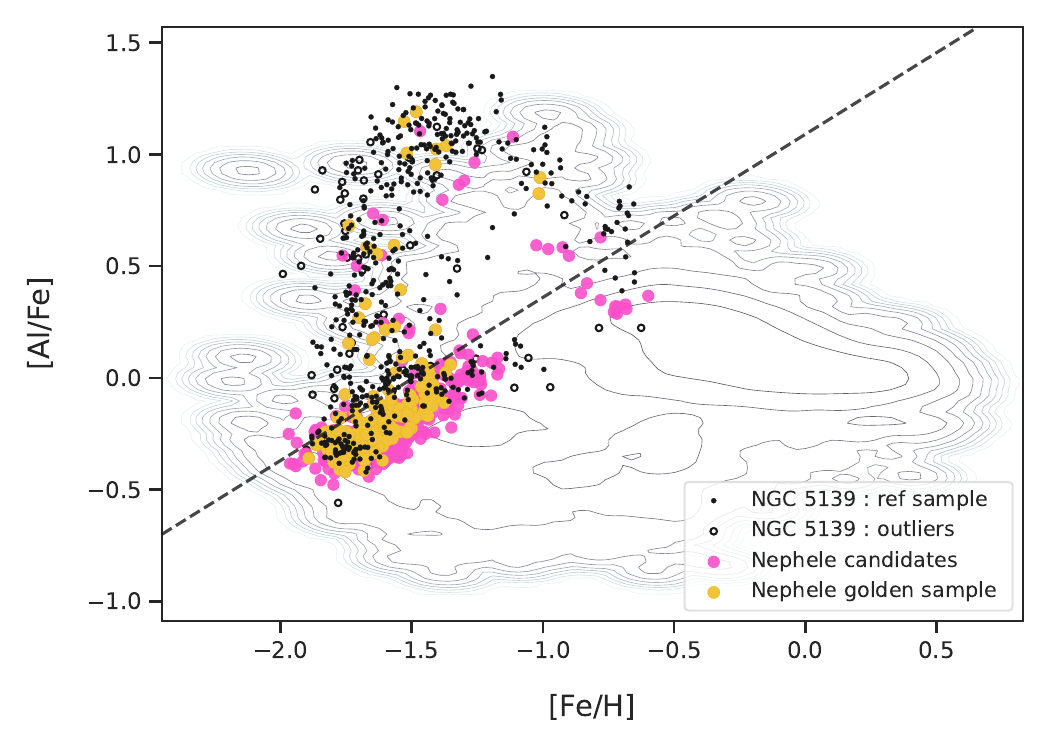}
\includegraphics[clip=true, trim = 2mm 15mm 0mm 1mm, width=0.33\linewidth]{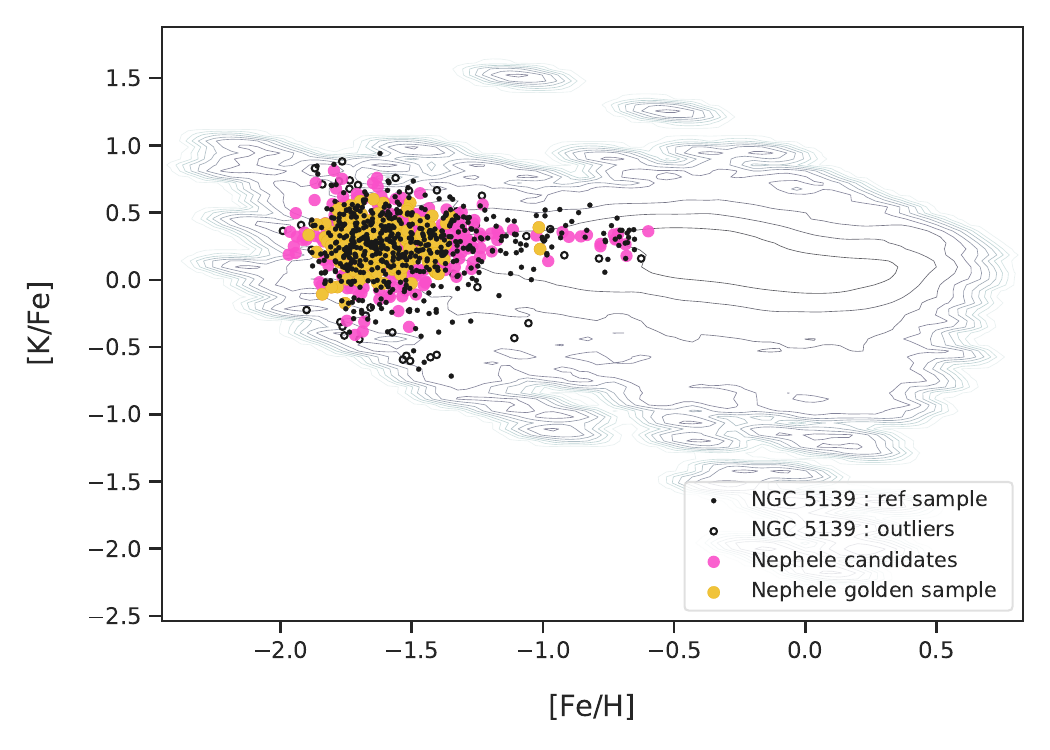}\par
\includegraphics[clip=true, trim = 1mm 0mm 0mm 1mm, width=0.33\linewidth]{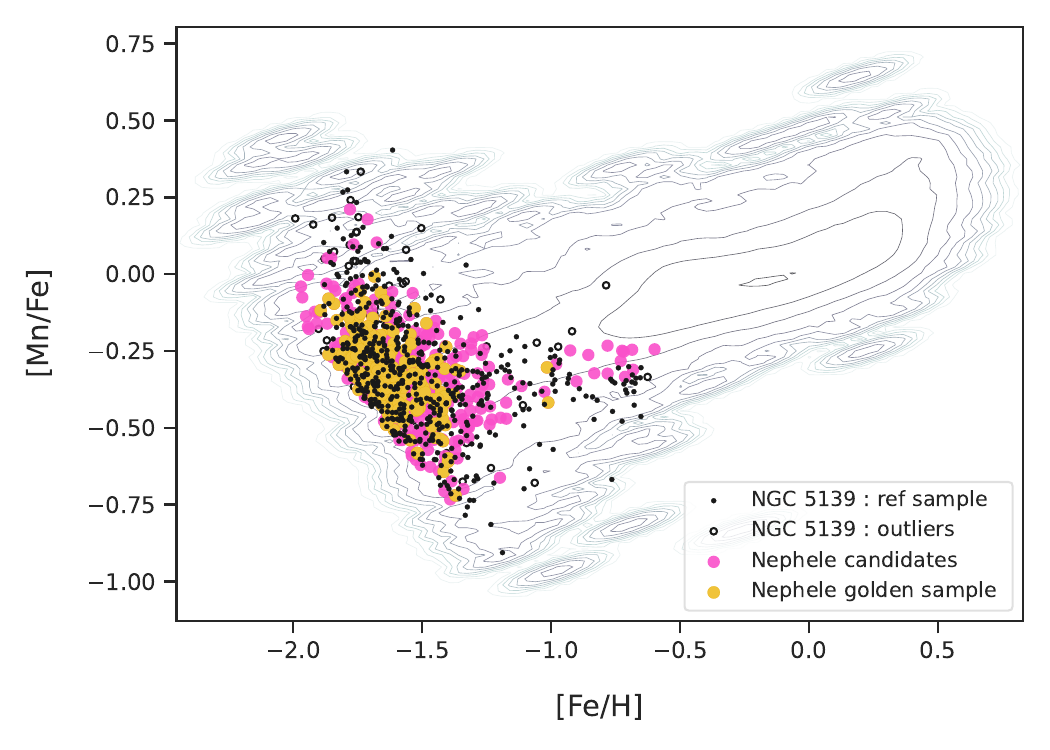}
  \caption{Chemical abundance relations for $\omega$~Cen members and APOGEE DR17 field stars. Filled black and coloured symbols represent respectively the $\omega$~Cen reference sample and the subset of field stars identified as chemically compatible with it by GMMChem. Outliers, defined as stars with a probability density below the 10th percentile threshold, are shown as empty black circles for $\omega$~Cen and as contours for field stars. The dashed line in the central panel indicates the division for our definition of Al-rich stars (see Eq.~\ref{eq:alrich}).}
              \label{gmm_field}%
    \end{figure*}

Figure~\ref{gmm_field} shows the results of the GMMChem in the 2D projections of abundance spaces for all the abundances used to build the model. The Nephele candidates, golden sample and outliers are highlighted, and the corresponding distribution of $\omega$~Cen stars (reference sample and outliers) is also reported for comparison. Candidate Nephele stars -- which we recall here are all field stars in the APOGEE~DR17 catalogue that satisfy the quality criteria defined in Sect.~\ref{obsdata} and which have a probability to be chemically compatible with $\omega$~Cen $P \geq 0.7$ -- follow closely the chemical patterns of $\omega$~Cen, up to its most metal-rich edges. Most of these candidate stars, however, redistribute in the high [Mg/Fe], low [Al/Fe] region where the bulk of $\omega$~Cen stars are (Fig.~\ref{gmm_field}). Note that this is not a natural consequence of the method used, because this latter requires that the stars be within the 8D volume defined by $\omega$~Cen chemical abundances, without imposing any constraints on the density of occupation of that volume. If we restrict the analysis to the Nephele golden sample (see also Fig.~\ref{gmm_field}), it follows even more closely the bulk of the chemical patterns of $\omega$~Cen, with a handful of stars still distributed in the low-Mg, high-Al sequences.

\begin{figure}\centering
  \includegraphics[clip=true, trim = 3mm 0mm 0mm 3mm, width=0.9\linewidth]{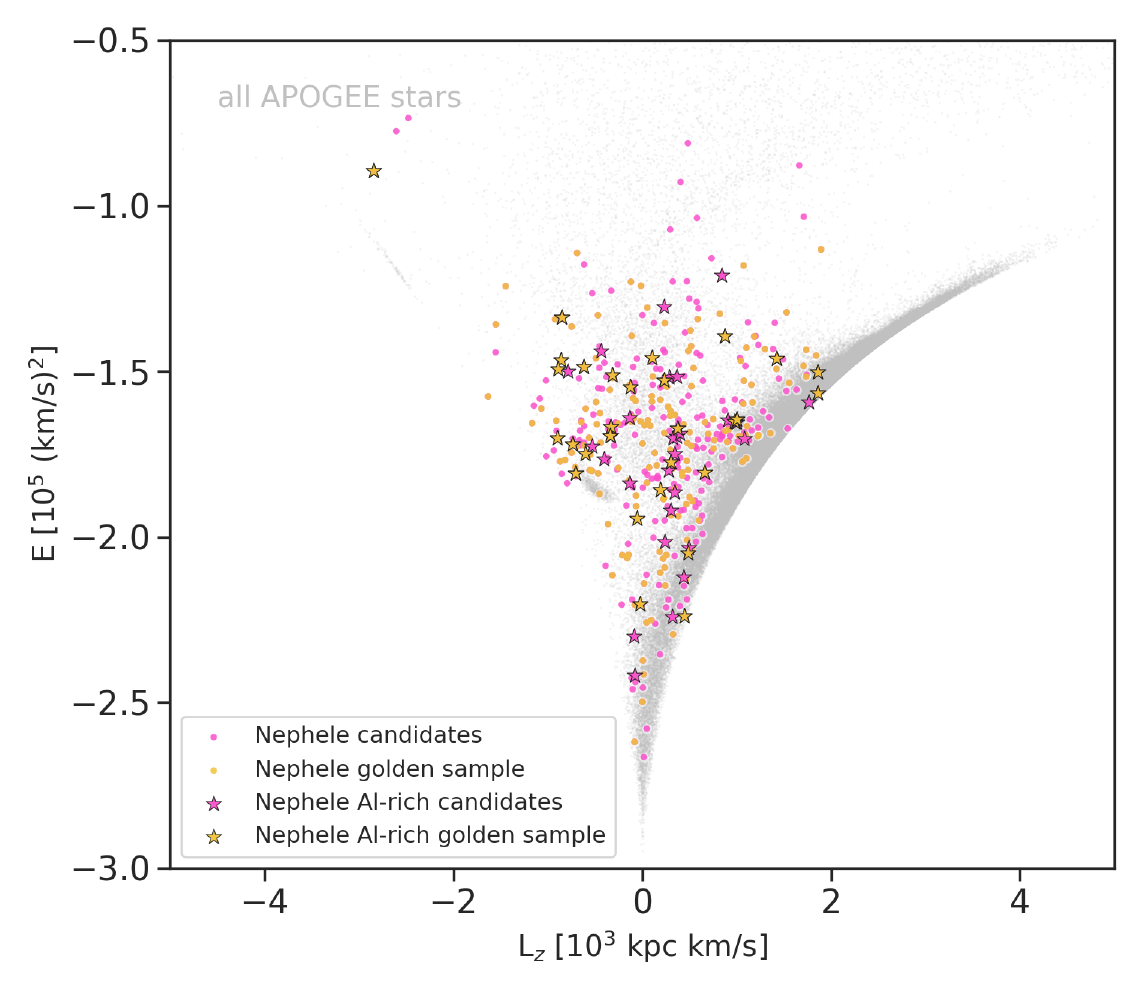}
  \caption{Distribution of Nephele candidates (magenta) and golden sample (golden) in kinematic spaces: orbital energy (E) versus the z component of the angular momentum (L$\rm_z$). Al-rich sub-samples (see definition in Eq. \ref{eq:alrich}) are shown as star symbols. For comparison, the distribution of all APOGEE stars is shown as grey points.}
              \label{e_lz}%
    \end{figure}

So far, our study has only used chemical information to find stars possibly lost from Nephele. It is therefore interesting to understand how these stars are redistributed in kinematic spaces, to see whether or not they show any consistency.
Spaces such E-L$_z$ space (that is orbital energy versus z component of the angular momentum)  have in fact been used for several years to identify the origin of stars and Galactic GCs (whether in situ or accreted), and to reconstruct the accretion history of our Galaxy (see the review by \citet{smith2016} and more recent works by \citet{kruijssen20, massari2019}, among many others) although this approach has been questioned on several occasions  \citep{jeanbaptiste17, pagnini2023, khoperskov23a, khoperskov23b, khoperskov23c, mori24, boldrini2025, mori2025}. In Fig.~\ref{e_lz}, we show the distribution in the E-L$_z$ space of Nephele candidate members and stars of the golden sample. It is interesting to see that both groups show an extended range of orbital energies and angular momenta, as expected if Nephele was massive enough at the time of its merger with the Milky Way \citep[i.e. mass ratio of about 1:10, see][]{jeanbaptiste17}, or if other events occurred afterwards to reshuffle the orbits of the accreted stars \citep[see, for example, the recent works by][and by Guillaume et al, in prep]{weerasooriya2025}. The fact that even the golden sample -- i.e. stars which have a probability larger than 99\% to be compatible with $\omega$~Cen -- shows such a broad extension, strengthens this result, which can be hardly interpreted as due to polluters (i.e. stars only marginally compatible with the $\omega$~Cen progenitor).
 Note that the extended distribution of  Nephele debris in the E-L$_z$ space has been already highlighted by \citet{anguiano2025}. We provide a more detailed comparison between our results and theirs in the Discussion section. One additional element that it is worth citing when discussing Fig.~\ref{e_lz} is that some of the Nephele debris, some of its golden sample stars included, show very prograde orbits, lying in the region of the E-L$_z$ space occupied by MW stars with disc-like kinematics.  This result also applies to Al-rich stars, which  have been selected as stars, among the  Nephele candidates, which satisfy the following relation (see also Fig.~\ref{gmm_field}, middle row, middle panel): 
\begin{equation}
\label{eq:alrich}
\rm [Al/Fe] > 0.73 \times [Fe/H] + 1.09.
\end{equation}
 These stars could be  signatures of the tidal disruption of GCs likely accreted and/or formed in situ \citep[][]{fernandeztrincado2020}. Indeed, it has been shown that this population could have been dynamically ejected in different orbital configurations from GC systems at similar metallicity, or possibly a massive system similar to $\omega$~Cen \citep[][]{meza2005, majewski2012}. \\

To summarise: so far, we have identified stars that are highly likely to have origins linked to the progenitor of $\omega$~Cen. However, since we have shown in \citet{pagnini2025} that some Galactic GCs also share part of $\omega$~Cen's chemical patterns, some of the stars identified so far as Nephele candidate stars (or golden sample stars) could actually be stars lost from these clusters and now part of their streams. Some stars could also be associated with the streams of the $\omega$~Cen GC itself.
 In the next section, we will therefore focus on understanding which of the Nephele candidates may in fact be stars linked to streams from the GCs of the Nephele family.

\subsection{From Nephele stars in the field to stellar streams of Nephele’s family
clusters}
\label{ocen_stream}

In the previous Section, we have discussed how stars chemically compatible with $\omega$~Cen are redistributed in the E-L$_z$ plane. 
GCs that we associated with $\omega$~Cen in \citet{pagnini2025} -- namely NGC~6205,  NGC~6254, NGC~6273, NGC~6656, NGC~6809 and NGC~6752 -- are also in the middle of this distribution. Which of Nephele's stars can be linked to the streams of these clusters? To answer this question, we used the GMMKin model  to determine which of Nephele's stars are kinematically compatible with the clusters in question. For this, we used the simulations in \citet{ferrone2023}, which offer predictions on the kinematics of the streams. Since the clusters are significantly less massive than dwarf galaxies, we expect them to conserve their orbital energies and angular momenta over timescales of a few billion years if no significant changes to the Galactic gravitational potential occur during this time. It should also be noted that among the simulations made by \citet{ferrone2023}, we only use those in which the streams develop in axisymmetric potentials, which therefore allow the conservation of the z component of the angular momentum.

\begin{figure}\centering
\includegraphics[clip=true, trim = 10mm 65mm 10mm 60mm, width=\linewidth]{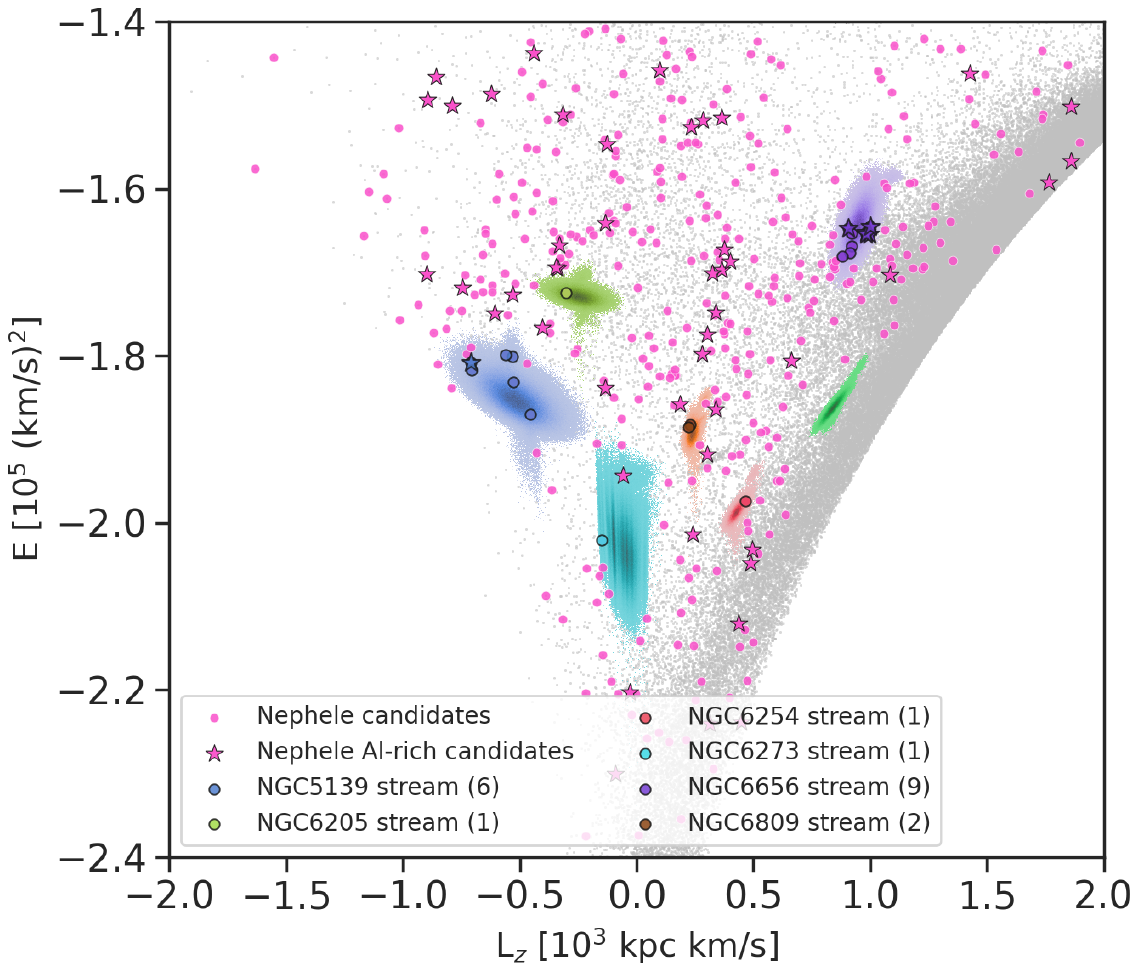}
  \caption{Distribution of \textit{simulated} Nephele's GCs and \textit{observed} field stars in the E-L$\rm_z$ space. All Nephele candidates and the Al-rich subsample (respectively as magenta points and magenta star symbols) are compared to simulated Nephele's GCs (colour-coded maps). Stars that result also kinematically compatible with Nephele's GCs according the GMMKin are identified as colour-coded symbols with their number reported in parentheses. For comparison, in both panels, the distribution of all APOGEE stars is shown as grey points.}
              \label{e_lz_sims}%
    \end{figure}

 Given these premises, Figure~\ref{e_lz_sims} shows the stars that, according to the GMMKin, are highly likely to be linked to the streams of the aforementioned clusters, and how they are redistributed in the E-L$_z$ plane. Table~\ref{stream_ID} in Appendix~\ref{app:chem_compatible} shows the APOGEE identifiers of these stream stars, together with their associated parent cluster. The Table also contains a third column, where we add additional details (for example, if the star is an interloper and not a robust stream candidate, see Sect.~\ref{neph_GCstreams} for an explanation and discussion on this point). The corresponding chemical abundance relations for \textit{Nephele's} GC members and their associated field stars selected by GMMKin are shown in the abundance planes of Figs.~\ref{ngc5139_chem}, \ref{ngc6205_chem}, \ref{ngc6254_chem}, \ref{ngc6273_chem}, \ref{ngc6656_chem}, and \ref{ngc6809_chem}, illustrating each GC and its debris.
 Our GMMKin finds 6 stars that are kinematically compatible with $\omega$~Cen, 1 with NGC~6205, 1 with NGC~6254, 1 with NGC~6273, 9 with NGC~6656, 2 with NGC~6809, and none with NGC~6752. When we restrict ourselves to Al-rich stars only (Fig.~\ref{e_lz_sims}), we find one star compatible with $\omega$~Cen, and five with NGC~6656. Overall, Al-rich stars constitute only about $17\%$ of the Nephele candidates (81 out of 470), whereas they account for $6/20 \simeq 30\%$ of the stars that are also kinematically associated with the streams of Nephele's GCs in our default GMMKin selection. Thus Al-rich stars are at least as frequent, and in fact somewhat enhanced, among the GC-stream candidates compared to the full Nephele sample, even though their absolute numbers remain small because they are relatively rare in APOGEE and must additionally satisfy the stringent GMMKin kinematic criteria (see also Appendix~\ref{app:chem_compatible}).

\begin{figure}
\centering
\includegraphics[clip=true, trim = 4mm 4mm 4mm 4mm, width=\linewidth]{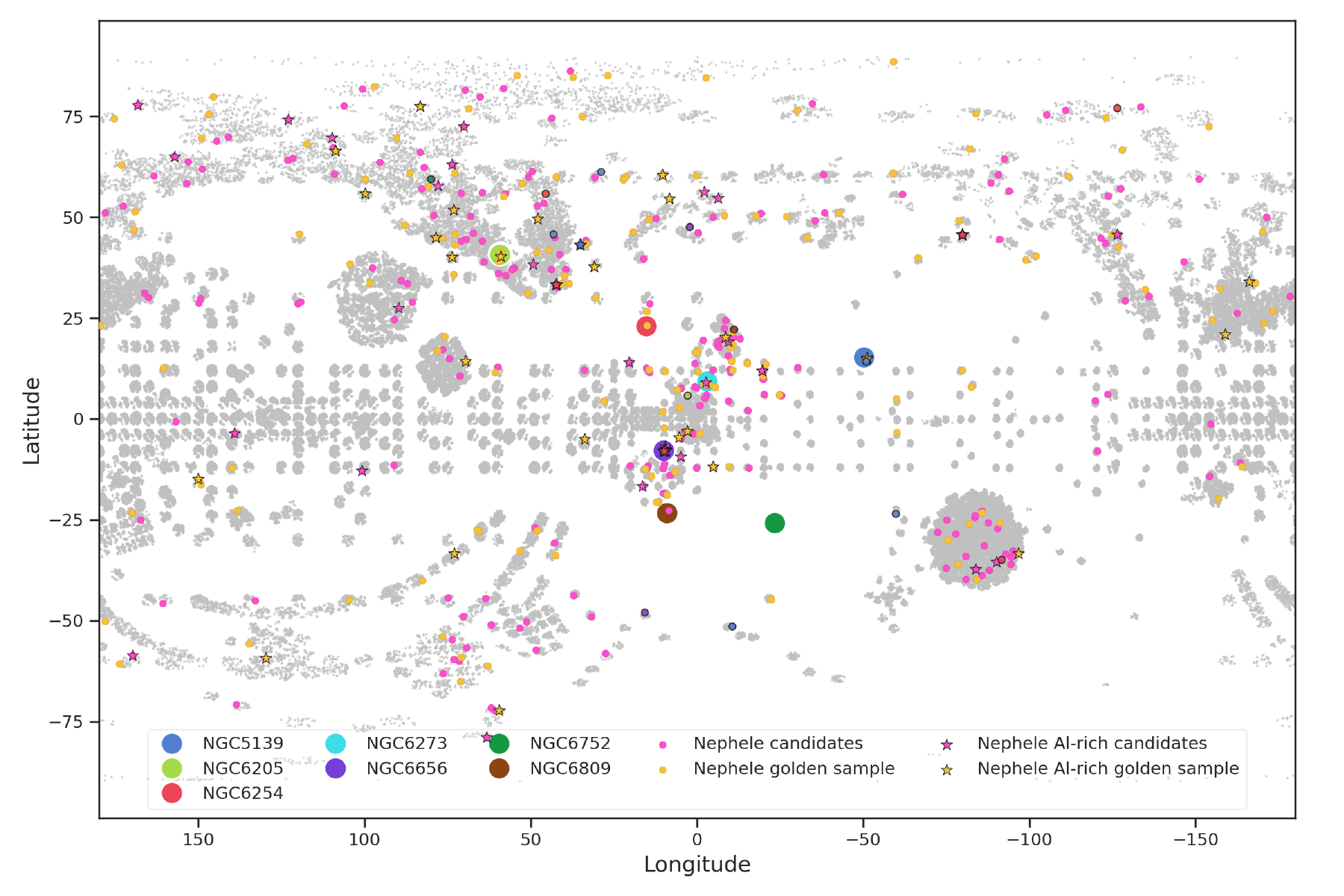}
\caption{Nephele's distribution in the sky. Nephele candidate stars (magenta symbols), its golden sample (golden symbols), $\omega$ Centauri and its family of GCs as identified in \citet{pagnini2025} (colour-coded circles), and their likely stream stars (colour-coded symbols) are shown together with all APOGEE stars (grey points).}
\label{Nephele_longlat}
\end{figure}

 Note that some of the stars found by the GMMKin method may actually be part of the clusters themselves. For example, we find one star compatible with $\omega$~Cen and four stars (3 of which are also Al-rich) compatible with NGC~6656 that have a \texttt{VB\_PROB}= -999 \rm thus, in the selection presented in Sect.~\ref{obsdata}, they are kept among the field population. According to the RV-based membership presented in \citet{schiavon2023}, instead, their probability of belonging to their corresponding cluster is higher than 0.97 (see Tab.~\ref{stream_ID}). The fact that these stars are retrieved as being associated with the cluster progenitor can be seen as a sanity check of our adopted procedure.
Finally, Fig.~\ref{Nephele_longlat} reports the distribution in the sky of Nephele candidate stars, and of its golden sample, together with $\omega$~Centauri, its family of GCs, as identified in \citet{pagnini2025}, and their likely stream stars. The distribution of these stars is clearly influenced by the APOGEE footprint. However, it is interesting to see how Nephele's stars are  distributed across the entire sky, and how the stars in the streams can also be at large angular distances from the clusters we have associated them with, thus demonstrating the full potential of such an analysis.

\section{Discussion}\label{discussion}

The results presented in this paper are new, to our knowledge, in several respects.
A not negligible fraction of field stars at low metallicities is chemically compatible with $\omega$~Cen (16 \% of field stars with [Fe/H]$\le -1.6$ in APOGEE DR17 results indeed compatible with $\omega$~Cen with P $\geq$ 0.7). This discovery reveals the importance of the Nephele accretion in the early history of our Galaxy and complements the findings in \citet{pagnini2025} regarding the existence of at least six galactic GCs - NGC~6656, NGC~6752, NGC~6254, NGC~6809, NGC~6273 and  NGC~6205 - all associated with $\omega$~Cen and its progenitor galaxy. The stars and GCs associated with this accretion event are redistributed over a wide range of similar energies and angular momenta, suggesting that Nephele must have been massive relative to the Milky Way at the time of its accretion, and/or that other processes intervened after its accretion to redistribute its stars in the E-L$_z$ plane. In the following of this discussion, we start by placing our results in the context of previous works in the literature, focusing in particular on Nephele field stars, their link with Gaia Sausage Enceladus (hereafter GSE, Sect.~\ref{neph_gse}) and how our GMMChem selection  compares with that presented by \citet{anguiano2025} (see Sect.~\ref{neph_anguiano}). We then move to discuss the results concerning new candidate members of GC stellar streams and the advances with respect to previous works (see Sect.~\ref{neph_GCstreams}).  

\subsection{Nephele and Gaia Sausage-Enceladus}\label{neph_gse}

Nephele candidate stars redistribute over energies and angular momenta intervals usually adopted to define other accretions events, such as GSE, Heracles, Helmi Stream and Thamnos (see our  Fig.~\ref{e_lz} and Fig.~4 in \citet{horta2023} for comparison). The possibility that Thamnos contains stars chemically compatible with $\omega$~Cen has been recently discussed by \citet{mori2025}. In the following we rather comment on the overlap between GSE and Nephele stars, first because some works have suggested that $\omega$~Cen may be the nucleus of GSE which -- if confirmed  -- would imply that Nephele and GSE are the same progenitor galaxy, and secondly because GSE has been so far identified as one of the major ancient building block of our Galaxy \citep{nissen2010, belokurov2018, haywood2018, helmi2018}, making necessary to compare other ancient accretions with it.  

In previous studies aiming to identify the progenitor system of GCs in the Milky Way based on kinematic criteria, $\omega~$Cen has been proposed as the former nuclear star cluster (NSC) of the GSE accretion event \citep[e.g.][]{massari2019}. Interestingly, the set of clusters typically linked to $\omega~$Cen through such kinematic associations does not coincide with those we attribute to the Nephele progenitor. Nonetheless, one or two clusters from the Nephele group—specifically NGC 6205 and NGC 6656—have been associated with GSE in the literature \citep[][]{massari2019, callingham2022}, suggesting some degree of overlap or ambiguity in progenitor assignments.

The distribution in E-L$_z$ space of Nephele's field stars and clusters shown in Figures~\ref{e_lz}, \ref{e_lz_sims} suggest an overlap in kinematic spaces between these two systems. Indeed, for example in \citet{horta2023}, GSE is defined in the E-L$_z$ space as a substructure having $\rm | L_z | < 0.5 \,(\, \times\, 10^3\, kpc\, km\, s^{-1} )$ and $\rm-1.6 < E < -1.1 \,(\, \times \, 10^5\, km^2\, s^{-2})$. Such an overlap would not be surprising if Nephele and Gaia Sausage Enceladus were both massive galaxies relative to the MW, at the time of their accretion. Simulations indeed show that even independent systems, once accreted onto the MW, can be redistributed in similar regions of the kinematic spaces \citep[see e.g.][]{pfeffer2018, khoperskov23a, pagnini2023, mori24}. 

\begin{figure}\centering
\includegraphics[clip=true, trim = 3mm 3mm 3mm 4mm, width=\linewidth]{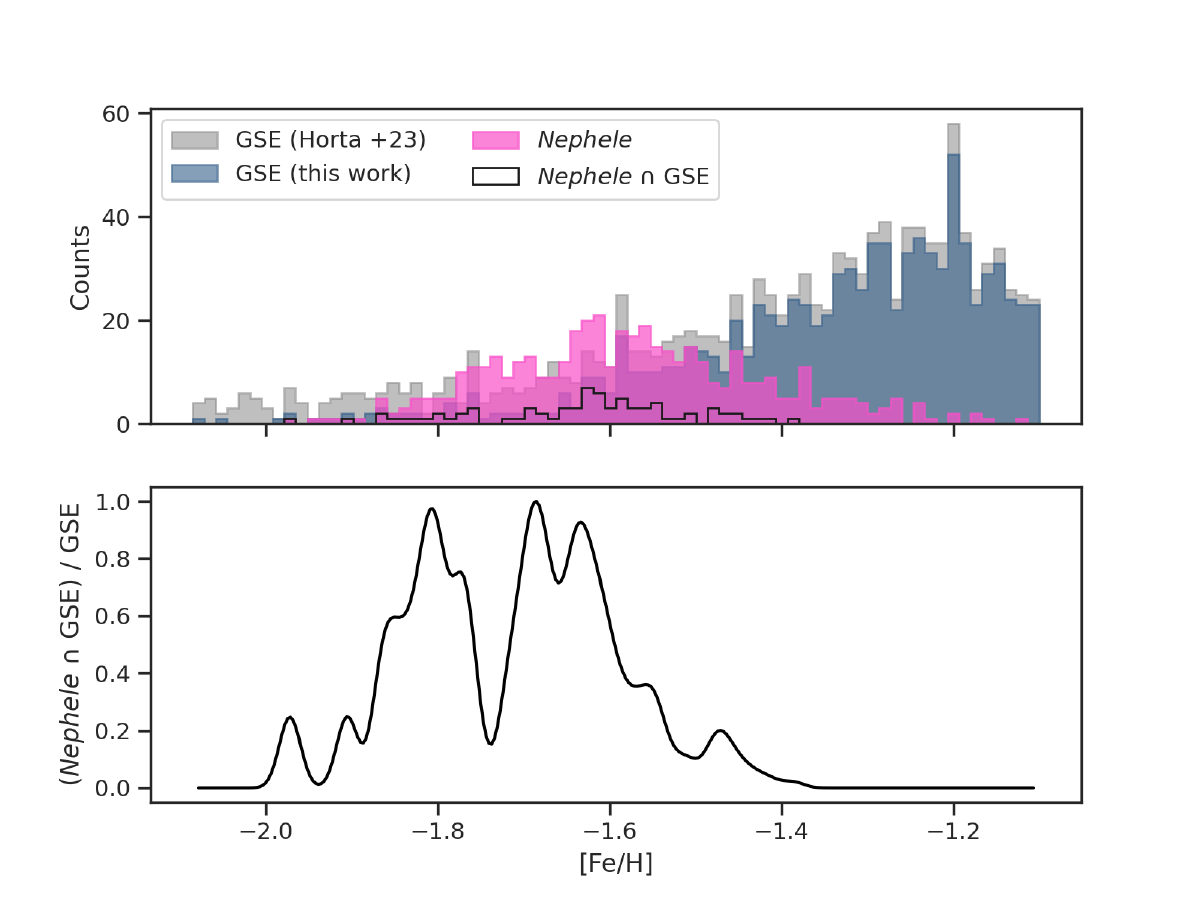}
  \caption{Metallicity distribution functions (MDFs) of Nephele and Gaia Sausage-Enceladus (GSE). \textbf{Top panel:} The MDFs of GSE as selected in \citet{horta2023} (in grey), the same definition of GSE plus our cuts in chemical abundances (see Sect.~\ref{discussion}, in dark grey), Nephele's candidates (in magenta), and Nephele's candidates that fall within the GSE selection (in black).
  \textbf{Bottom panel:} Contribution of Nephele among stars of GSE for each [Fe/H] bin. }\label{mdf_GSE}
 \end{figure}
 
The upper panel of Figure~\ref{mdf_GSE} shows the metallicity distribution function (MDF) of Nephele's candidates, in comparison with the GSE population as defined by \citet{horta2023}. Additionally, we include a refined selection in which we retain only stars with all abundance flags equal to zero for the elements used in our GMMChem analysis (see Sect.~\ref{GMMChem}). In this way we are consistent with the selections made for the Nephele field stars and can thus compare the two datasets. In the bottom panel of Fig.~\ref{mdf_GSE} we have indeed calculated, for each bin of [Fe/H], the contribution of Nephele's field stars among those of GSE. As we can see, the importance of Nephele increases as [Fe/H] decreases, reaching a maximum for -1.9 < [Fe/H] < -1.6, where the peak of $\omega~$Cen MDF is actually found \citep[see][]{pagnini2025}. This finding implies that the contamination of Nephele on the GSE sample may be significant up to [Fe/H] as high as $\sim -1.6$.  Such convergence raises the question of whether GSE and Nephele could in fact be one and the same (see a more detailed comparison of their abundance trends in Appendix~\ref{app:nephele_gse}). Few lines of evidence argue against this interpretation. In \citet{pagnini2025}, we showed that the chemical abundance patterns of $\omega~$Cen are fundamentally incompatible with those of the GCs traditionally associated with GSE—namely NGC 1851, NGC 1904, NGC 2298, and NGC 2808 \citep[see also][]{myeong2018, callingham2022}. These clusters lack the complex chemical signatures characteristic of $\omega~$Cen. Conversely, the clusters we associate with Nephele—NGC 6205, NGC 6273, NGC 6656, NGC 6752, NGC 6809, and NGC 6254—display abundance patterns that align closely with those of $\omega~$Cen across a multi-dimensional chemical space. This chemical coherence, not seen in the GSE-linked clusters, supports the interpretation that $\omega~$Cen and its associated system formed in a distinct progenitor. That said, the existence of Nephele field stars in common with the kinematic definition of GSE indicates that the overlap between these two systems is not negligible. This raises the possibility that some stars currently classified as part of GSE—especially those at lower metallicities—could in fact originate from Nephele, or that the two had similar accretion processes dynamically mixing in phase space. Importantly, it also underscores the challenge of defining GSE itself \citep[][]{carrillo2024}. The selection adopted in this work \citep[following][]{horta2023} is based purely on kinematics, but several other definitions in the literature incorporate additional chemical constraints \citep[e.g.][]{myeong19, naidu2020, feuillet2020, das2020}, often emphasising low-$\alpha$ and metal-poor populations on highly eccentric orbits. These varying criteria lead to partially overlapping but not identical stellar samples, complicating direct comparisons and can significantly affect which stars are attributed to GSE and how its global properties are interpreted. 
Finally, one last point for which -- based on our current knowledge -- we tend to rule out a link between the origin of $\omega$~Cen and that of GSE is the following.
 The NSCs of the MW and of Sgr  have chemical patterns similar to those of the host galaxy \citep{ryde2025, nandakumar2025, pagnini2025}. Although the reason for this similarity is not yet clear, current observational data (but note, based on only two NSCs) seem to indicate that the chemical patterns of NSCs are representative of those of the host galaxy. This is not true for $\omega$~Cen and GSE, for which - even recently - it has been shown that the chemical overlap is minimal \citep[see][]{mori2025}. If $\omega$~Cen was the NSC of GSE, based on the results of \citet{pagnini2025, ryde2025, nandakumar2025} we would expect to see a chemical similarity between their chemical patterns which is not observed.

Altogether, the chemo-dynamical properties of the Nephele debris and its associated GCs point to a progenitor in the massive-dwarf regime. The metallicity distribution of $\omega$~Cen and of the high-probability Nephele debris peaks at $\mathrm{[Fe/H]} \simeq -1.6$. Inserting this value into the mass--metallicity relation for accreted systems derived by \citet[][their Eq.~3]{Naidu2022} yields a characteristic stellar mass $M_\star \approx 3 \times 10^{7}\,M_\odot$, with the intrinsic scatter implying a plausible range of roughly $M_\star \sim 10^{7} \text{--} 10^{8}\,M_\odot$. A progenitor stellar mass of order $10^{8}\,M_\odot$ for the system hosting $\omega$~Cen is also consistent with dynamical and chemical evolution models for its putative parent dwarf galaxy \citep[e.g.][]{bekki03}. This places Nephele well above the regime of ultra-faint dwarfs and low-mass classical dSphs, and into the domain of relatively massive accreted systems. The number of massive GCs we associate with Nephele is also compatible with the observed diversity of GC systems in dwarf galaxies at similar stellar masses: the sample compiled by \citet{eadie22} shows that galaxies with $M_\star \sim 10^{7}$--$10^{8}\,\mathrm{M_\odot}$ span a wide range of GC specific frequencies, from systems with no known GCs to others hosting rich GC systems with $N_{\rm GC} \sim 10$ \citep[see Fig.~1 in][]{eadie22}.\footnote{For the Local Group dwarf galaxies in the \citet{eadie22} sample at $M_\star \sim 10^{7}$--$10^{8}\,\mathrm{M_\odot}$, the majority have $N_{\rm GC} \approx 10$.} In this context, a progenitor with $\omega$~Cen plus a handful of additional massive clusters lies toward the high end of the observed distribution. The robustness of the individual GC memberships is discussed in detail in \citet{pagnini2025}; here we focus on the aggregate chemo-dynamical picture, noting that some level of field contamination (e.g. in situ halo or GSE debris) in our debris sample is expected but does not affect the interpretation of Nephele as the relic of a relatively massive, $\omega$~Cen-like accretion event.

  \subsection{Nephele and the $\omega$~Cen debris studied by \citet{anguiano2025}}\label{neph_anguiano}

The recent study by \citet{anguiano2025} also analysed APOGEE~DR17 data to identify stars originally associated with $\omega$~Cen that now orbit within the Galactic field. However, there are differences in our approaches. \citet{anguiano2025} apply different quality selection cuts to the  APOGEE~DR17 catalogue -- among others, they do not restrict the search for $\omega$~Cen debris to giants, as we did, but included also dwarfs in their sample. They make use of a different method from ours -- a neural network
trained on APOGEE observations of $\omega$~Cen's core, while we use a GMM. They use a set of chemical abundances to train the neural network which is different from the set we used for the GMMChem. Even if their conclusions on the broad distribution of $\omega$~Cen debris in the E-L$_z$ space are similar to ours, it is still worth understanding how much the two methods produce the same results. \\

\begin{figure*}\centering
\includegraphics[clip=true, trim = 0mm 0mm 0mm 0mm, width=0.3\linewidth]{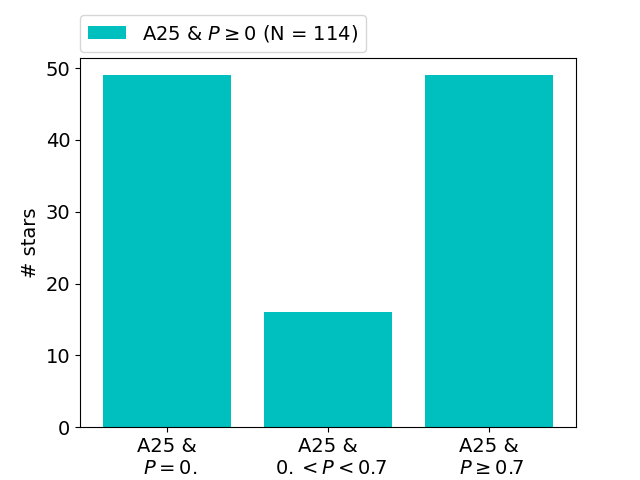}
\includegraphics[clip=true, trim = 0mm 0mm 0mm 0mm, width=0.3\linewidth]{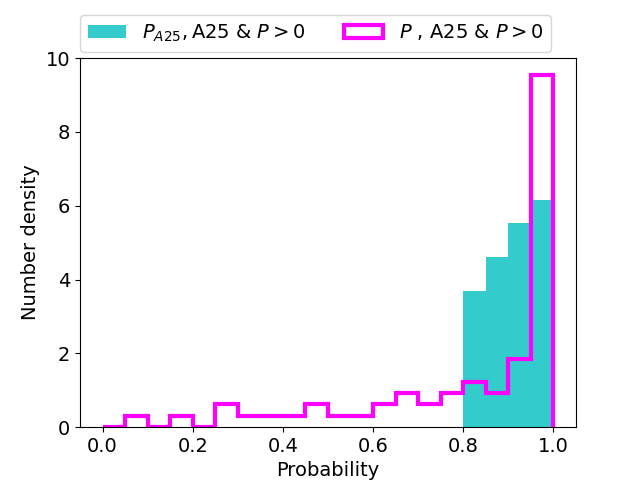}
\includegraphics[clip=true, trim = 0mm 0mm 0mm 0mm, width=0.3\linewidth]{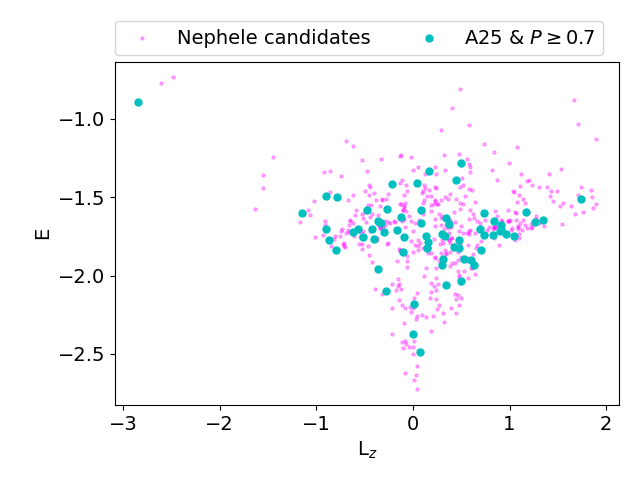}\\

\includegraphics[clip=true, trim = 0mm 0mm 0mm 0mm, width=0.3\linewidth]{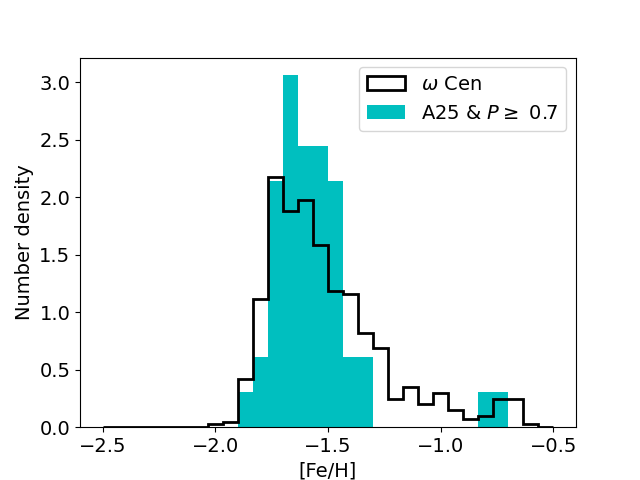}
\includegraphics[clip=true, trim = 0mm 0mm 0mm 0mm, width=0.3\linewidth]{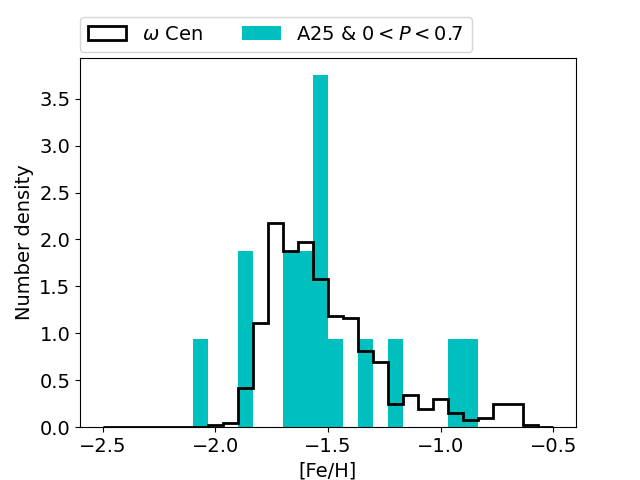}
\includegraphics[clip=true, trim = 0mm 0mm 0mm 0mm, width=0.3\linewidth]{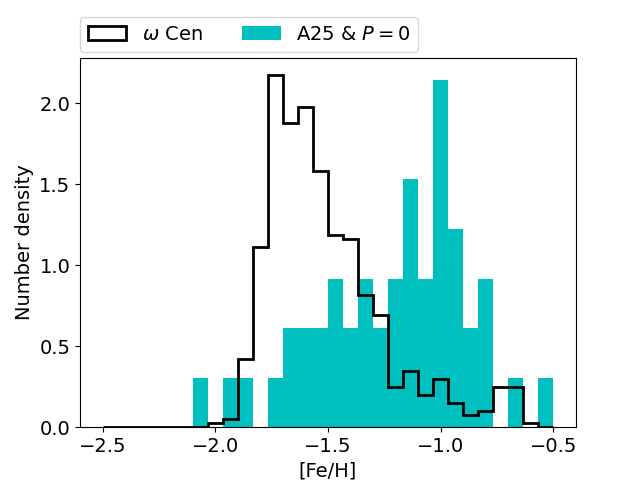}

\includegraphics[clip=true, trim = 0mm 0mm 0mm 0mm, width=0.3\linewidth]{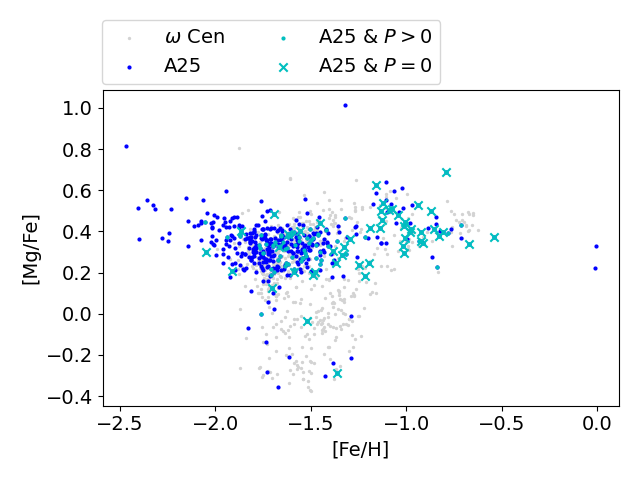}
\includegraphics[clip=true, trim = 0mm 0mm 0mm 0mm, width=0.3\linewidth]{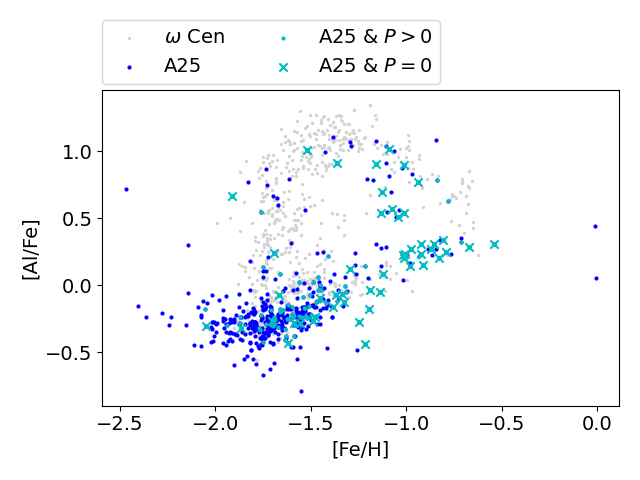}
\includegraphics[clip=true, trim = 0mm 0mm 0mm 0mm, width=0.3\linewidth]{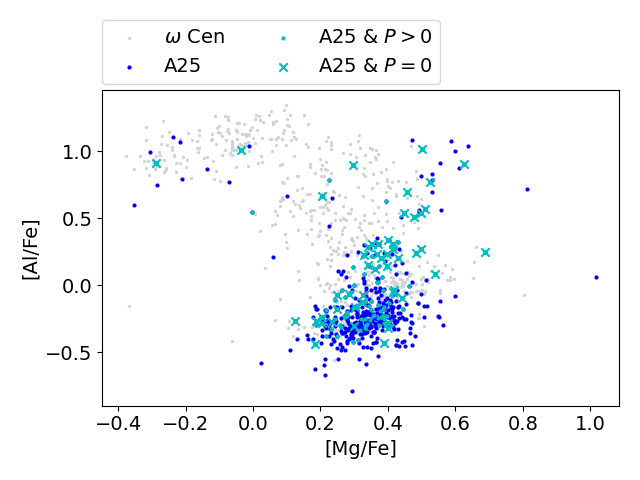}

  \caption{Comparison with stars in \citet{anguiano2025} (A25). \textbf{Top row, left panel:} Number of stars in the A25 sample that, according to our analysis, have a null ($P=0$), not null ($0<P<0.7$) or high ($P\geq 0.7$) probability of being chemically compatible with $\omega$~Cen. \textbf{Top row, middle panel:} Distribution of probabilities $P$ (magenta) of the stars of A25's sample which, according to our analysis, have a not null probability of being chemically compatible with $\omega$~Cen, compared to the analogue distribution of probabilities $P_{A25}$ (cyan) provided by A25. It can be noted that part of the stars which have a high $P_{A25}$ turns out to have a low $P$, according to our GMMChem model. \textbf{Top row, right panel:} Distribution in the E-L$_z$ plane of Nephele candidate members ($P \geq 0.7$, magenta), compared to that of stars in A25 (cyan) which have also a high $P$, according to our analysis, of belonging to Nephele.  \textbf{Middle row, left panel:} Normalised [Fe/H] distribution of A25's stars which are classified as Nephele candidate members ($P \geq 0.7$) also by our GMMChem model (cyan). For comparison, the normalised [Fe/H] distribution of $\omega$~Cen stars is also shown. \textbf{Middle row, middle panel:} Same for A25's stars which have a moderate probability ($0 < P < 0.7$) of being chemically compatible with $\omega$~Cen, according to our GMMChem model. For comparison, the normalised [Fe/H] distribution of $\omega$~Cen stars is also shown. \textbf{Middle row, right panel:} Same for A25's stars which have a null probability $P$ of  being chemically compatible with $\omega$~Cen, according to our GMMChem model. For comparison, the normalised [Fe/H] distribution of $\omega$~Cen stars is also shown. \textbf{Bottom row, left panel:}  Distribution in the [Mg/Fe] vs [Fe/H] plane of stars in the A25's sample. Blue points indicate A25's stars which have not been analysed in our work, cyan points indicate A25's stars for which a probability $P$ has been assigned by our GMMChem model. Cyan crosses indicate A25's stars which have a $P=0$, according to our GMMChem model. The corresponding distribution of $\omega$~Cen stars in this plane is also shown (light grey points). \textbf{Bottom row, middle panel:} Same for [Al/Fe] vs [Fe/H]. \textbf{Bottom row, right panel:} Same for [Al/Fe] vs [Mg/Fe].}\label{comp_ang}
 \end{figure*}

Before commenting on similarities and differences in our results, it is important to understand the size of the data sets we will be comparing. \citet{anguiano2025} provide a sample of 463 stars which have a probability $P_{A25} \geq 0.8$ to be chemically compatible with $\omega$~Cen according to their model.  With our GMMChem, we find 1441 stars which have a $P > 0$ to be chemically compatible with $\omega$~Cen,  470 of which have a probability of chemical compatibility $P \geq 0.7$ and constitute our candidate members. While the two samples (i.e. $P_{A25}\geq 0.8$ and $P \geq 0.7$) have comparable sizes, their common APOGEE identifiers -- as we checked -- are only 49. This number increases to 65 if we relax our probability threshold and we check -- among our whole sample of stars with $P > 0$ -- how many are also in the \citet{anguiano2025} sample. The limited overlap between the two samples may be due to different reasons. The methodology adopted is clearly not the same. Moreover, the fact that the initial quality cuts applied to the APOGEE catalogue are also different (see our Sect.~\ref{obsdata} and their Sect.~3.1) implies that of the 463 stars in the  \citet{anguiano2025} sample only 114 satisfy our initial selection cuts.  So only for 114 stars out of the 463 of \citet{anguiano2025} we can effectively make a comparison with our work. Taking this normalisation into account, 49/114 stars, i.e. 43\% of the sample, have a high probability in both works of being chemically compatible with $\omega$~Cen (respectively $P_{A25}\geq 0.8$ for \citet{anguiano2025}, and $P \geq 0.7$ for us), (65-49)/114 stars, i.e. 14\% of the sample, have a high probability for \citet{anguiano2025} ($P_{A25} \ge$0.8) but $0 < P < 0.7$ for us, and finally (114-65)/114 stars, i.e. 43\% of the sample,  have a high probability of compatibility for \citet{anguiano2025} but a null probability for us (see Fig.~\ref{comp_ang}, top left panel). 

Stars in the \citet{anguiano2025} sample which, according to our analysis, have a not null probability $P$ to be chemically compatible with $\omega$~Cen span the whole range of positive $P$: while we observe a peak at $P\sim 1$, we notice also that there is a long tail in the probability distribution that extends up to very low $P$ values (top row-middle panel). Nephele candidate members (i.e. stars with $P\geq 0.7)$ which are also in the \citet{anguiano2025} sample show a [Fe/H] distribution which peaks where the peak of the  $\omega$~Cen distribution is found (see Fig.~\ref{comp_ang}, second row, left panel). As the probability $P$ of chemical compatibility with $\omega$~Cen decreases,  the [Fe/H] distribution of the sample common to us and \citet{anguiano2025} tends to  shift towards higher [Fe/H] values (in particular the mean of the distribution, see Fig.~\ref{comp_ang}, second row, middle and right panels). Remarkably, stars that according to \citet{anguiano2025} have a high probability to belong to $\omega$~Cen debris, but that according to our analysis have a low probability ($P < 0.7$) of a common origin with the cluster,  show a [Fe/H] which peaks at values $\sim -1$, well beyond the $\omega$~Cen [Fe/H] peak (Fig.~\ref{comp_ang}, second row, right panel). This shows that outside the bulk of $\omega$~Cen, our methods start to differ. 

As a further comparison, in Fig.~\ref{comp_ang} (bottom row), we have reported the distribution in the [Mg/Fe]-[Fe/H], [Al/Fe]-[Fe/H], and [Mg/Fe]-[Al/Fe] planes of the whole set of stars in \citet{anguiano2025}'s sample, and those in \citet{anguiano2025}'s sample for which we have estimated a probability $P$, and we compare these two distributions with that of $\omega$~Cen itself, for each of these planes. It is interesting to see how there are stars, at the periphery of the distributions of these abundance planes, that \citet{anguiano2025} assign as $\omega$~Cen debris but that for us have zero probability of being associated with $\omega$~Cen - these are stars that the GMMChem rejected every time.
 For many other stars, our initial selection did not include them in the analysed set. Finally, note how some of \citet{anguiano2025}'s stars, to which our method attributes zero probability of belonging to $\omega$~Cen, are located in a region of the [Al/Fe]-[Mg/Fe] plane where no $\omega$~Cen stars are present. 
 
 To conclude this comparison, it is interesting to discuss how stars with high chemical compatibility with $\omega$~Cen according to both \citet{anguiano2025}'s and  to our analysis are redistributed in the E-L$_z$ plane. It can be seen that, compared to the extended energy distribution of the entire sample, the latter are mostly redistributed in an energy ‘band’ between $-2 \lesssim E \lesssim -1.5$ covering the whole range of angular momenta. It cannot be ruled out that this is indeed the region with the highest probability of containing $\omega$~Cen debris, at least in the APOGEE~DR17 dataset. 

Overall, both methods recover a similarly broad distribution of Nephele debris in the E-L$_z$ plane and agree well for the core of the $\omega$~Cen chemical locus; the stars that are selected by one method and not by the other are mainly objects near the edges of this locus, where the different parent samples, input abundance sets and the more flexible neural-network decision boundary of \citet{anguiano2025} naturally lead to a less conservative selection than our explicitly probabilistic GMM.

\subsection{Additional comments on candidate stars in GC streams of Nephele's family}\label{neph_GCstreams}

An additional outcome of our analysis is the identification of six field stars in the APOGEE sample that exhibit both chemical and kinematic properties consistent with $\omega$~Cen and its predicted tidal stream. Notably, one of these stars is Al-rich, suggesting being a second-generation star with GC origin. These stars are spatially and chemically distinct from the Fimbulthul stream \citep[][]{ibata19b}, which lies outside the APOGEE footprint (see Fig.~\ref{Nephele_longlat}). This provides the first evidence, to our knowledge, of additional $\omega$~Cen debris beyond the known Fimbulthul members. Independent evidence for extended $\omega$~Cen debris also comes from optical spectroscopic surveys.
Using GALAH DR2 and \emph{Gaia}, \citet{simpson2020} chemically tagged two Fimbulthul stream stars to $\omega$~Cen on the basis of their Na--Al enhancement and $s$–process abundances. In addition, \citet{lind2015} and \citet{fernandeztrincado2015} identified further halo stars and RR~Lyrae populations, from other high–resolution spectroscopic data sets, whose kinematics and detailed chemistry are consistent with being debris from $\omega$~Cen or its progenitor. More recently, \citet{youakim2023} exploited GALAH~DR3 and \emph{Gaia}~eDR3 with an unsupervised clustering algorithm to find 18 candidate $\omega$~Cen debris stars spread over more than $50^\circ$ on the sky. Although the overlap between the GALAH and APOGEE samples is limited by the different footprints and selection functions of the two surveys, the GALAH–based candidates occupy a similar metallicity and light–element abundance regime to the $\omega$~Cen locus, and where available their orbits are retrograde and halo–like.

Beyond candidate stars of $\omega$~Cen stream, some are found to have kinematics compatible with the predicted streams of NGC~6205, NGC~6254, NGC~6273, NGC~6656, and NGC~6809. For all these clusters the detection of streams has so far been limitedly successful and mostly confined to the periphery of these clusters: in the case of NGC~6205 extra-tidal stars have been identified but only in the cluster surroundings \citep[see, for example][]{cheng24}; for NGC~6254, very recently, some tidal distortions have been found in the cluster outskirts, by analysing Euclid Early Release Observations \citep[see][]{massari24}; only in the case of NGC~6656, a star with radial velocity, [Fe/H] and $\alpha$-elements compatible with the cluster has been found at projected distances of about 10~degrees from it, in the RAVE data \citep{kunder14}. Finally the search for extra-tidal stars around NGC~6809, to date, has not led to any detection \citep{piatti21b}. Our work potentially opens the possibility to increase the number of tidal debris associated to these clusters significantly, also extending these discoveries to more extended distances from the clusters themselves.  

Despite the robustness of our selection method, our results are not immune to caveats. We recall that we made two selections in order to search for potential candidate stars from the streams of the clusters. GMMChem allowed us to find all the stars chemically compatible with $\omega$~Cen (the Nephele candidate stars); GMMKin allowed us to find the stars that, based on their $E$ and L$_z$, are located in kinematic regions where we would expect to find stars from the streams of these clusters (in accordance with \citet{ferrone2023} models).  If we compare the chemical abundances of these stars with those of the clusters to which they are associated, we can see that for some of them the abundances are not comparable to those of the clusters with which they share similar E and L$_z$. This is indeed the case of NGC~6656 (see Appendix~\ref{app:chem_compatible}, Fig.~\ref{ngc6656_chem}). Of course, these stars can be removed by operating an additional selection on [Fe/H], to restrict the range to that of the cluster under consideration. Applying this further selection, the number of stars in the streams of NGC 6205, NGC 6254, NGC 6273, NGC 6656, and NGC 6809, reduces, respectively, to {1, 1, 1, 6, 1.} However, it is interesting to understand the nature of these interlopers, and to which population they may be related to. By definition, these stars are chemically compatible with $\omega$~Cen (first step of our selection procedure) and compatible, in terms of E and L$_z$, with the streams of some of the clusters of the Nephele's family, without however sharing their chemical abundances. The simplest interpretation is that these are stars from the Nephele's field in which these clusters were and are immersed. In fact, Fig.~\ref{e_lz} shows that the Nephele's candidate stars pervade the space of energies and angular momenta surrounding the clusters. Some of them can therefore be found on E and L$_z$ similar to those of the GC streams associated with this accretion and enter the selection of candidate streams.This result highlights how delicate the search for stars associated with a stream (and/or an accretion) is, and how - even when applying various selection procedures - the possibility of contamination by other populations cannot be ruled out and must in fact be verified.  

A further limitation of our analysis arises from the fact that APOGEE is not an all-sky survey, and we are therefore constrained by its restricted spatial coverage (see Fig.~\ref{Nephele_longlat}). Nonetheless, this survey remains fundamental to our study, as it enables a consistent comparison between the chemical abundances of field stars and those of GC stars observed and processed through the same reduction pipeline. However, APOGEE primarily targets specific regions of the Galaxy, particularly the innermost parts, where the identification of stellar streams is particularly challenging. In these regions, the short dynamical timescales cause unbound stars to quickly disperse, rapidly erasing their phase-space coherence. 

At the same time, regions such as that occupied by Fimbulthul fall outside the APOGEE footprint, limiting our ability to perform additional validation checks of our method.  

It is also important to note that, since our selection is limited to red giant stars in the APOGEE catalogue, and stars escaping from GCs are preferentially low-mass main sequence stars \citep[see, e.g.][]{baumgardt2003, kupper10}, our methodology could potentially identify many more escaped members if applied to more complete stellar sampling. In general, searching for stars in stellar streams will become feasible with the forthcoming wide-field spectroscopic surveys conducted with MOONS \citep[][]{cirasuolo2020, gonzalez2020}, WEAVE \citep[][]{jin2024}, and 4MOST \citep[][]{dejong2012}.

\section{Conclusions}
\label{conclusions}

In this study, we employed a chemical and dynamical tagging approaches to trace field stars that may be remnants of the Nephele accretion event—an hypothesised merger that brought $\omega~$Cen and a family of chemically similar GCs into the Milky Way. Using data from the APOGEE DR17 catalogue, we selected 470 stars chemically compatible with $\omega~$Cen in a 8D abundance space, and among them, 58 are aluminium-rich, consistent with second-generation stars found in GCs. Out of these 470 stars, 181 have a probability larger than 0.99 to belong to the Nephele accretion. Our main results can be summarised as follows.

The vast majority of the stars that we find to be chemically compatible with $\omega$~Cen distributes over a large range of the E-L$_z$ space, from prograde to retrograde orbits, suggesting that the accretion event which brought $\omega$~Cen into the Milky Way must have been relatively massive ($\sim$1:10 ratio at the time of the accretion), and/or that a significant redistribution in the E-L$_z$  space of stars from this accretion event has further occurred. Similar conclusions have been recently reached also by \citet{anguiano2025}.
The discovery that stars chemically compatible with $\omega$~Cen are distributed across such a vast region of the E-L$_z$ space accessible with APOGEE data, together with the identification in \citet{pagnini2025} of at least 6 GCs chemically compatible with $\omega$~Cen, shows that the accretion of $\omega$~Cen and Nephele constitutes a major, and perhaps hitherto insufficiently studied, event in the evolution of our Galaxy.

By combining this chemical tagging approach with predictions from the e-TidalGCs project, we also uncovered 6 stars that are both chemically and kinematically consistent with the predicted tidal stream of $\omega~$Cen, none of which belong to the previously discovered Fimbulthul stream, and additional stars associated with the tidal streams of NGC~6205, NGC~6254, NGC~6273, NGC~6656, and NGC 6809.

These results suggest that through this method we can trace tidal debris from GCs, extending stream detections beyond their immediate surroundings. We have seen for $\omega~$Cen and NGC~6656, that some identified candidates might be cluster members misclassified as field stars due to uncertainties in cluster membership probabilities, serving as a robustness check for our method.

We explored the relationship between Nephele and Gaia Sausage–Enceladus (GSE), showing that, by adopting for this latter the definition given by \citet{horta2023}, for [Fe/H]$\lesssim -1.6$ GSE is dominated by stars of the Nephele accretion. The chemical overlap observed at these metallicities indicates a non-trivial entanglement of their respective debris, and points to the need for a more refined, multi-dimensional definition of accretion structures such as GSE.

This work represents a step forward in chemically and kinematically identifying the disrupted remnants of a massive progenitor system, enriching our understanding of Galactic archaeology and the build-up of the Milky Way halo.

\section{Data availability}
Table \ref{table:Nephele_ID} is only available in electronic form at the CDS via anonymous ftp to \url{cdsarc.u-strasbg.fr} (130.79.128.5) or via \url{http://cdsweb.u-strasbg.fr/cgi-bin/qcat?J/A+A/}.

\begin{acknowledgements}
G.P. acknowledges the support from the Centre national d'études spatiales (CNES) through a postdoctoral fellowship. G.P., P.D.M., and P.B. are grateful to the "Action Thématique de Cosmologie et Galaxies (ATCG), Programme National ASTRO of the INSU (Institut National des Sciences de l’Univers) for supporting this research, in the framework of the project "Coevolution of globular clusters and dwarf galaxies, in the context of hierarchical galaxy formation: from the Milky Way to the nearby Universe" (PI: A. Lançon). P.D.M. and M.H. acknowledge the support of the French Agence Nationale de la Recherche (ANR), under grant ANR-13-BS01-0005 (project ANR-20- CE31-0004-01 MWDisc). P.B. and G.P. acknowledge finantial support by the IdEx framework of the University of Strasbourg. A.M.B. acknowledges funding from the European Union's Horizon 2020 research and innovation program under the Marie Sk\l{}odowska-Curie grant agreement No 895174. F.R. acknowledges support provided by the University of Strasbourg Institute for Advanced Study (USIAS), within the French national programme Investment for the Future (Excellence Initiative) IdEx-Unistra. O.A. acknowledges support from the Knut and Alice Wallenberg Foundation, the Swedish Research Council (grant 2019-04659), and the Swedish National Space Agency (SNSA Dnr 2023-00164). N.R. acknowledges support from the Swedish Research Council (grant 2023-04744) and the Royal Physiographic Society in Lund through the Stiftelsen Walter Gyllenbergs and M\''{a}rta och Erik Holmbergs donations. Funding for the Sloan Digital Sky Survey IV has been provided by the Alfred P. Sloan Foundation, the U.S. Department of Energy Office of Science, and the Participating Institutions. SDSS-IV acknowledges support and resources from the Center for High Performance Computing at the University of Utah. The SDSS
website is \href{}{www.sdss.org}. SDSS-IV is managed by the
Astrophysical Research Consortium for the Participating Institutions of the SDSS Collaboration including the Brazilian Participation Group, the Carnegie Institution for Science, Carnegie Mellon University, Center for Astrophysics | Harvard \&
Smithsonian, the Chilean Participation Group, the French Participation Group, Instituto de Astrof\'isica de Canarias, The Johns Hopkins University, Kavli Institute for the Physics and Mathematics of the Universe (IPMU) / University of Tokyo, the Korean Participation Group, Lawrence Berkeley National Laboratory, Leibniz Institut f\''ur Astrophysik Potsdam (AIP),  Max-Planck-Institut f\''ur Astronomie (MPIA Heidelberg), Max-Planck-Institut f\''ur Astrophysik (MPA Garching), Max-Planck-Institut f\''ur Extraterrestrische Physik (MPE), National Astronomical Observatories of China, New Mexico State University, New York University, University of Notre Dame, Observat\'ario Nacional / MCTI, The Ohio State University, Pennsylvania State University, Shanghai Astronomical Observatory, United Kingdom Participation Group, Universidad Nacional Aut\'onoma de M\'exico, University of Arizona, University of Colorado Boulder, University of Oxford, University of Portsmouth, University of Utah, University of Virginia, University of Washington, University of Wisconsin, Vanderbilt University, and Yale University. This work has made use of data from the European Space Agency (ESA) mission \textit{Gaia} (\href{}{https://www.cosmos.esa.int/gaia}), processed by the \textit{Gaia} Data Processing and Analysis Consortium (DPAC, \href{}{https://www.cosmos.esa.int/web/gaia/dpac/consortium}). Funding for the DPAC has been provided by national institutions, in particular the institutions participating in the \textit{Gaia} Multilateral Agreement. This work has made use of the computational resources obtained through the DARI grant A0120410154. 
\end{acknowledgements}

\bibliographystyle{aa}
\bibliography{aa}

\clearpage

\begin{appendix}
\section{Other elements}\label{app:other_elems}
\begin{figure*}\centering
\includegraphics[clip=true, trim = 0mm 0mm 0mm 0mm, width=.3\linewidth]{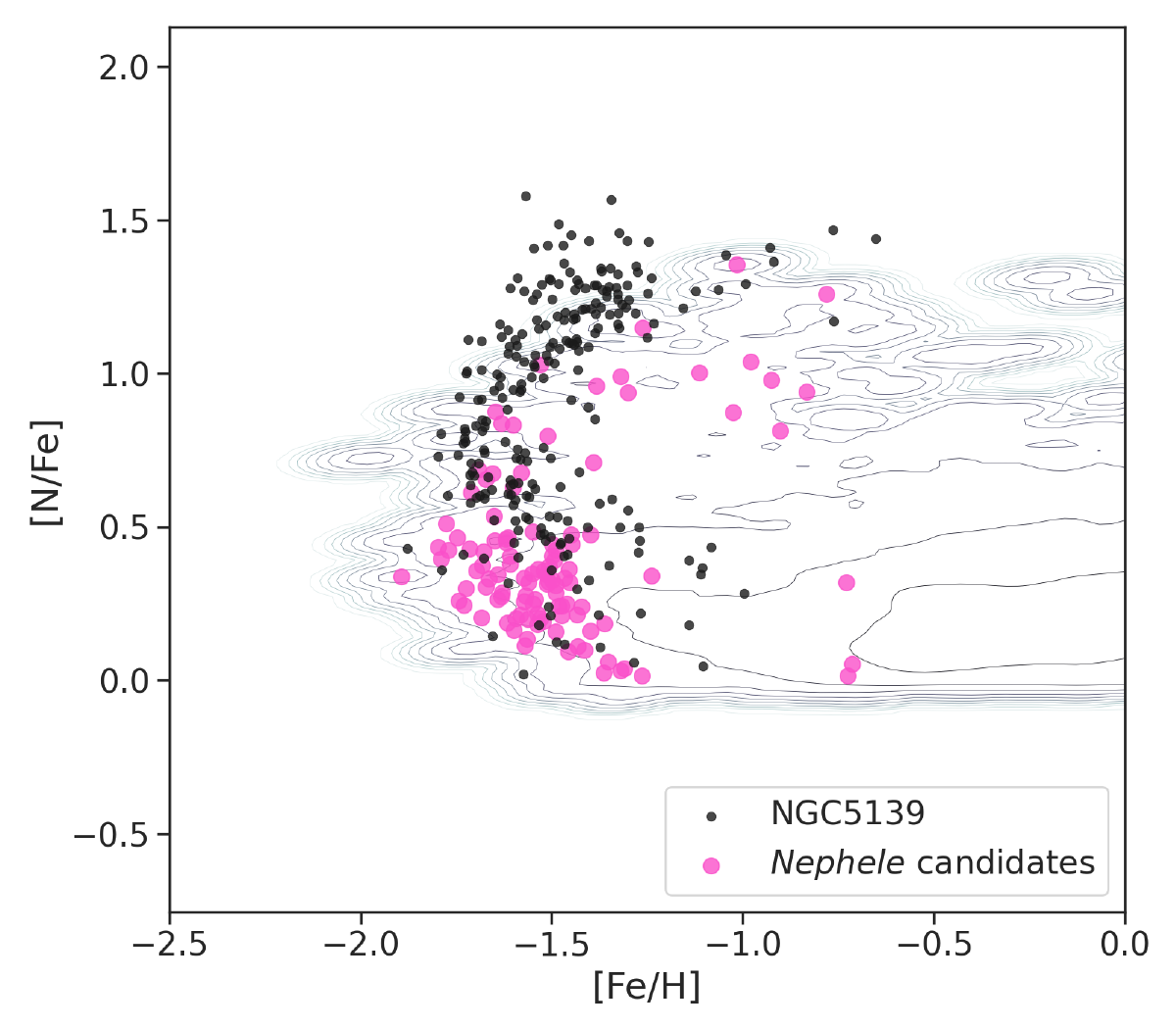}
\includegraphics[clip=true, trim = 0mm 0mm 0mm 0mm, width=.3\linewidth]{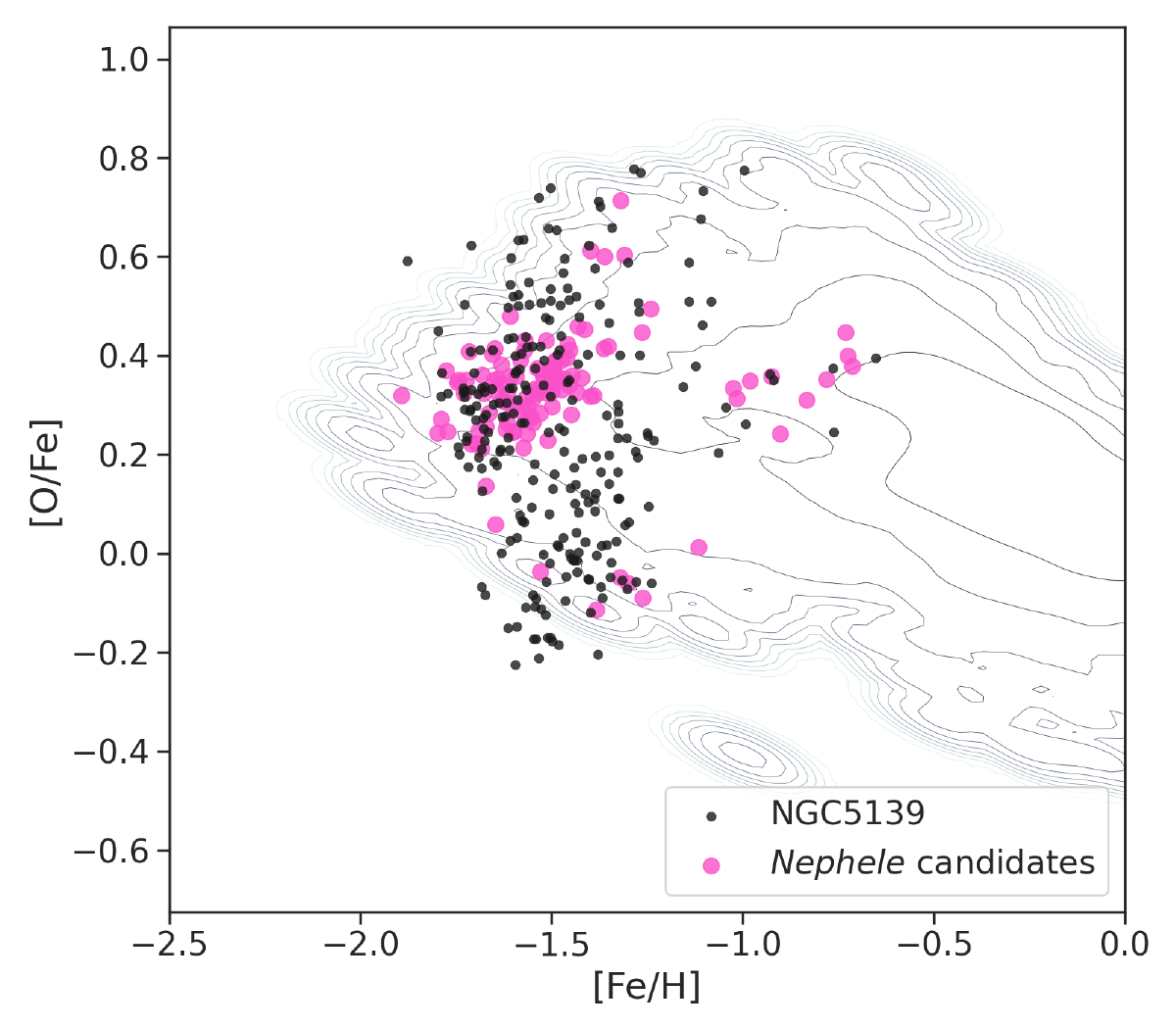}
\includegraphics[clip=true, trim = 0mm 0mm 0mm 0mm, width=.3\linewidth]{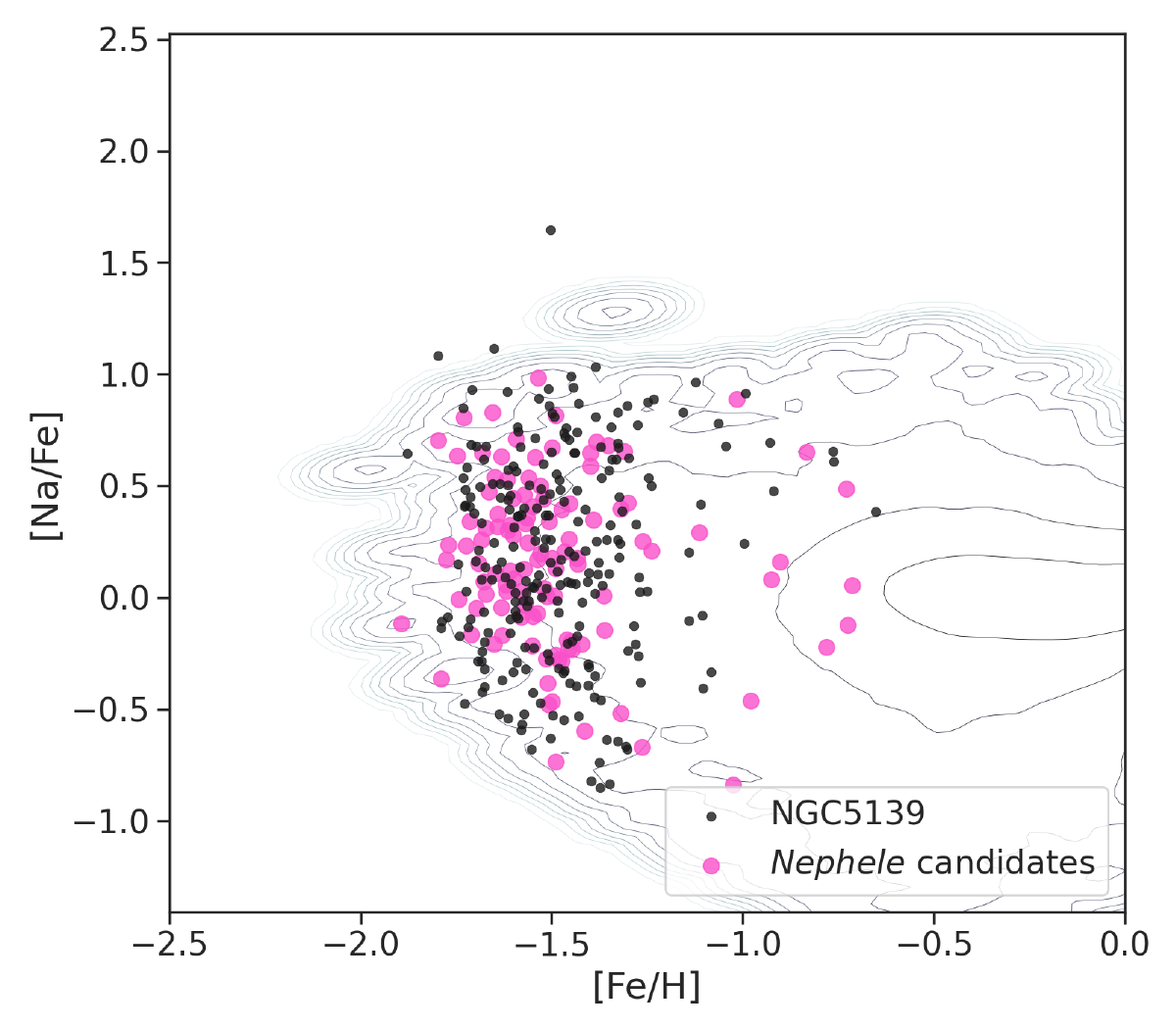}\par
\includegraphics[clip=true, trim = 0mm 0mm 0mm 0mm, width=.3\linewidth]{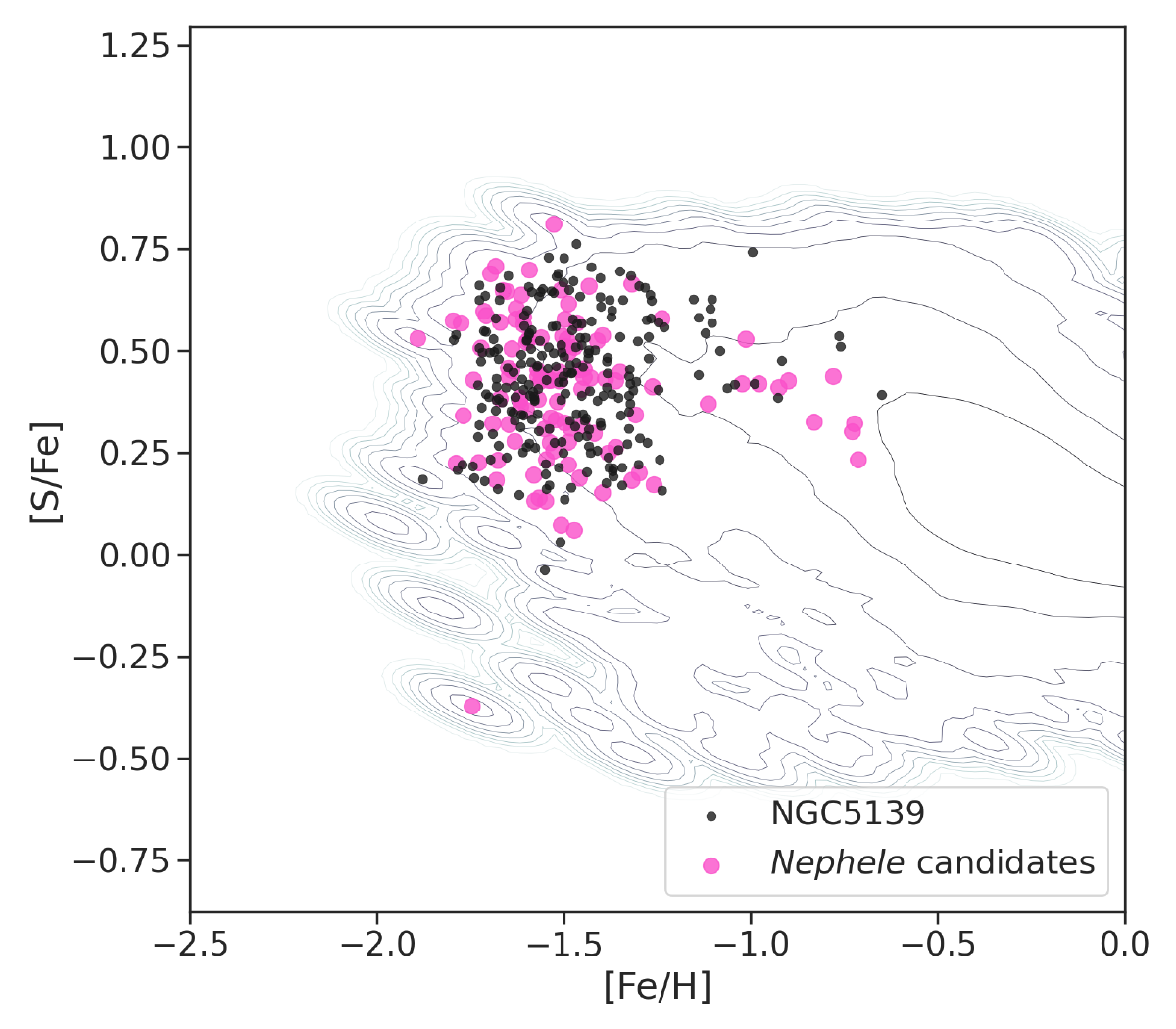}
\includegraphics[clip=true, trim = 0mm 0mm 0mm 0mm, width=.3\linewidth]{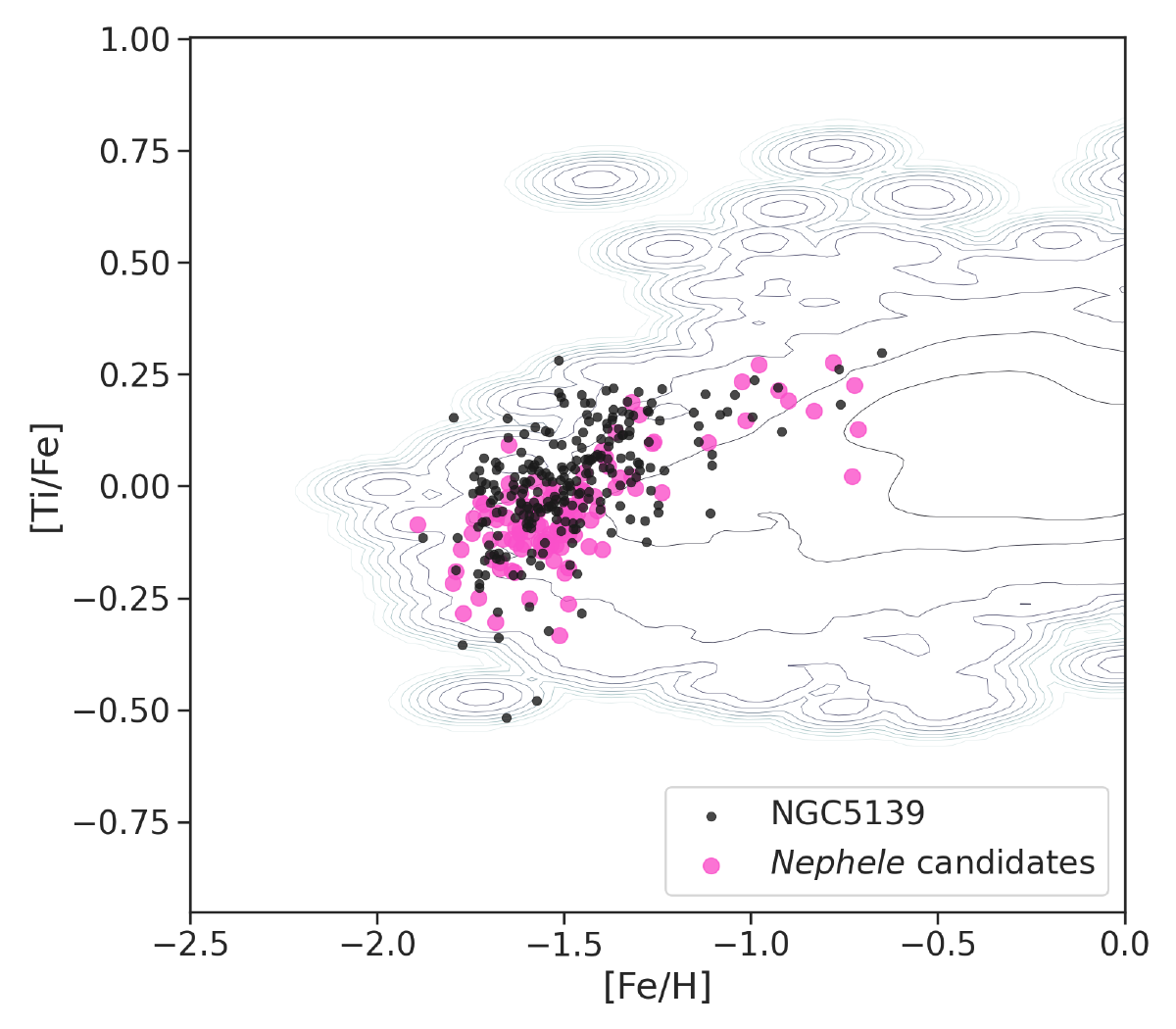}
\includegraphics[clip=true, trim = 0mm 0mm 0mm 0mm, width=.3\linewidth]{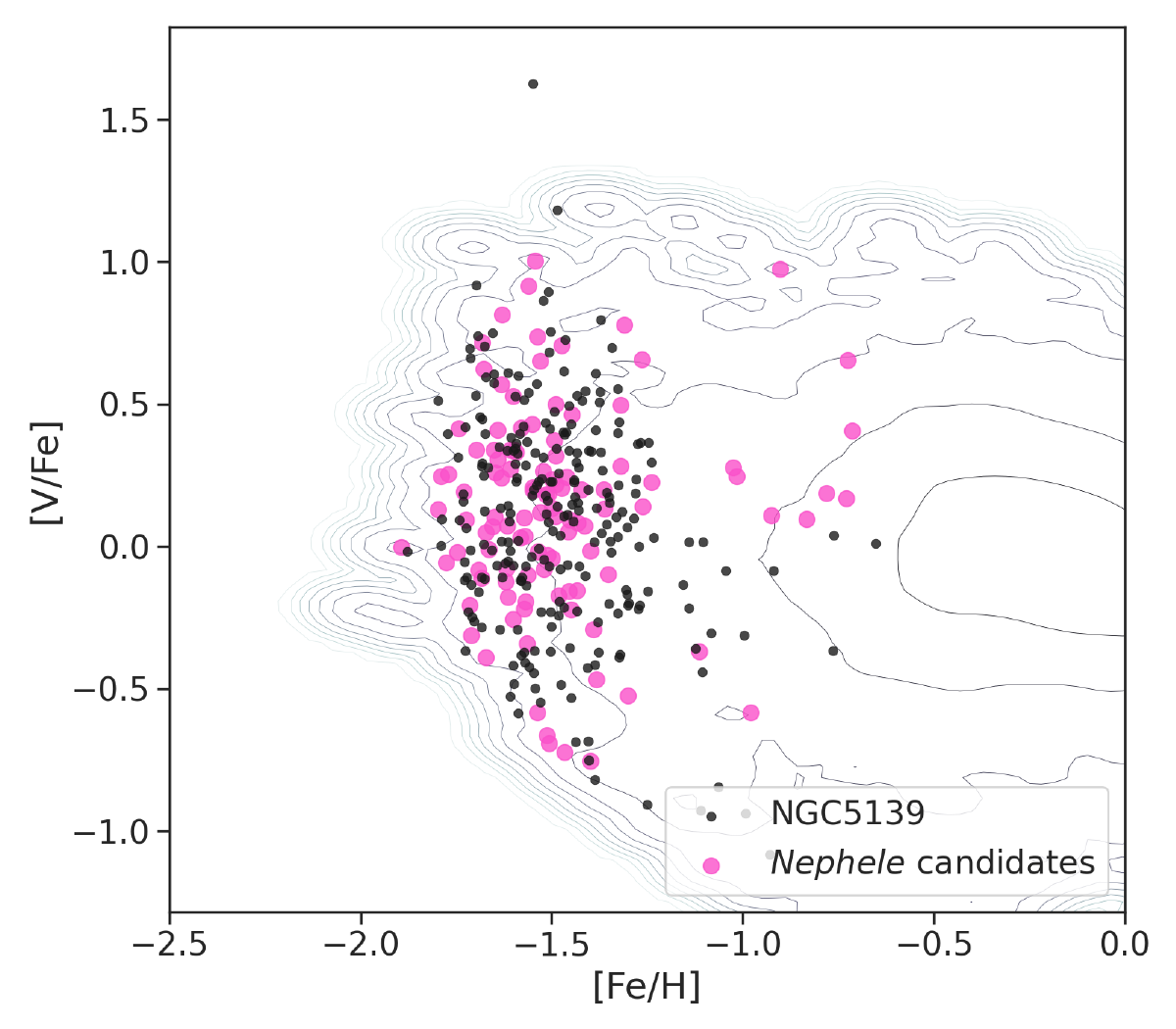}\par
\includegraphics[clip=true, trim = 0mm 0mm 0mm 0mm, width=.3\linewidth]{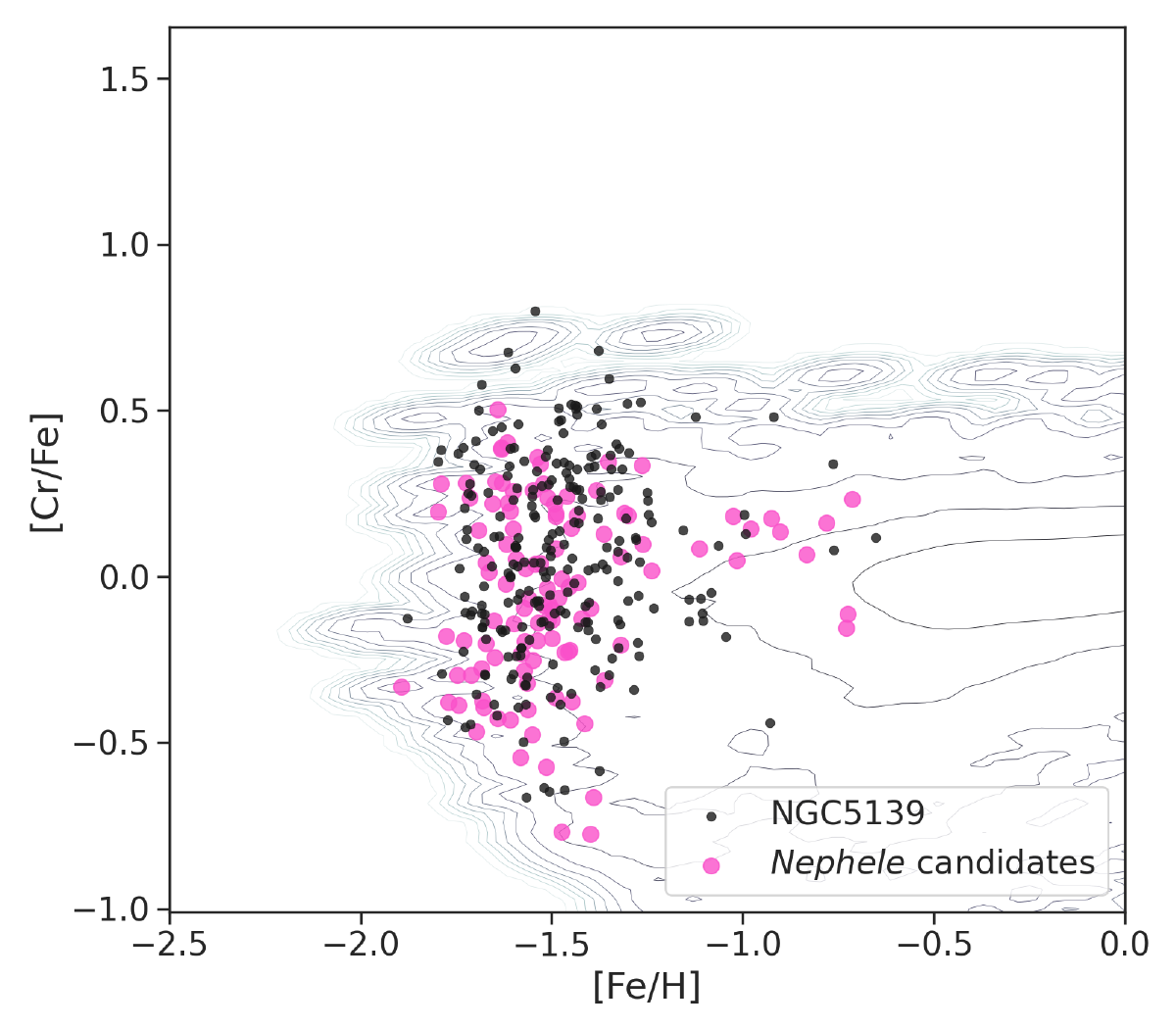}
\includegraphics[clip=true, trim = 0mm 0mm 0mm 0mm, width=.3\linewidth]{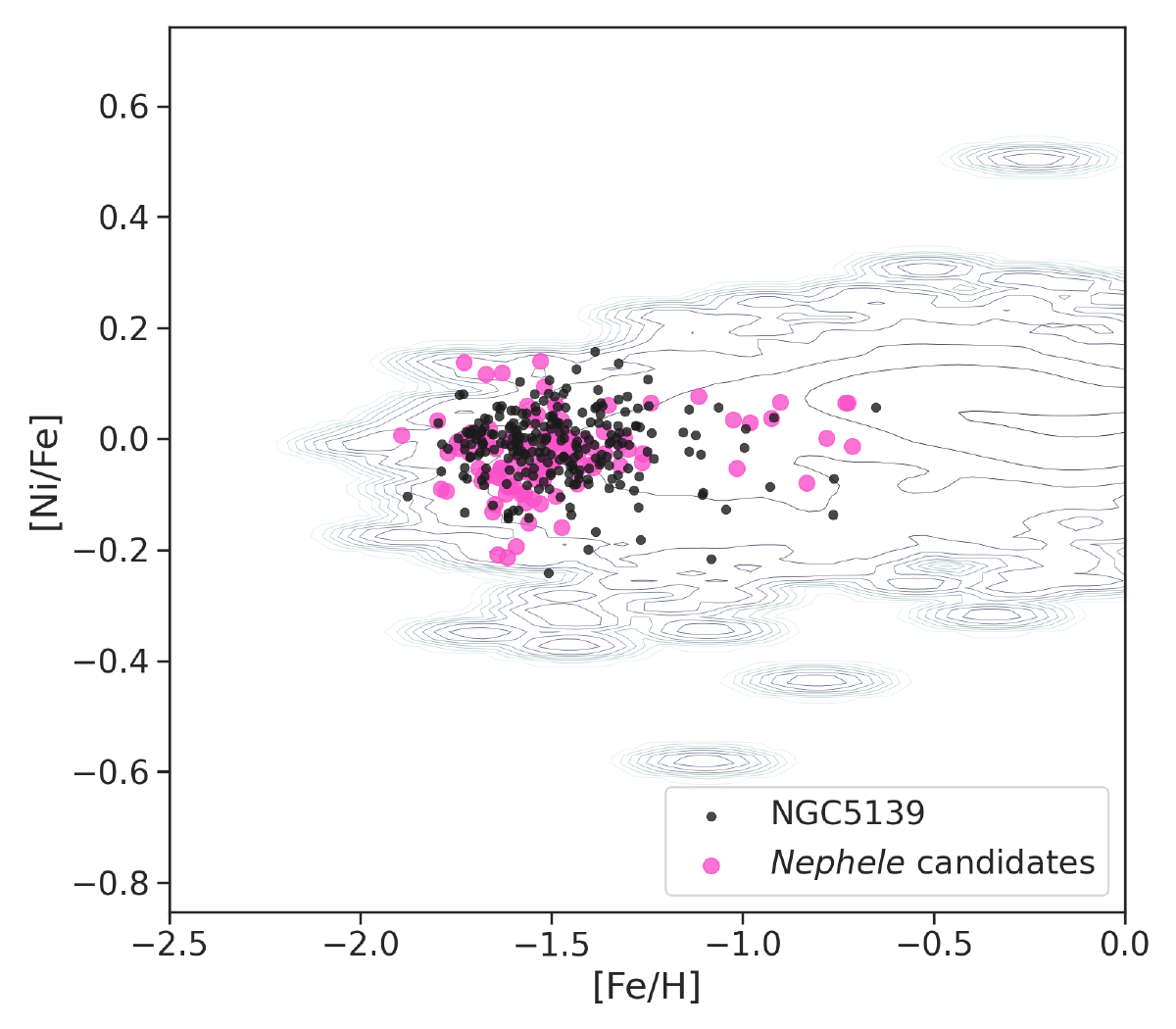}
\includegraphics[clip=true, trim = 0mm 0mm 0mm 0mm, width=.3\linewidth]{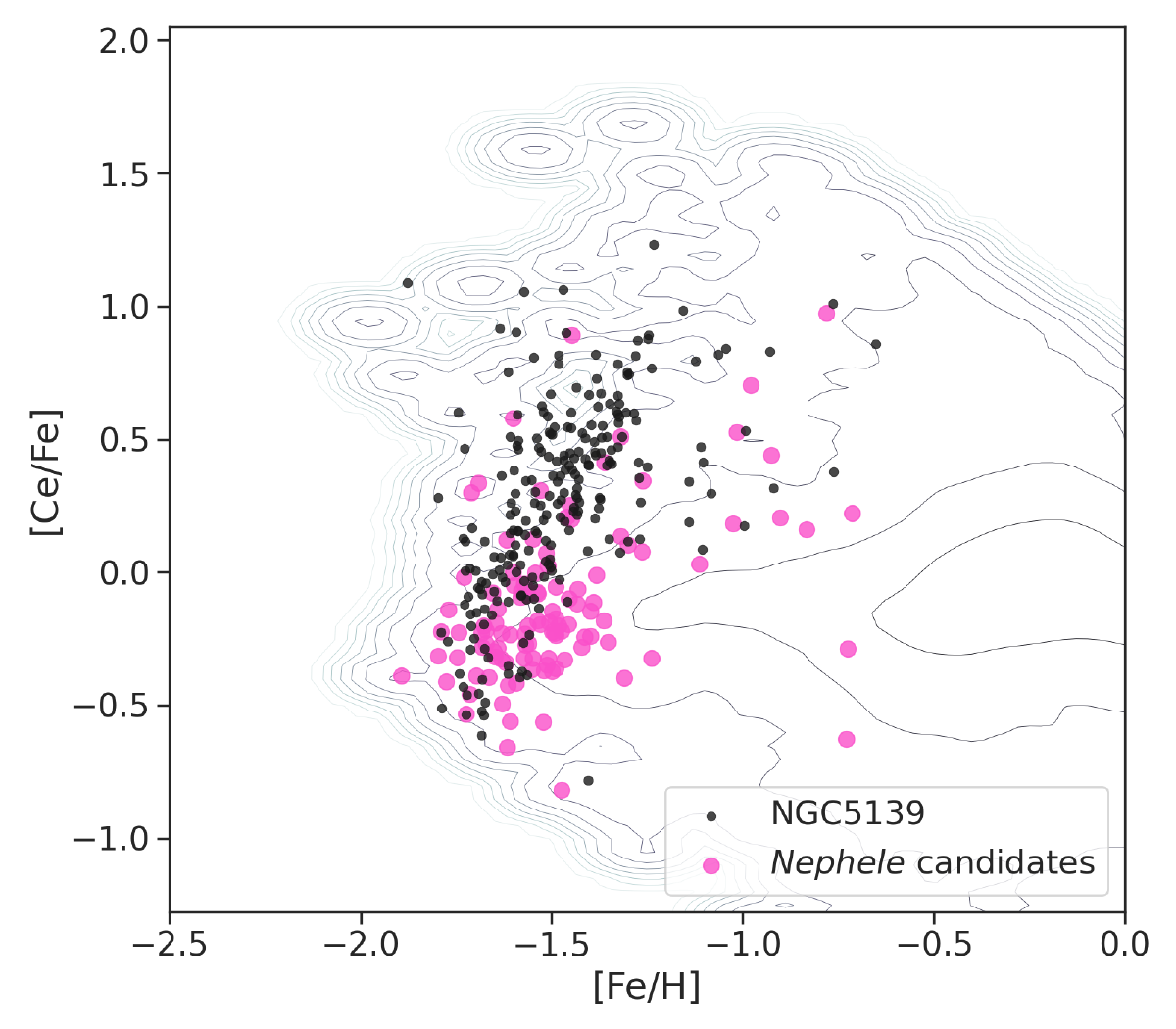}\par
\caption{[X/Fe] versus [Fe/H] planes of the field stars chemically compatible with $\omega$~Cen (magenta stars). Here, the different X elements are the others provided by APOGEE DR17 but NOT used in the GMM, namely: N, O, Na, S, Ti, V, Cr, Ni, and Ce. For comparison, the APOGEE sample of field stars and the distribution of stars belonging to $\omega$~Cen are also shown as respectively contours and black circles.}
\label{others}
\end{figure*}

Figure~\ref{others} shows the distribution of the field stars chemically compatible with $\omega~$Cen in [X/Fe] versus [Fe/H] planes where the different X elements are the others provided by APOGEE DR17 but not used in the GMMChem, namely: N, O, Na, S, Ti, V, Cr, Ni, and Ce. Although these abundances have been excluded \citep[see discussion in Sect~3 in][]{pagnini2025}, it is nevertheless interesting to see at least qualitatively how they compare with $\omega~$Cen (also shown). Overall, although these elements are not used in the GMMChem, in all these spaces the match between Nephele field stars and $\omega$~Cen is retrieved. The most important differences occur when examining the [N/Fe] vs [Fe/H] space: here most field stars are under-abundant compared to the bulk of $\omega$~Cen stars, having [N/Fe] $\lesssim$ + 0.5. In addition, the distribution in [O/Fe] vs [Fe/H] of Nephele stars turns out to be more clustered at high values of [O/Fe] ($\gtrsim$ +0.2), in contrast to that of $\omega$~Cen. Another element that seems to behave somewhat differently is Ce: for [Fe/H] $\lesssim$ -1.2 in fact, at a given metallicity Nephele stars turn out to be under abundant in [Ce/Fe] compared to $\omega$~Cen stars.

\section{Comparison of chemical abundances: Nephele's GCs versus their compatible stars}\label{app:chem_compatible}
\begin{figure*}\centering
\includegraphics[clip=true, trim = 3mm 0mm 0mm 3mm, width=0.33\linewidth]{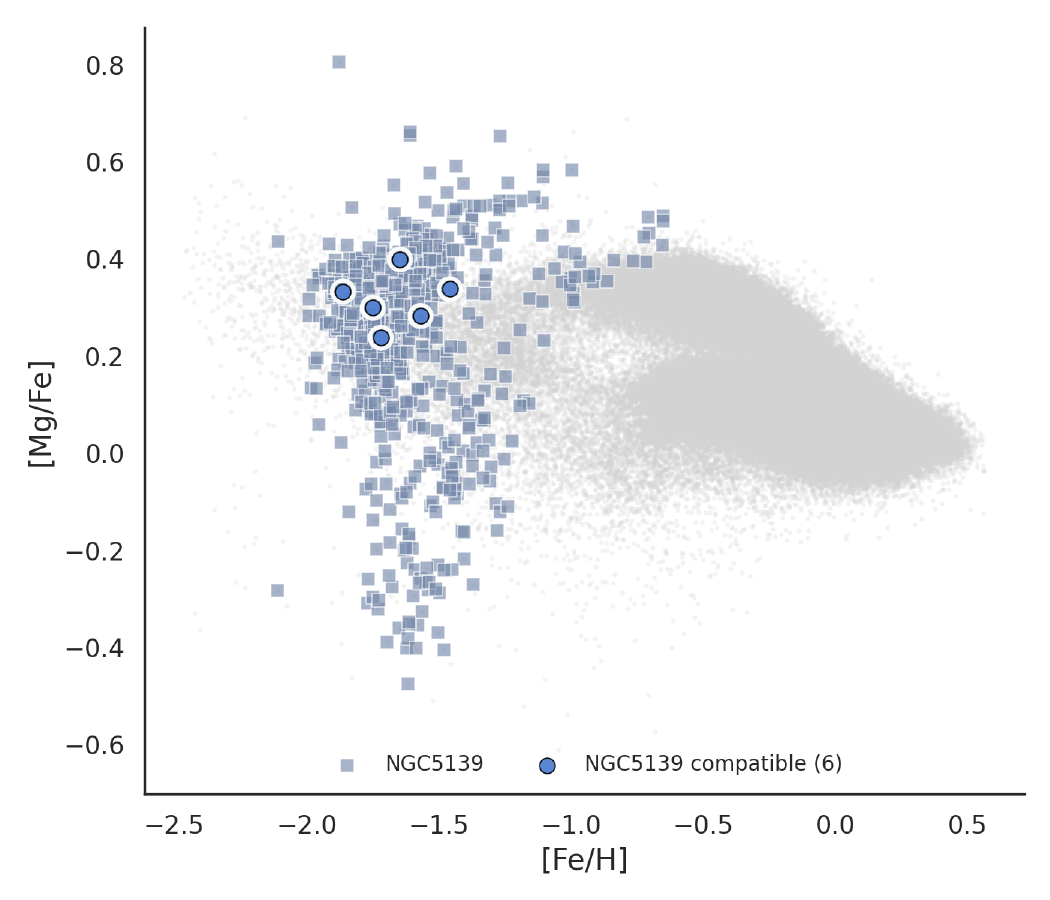}
\includegraphics[clip=true, trim = 3mm 0mm 0mm 3mm, width=0.33\linewidth]{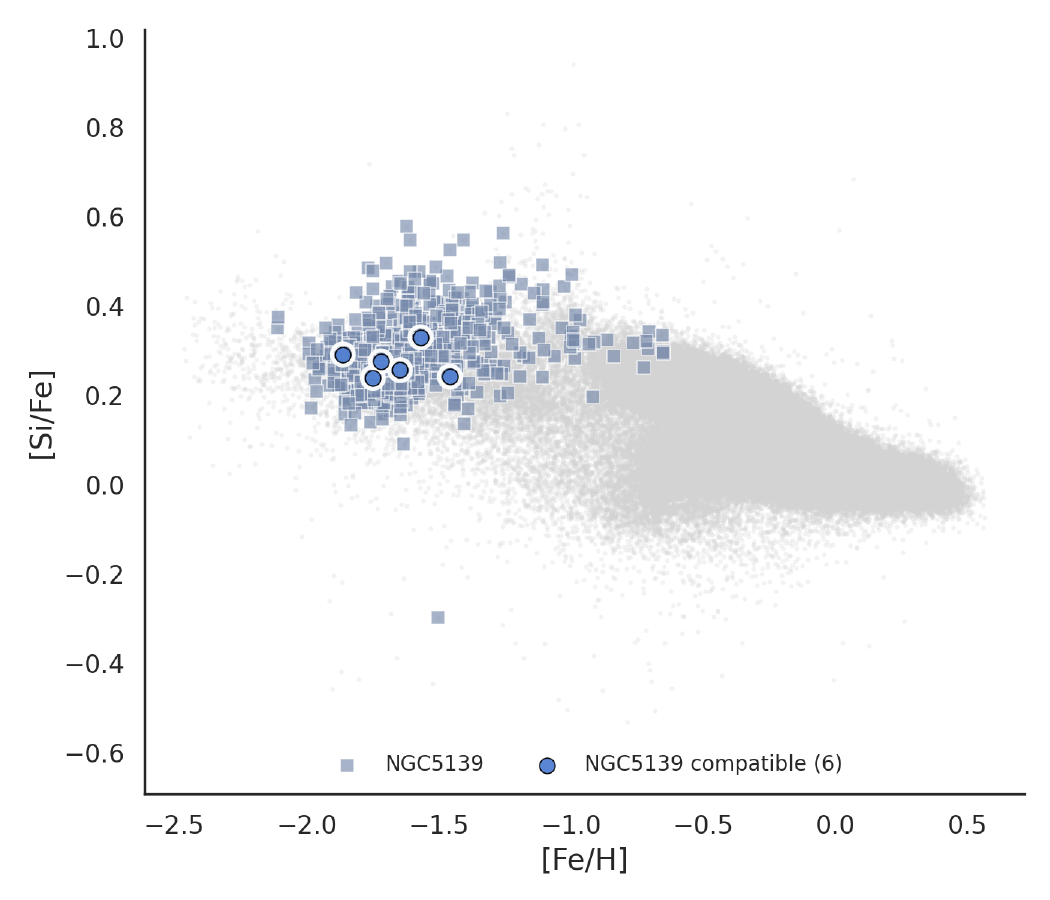}
\includegraphics[clip=true, trim = 3mm 0mm 0mm 3mm, width=0.33\linewidth]{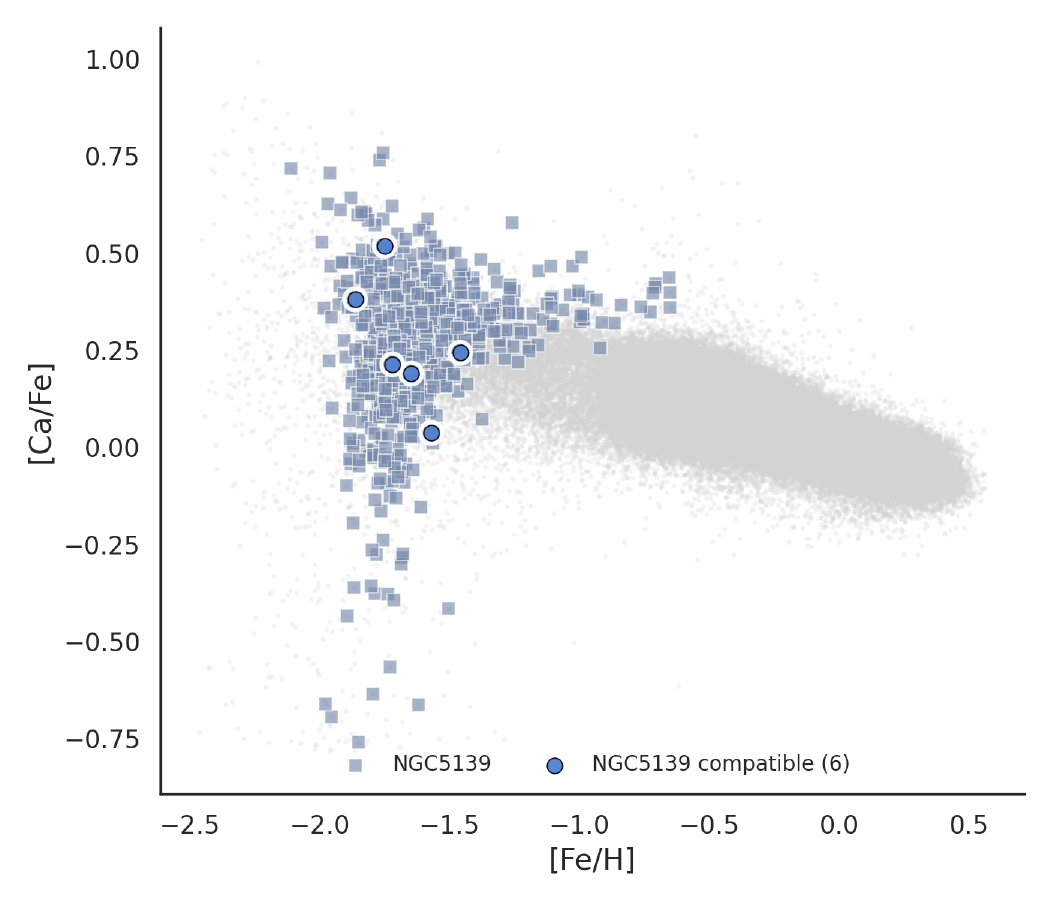}\par
\includegraphics[clip=true, trim = 3mm 0mm 2mm 0mm, width=0.33\linewidth]{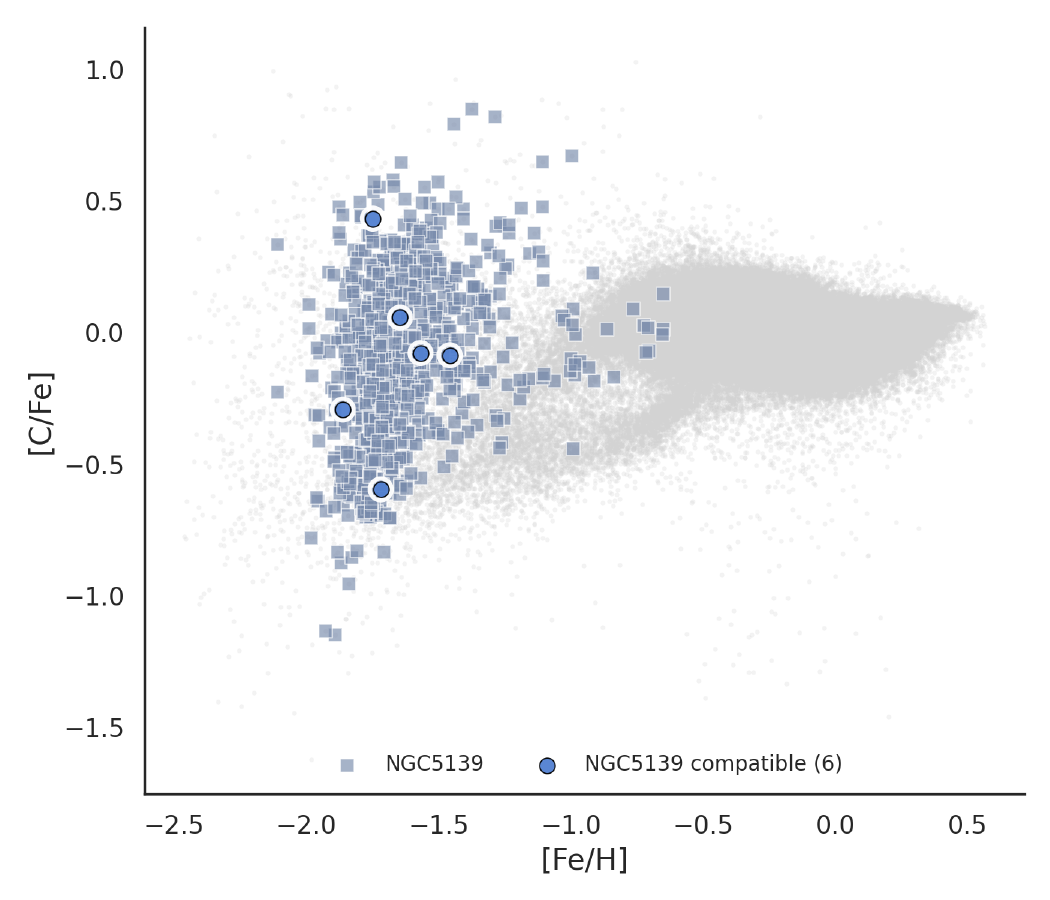}
\includegraphics[clip=true, trim = 2mm 0mm 0mm 1mm, width=0.33\linewidth]{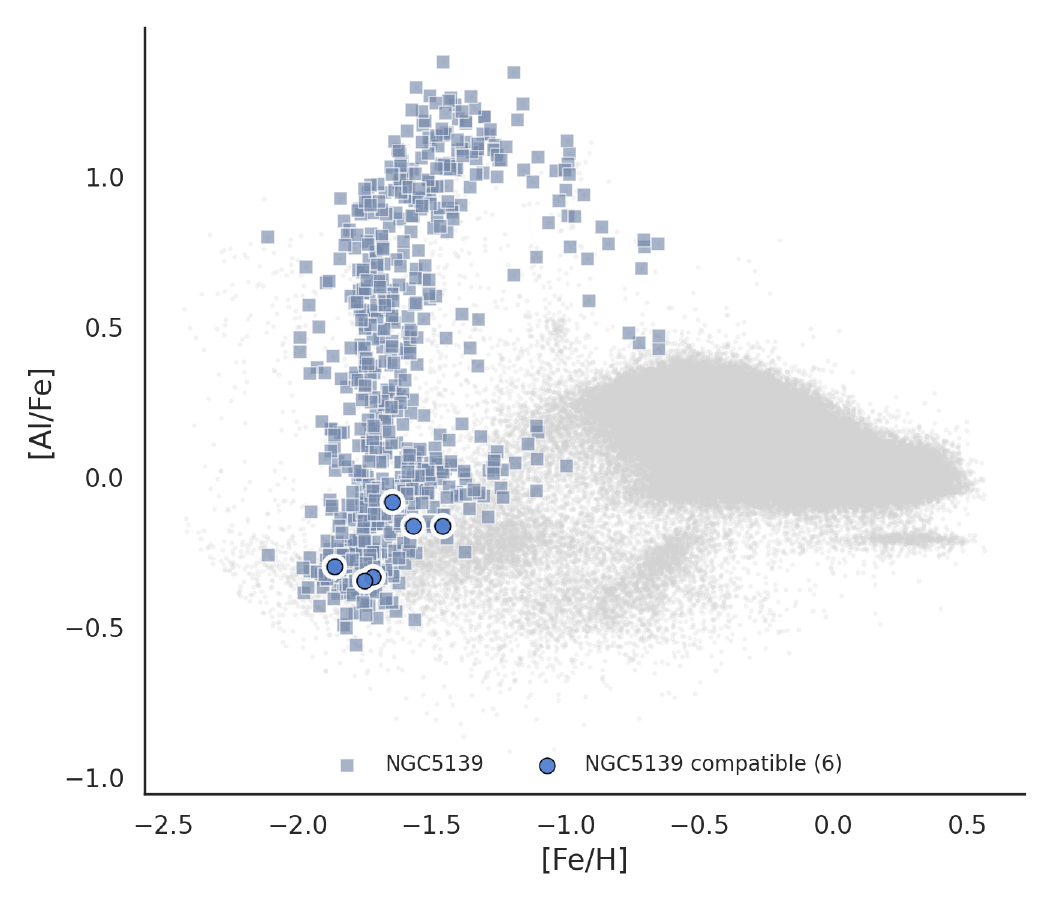}
\includegraphics[clip=true, trim = 2mm 0mm 0mm 1mm, width=0.33\linewidth]{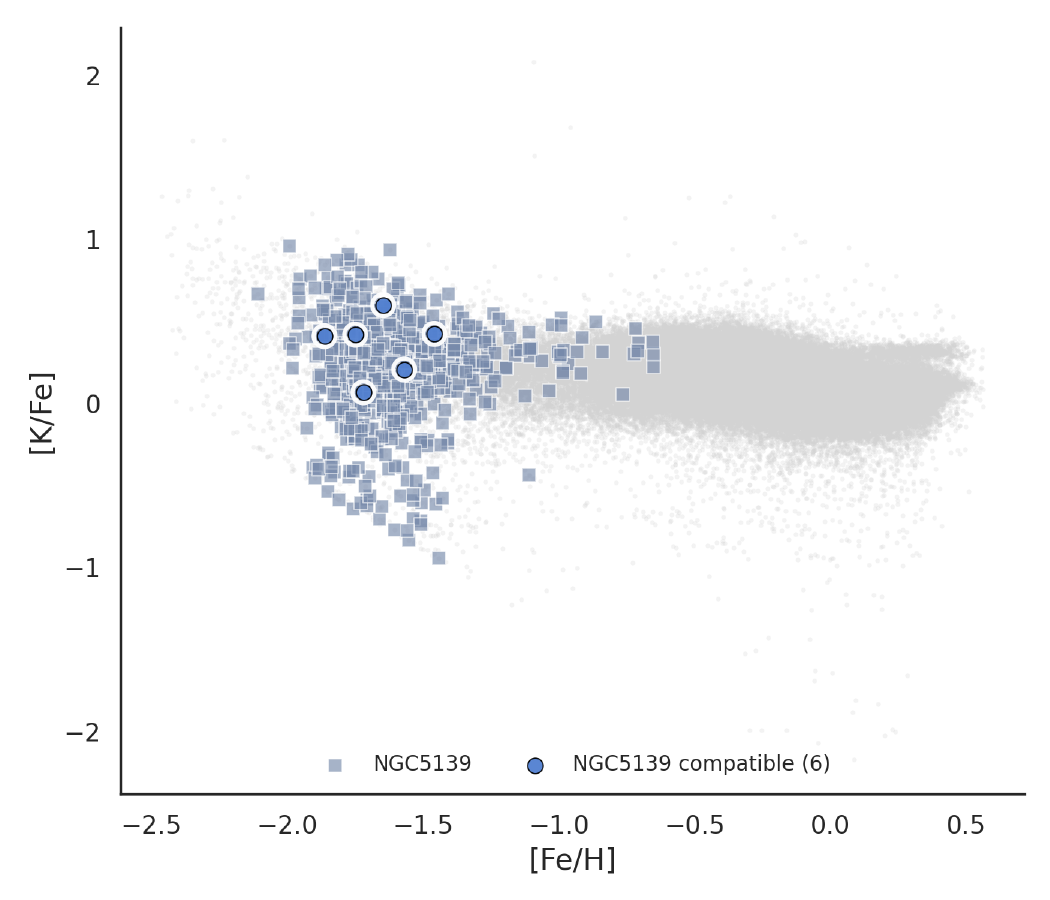}\par
\includegraphics[clip=true, trim = 1mm 0mm 0mm 1mm, width=0.33\linewidth]{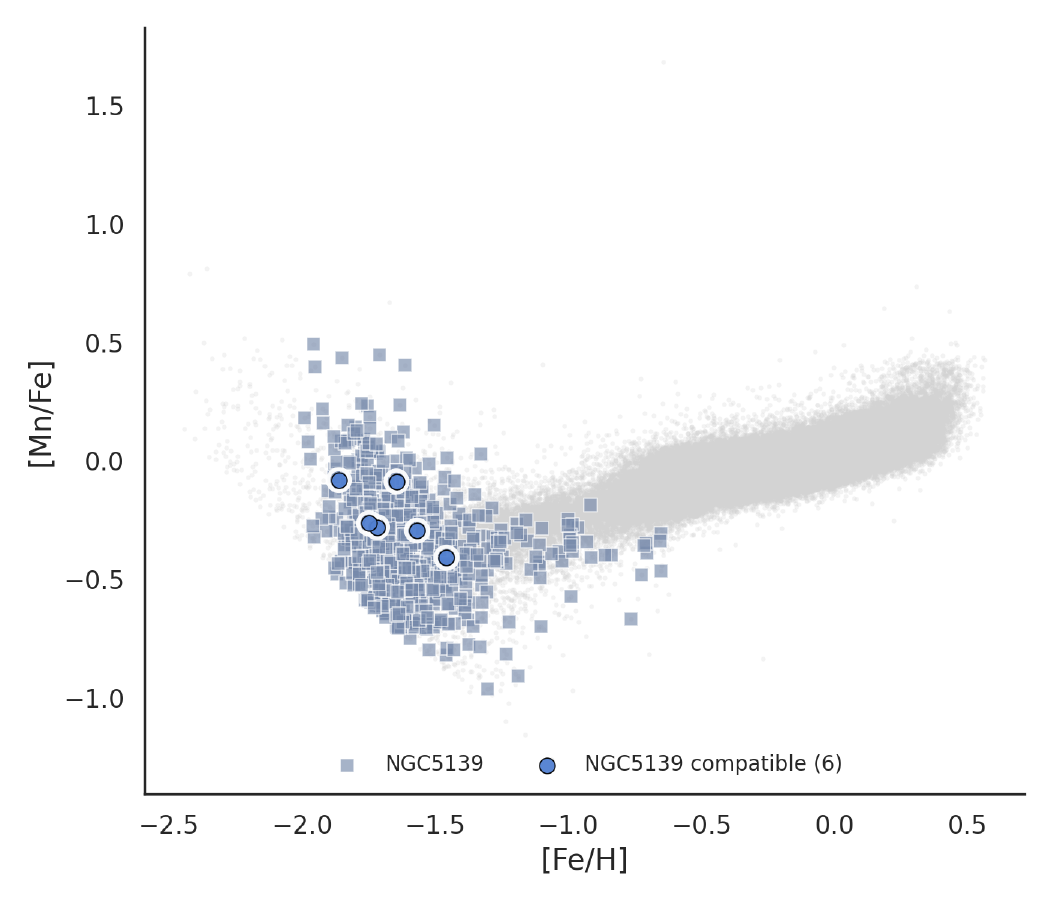}
  \caption{Chemical abundance relations for members of $\omega$~Cen (coloured squares) and field
stars that are chemically and kinematically compatible with it according to the GMMKin (coloured circles).  For comparison, all APOGEE field stars (grey points) are also shown.}
              \label{ngc5139_chem}%
    \end{figure*}

\begin{figure*}\centering
\includegraphics[clip=true, trim = 3mm 0mm 0mm 3mm, width=0.33\linewidth]{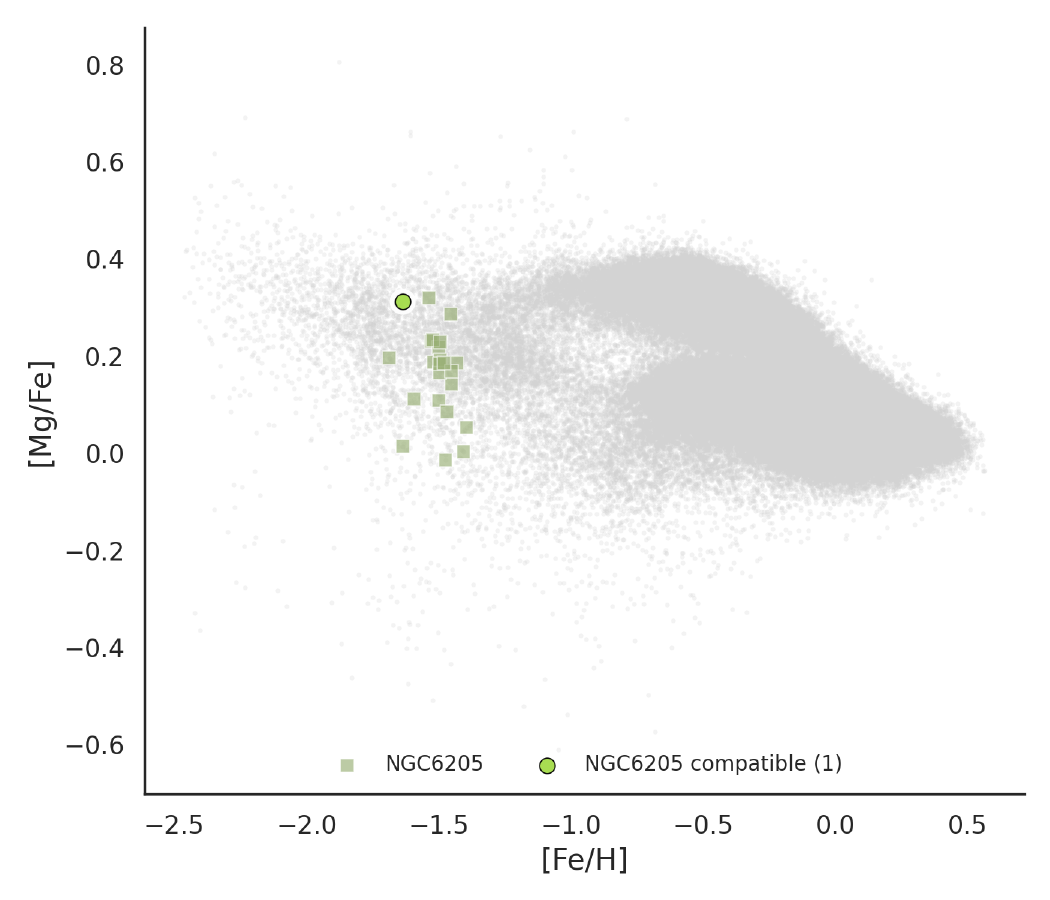}
\includegraphics[clip=true, trim = 3mm 0mm 0mm 3mm, width=0.33\linewidth]{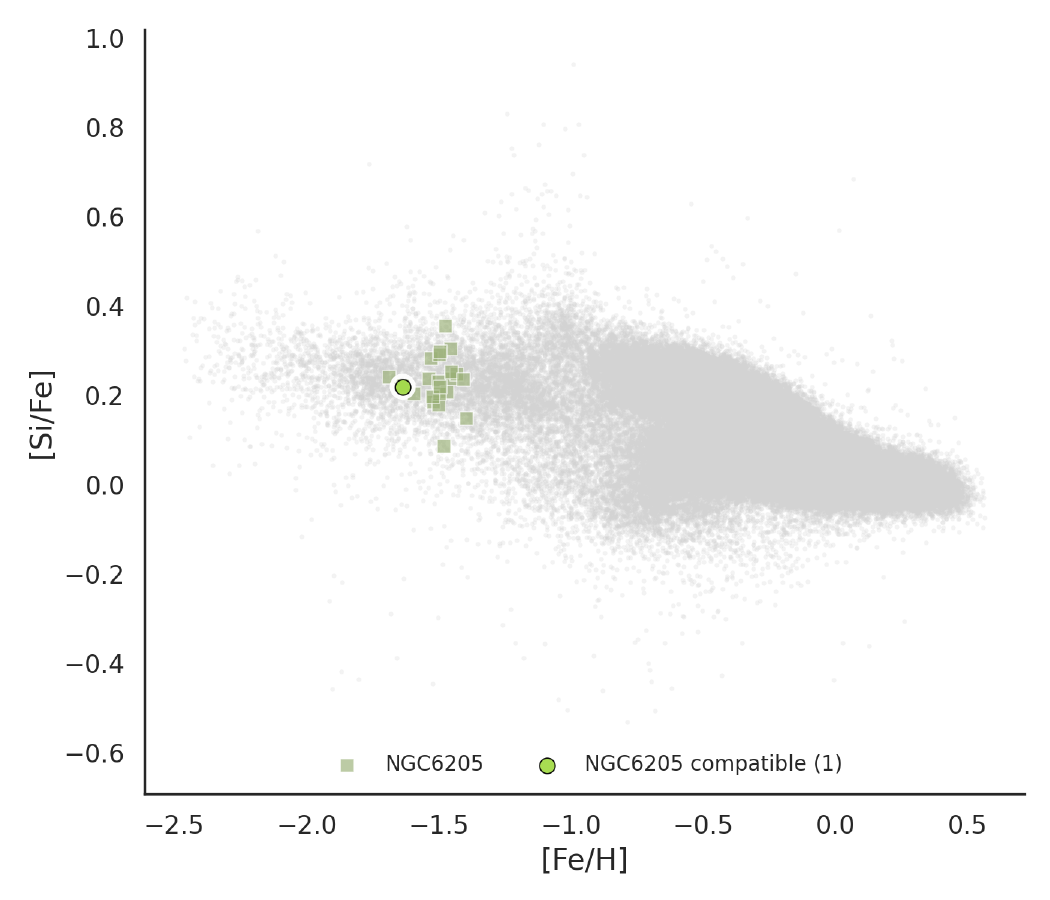}
\includegraphics[clip=true, trim = 3mm 0mm 0mm 3mm, width=0.33\linewidth]{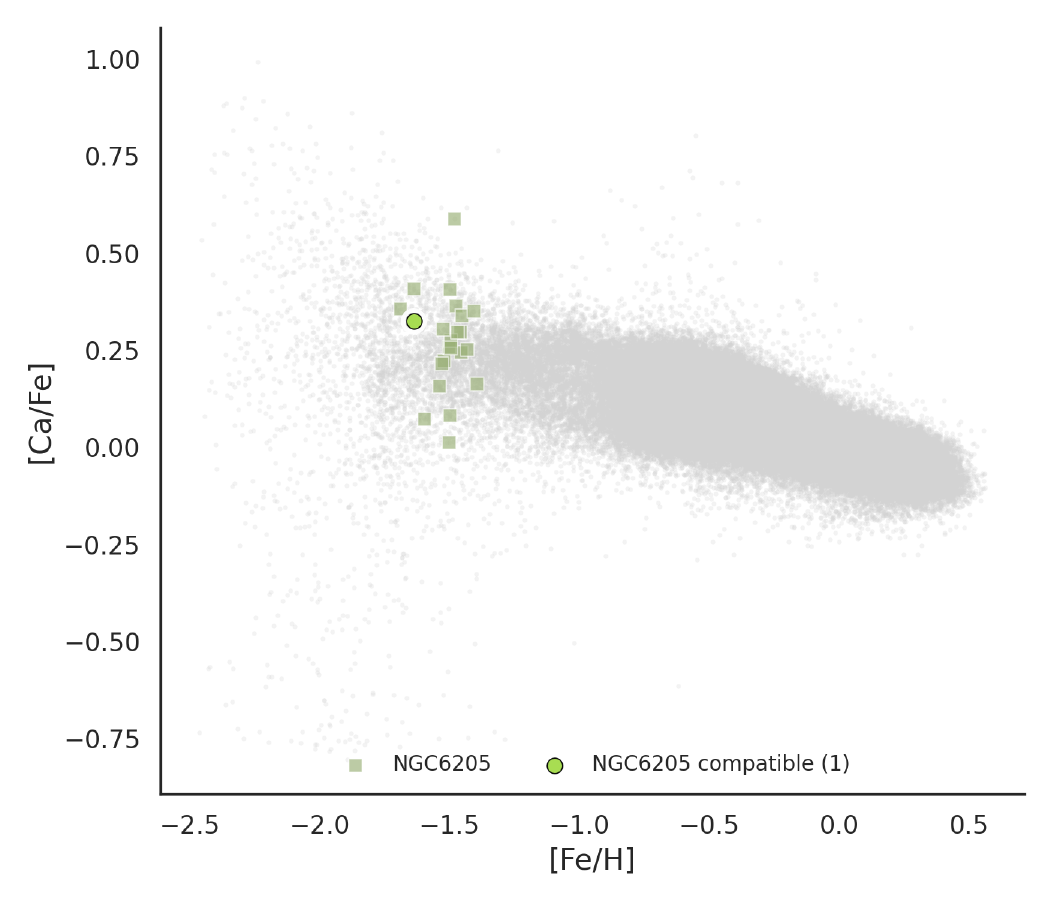}\par
\includegraphics[clip=true, trim = 8mm 60mm 10mm 70mm, width=0.33\linewidth]{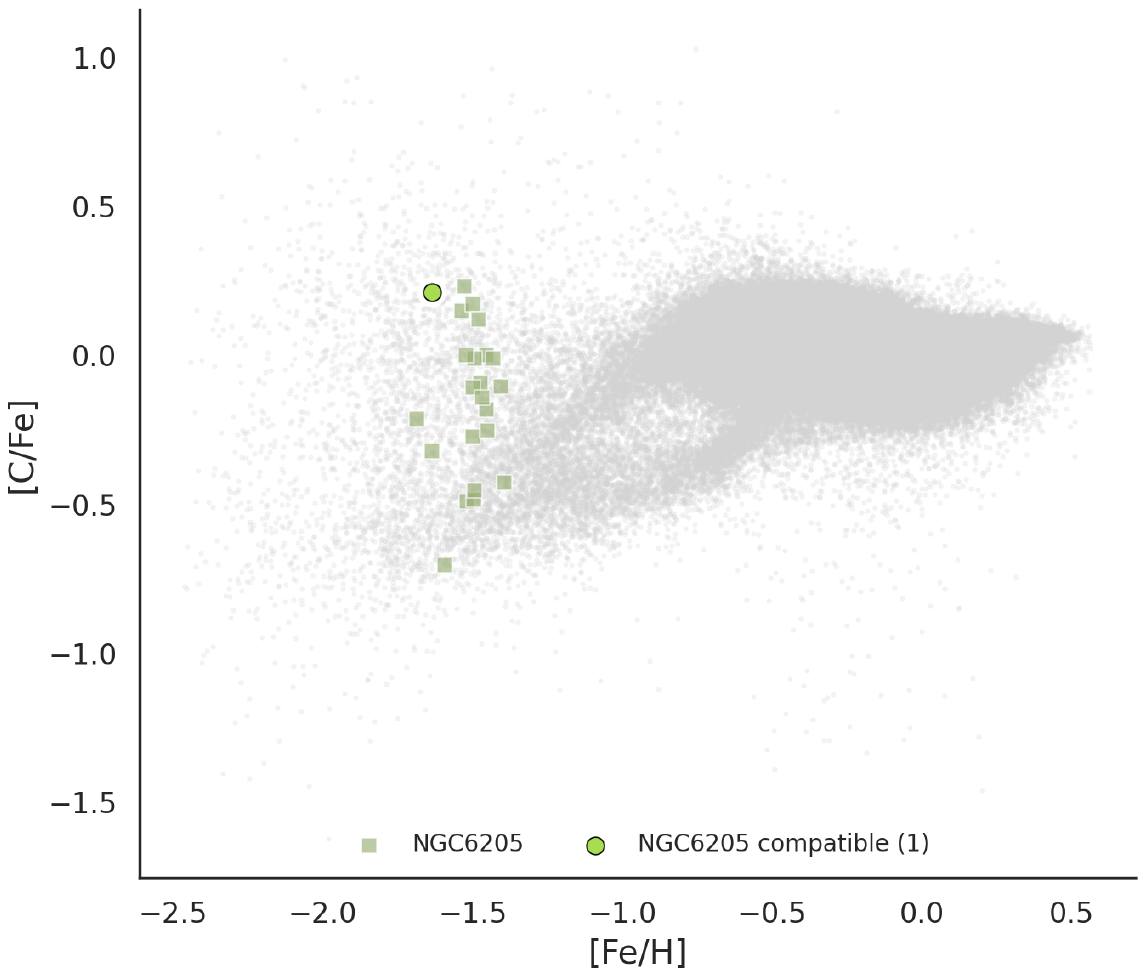}
\includegraphics[clip=true, trim = 2mm 0mm 0mm 1mm, width=0.33\linewidth]{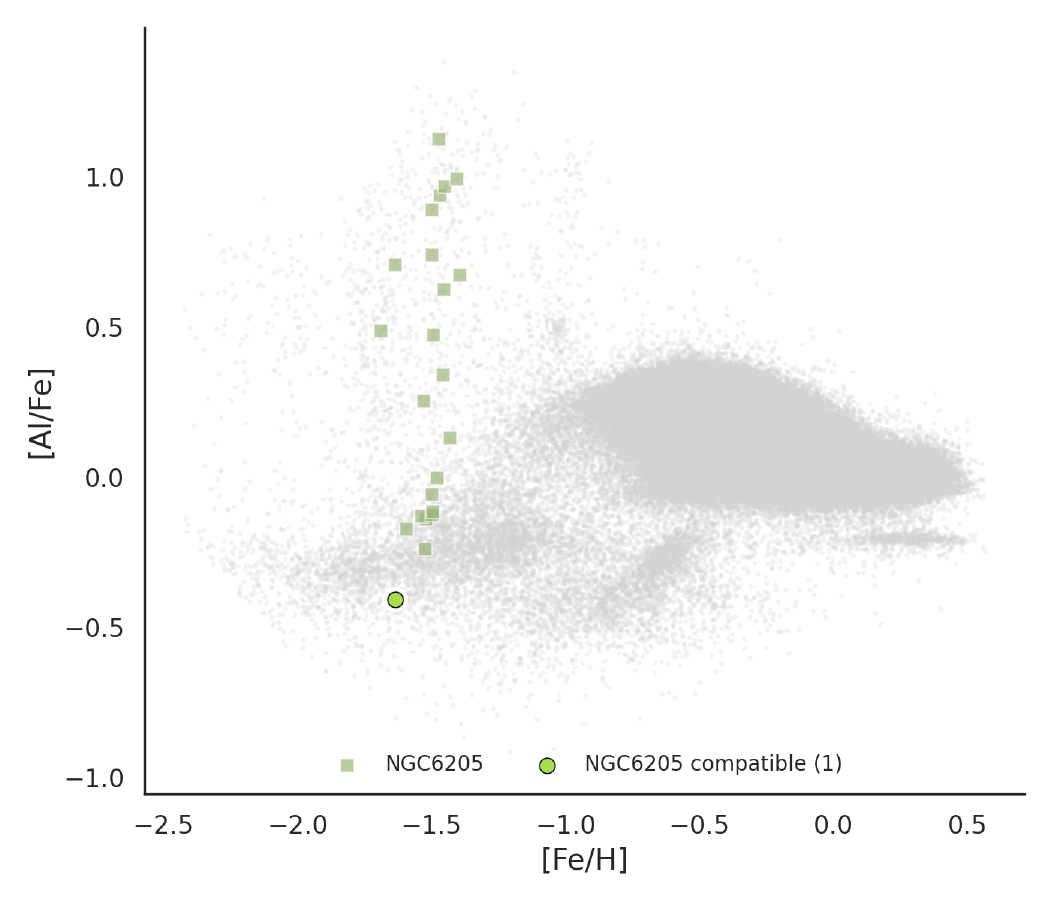}
\includegraphics[clip=true, trim = 2mm 0mm 0mm 1mm, width=0.33\linewidth]{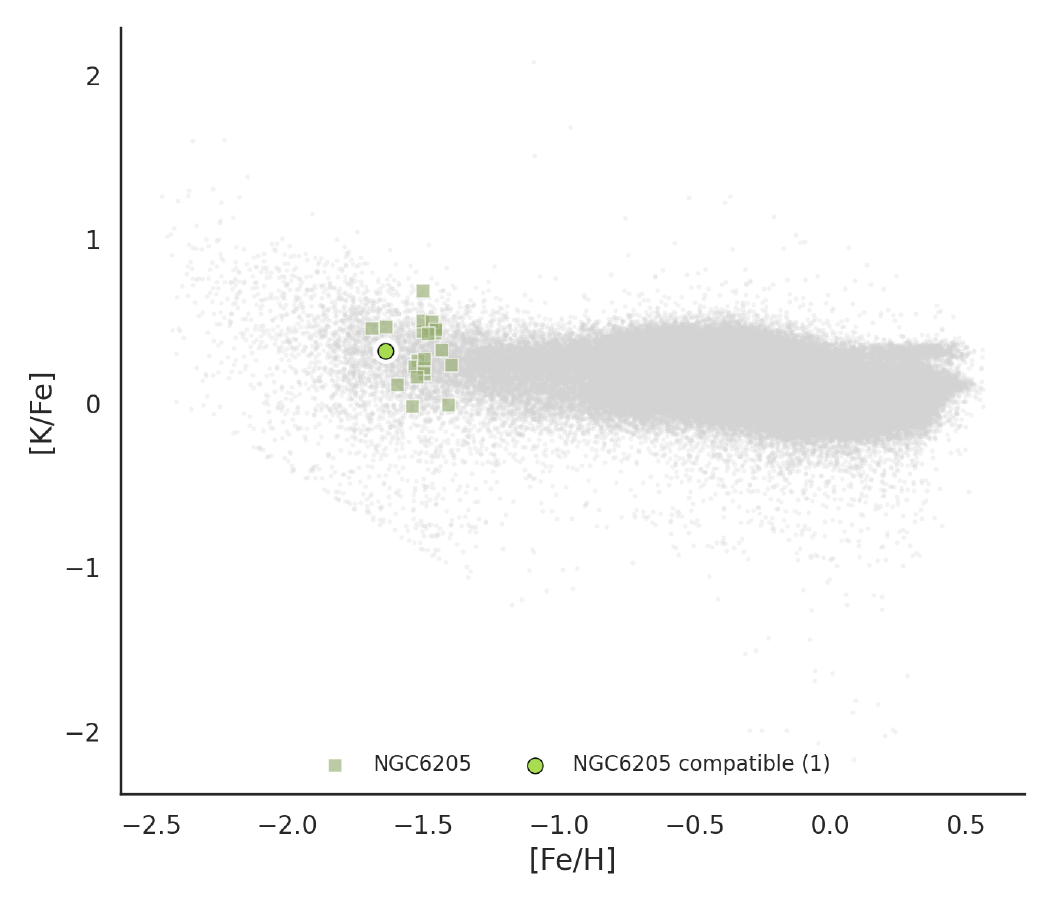}\par
\includegraphics[clip=true, trim = 1mm 0mm 0mm 1mm, width=0.33\linewidth]{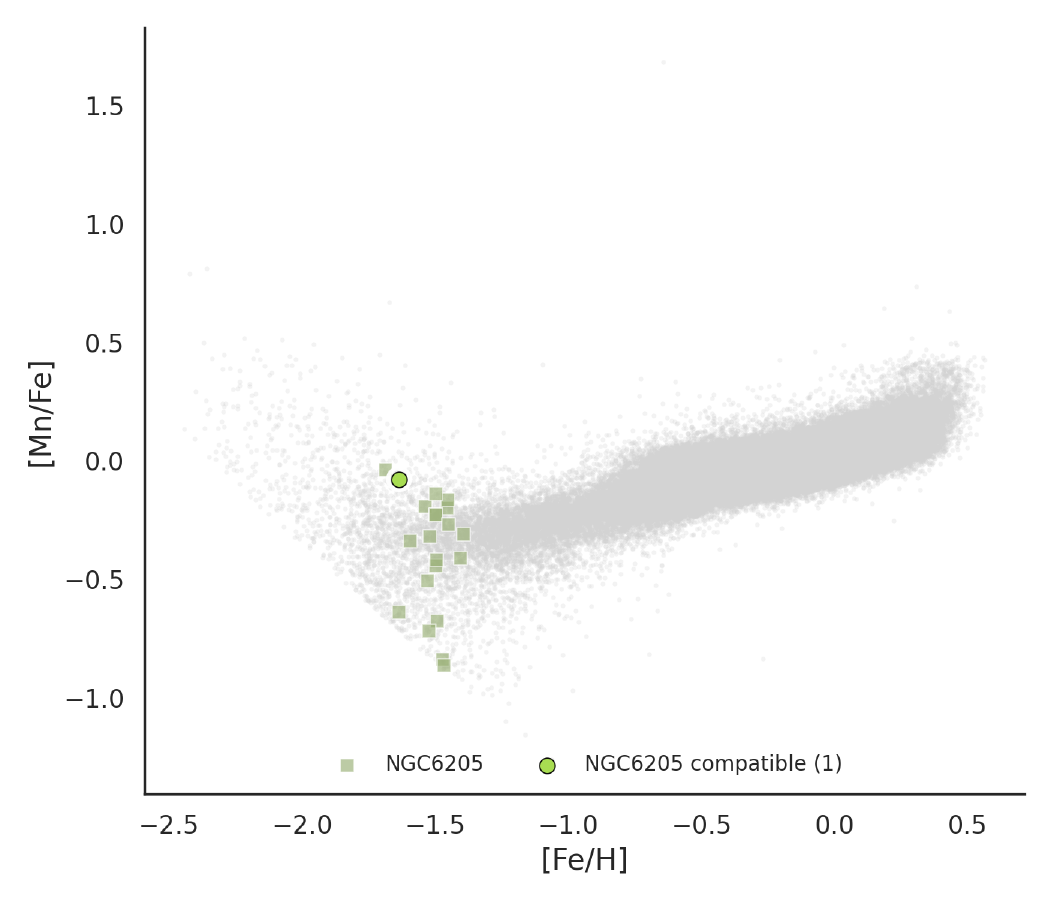}
  \caption{Chemical abundance relations for members of NGC~6205 (coloured squares) and Nephele candidates that kinematically compatible with NGC~6205 according to the GMMKin (coloured circles). Among these, stars outside the [Fe/H] range of the cluster under consideration are shown as empty circles. For comparison, all APOGEE field stars (grey points) are also shown.}
              \label{ngc6205_chem}%
    \end{figure*}

\begin{figure*}\centering
\includegraphics[clip=true, trim = 3mm 0mm 0mm 3mm, width=0.33\linewidth]{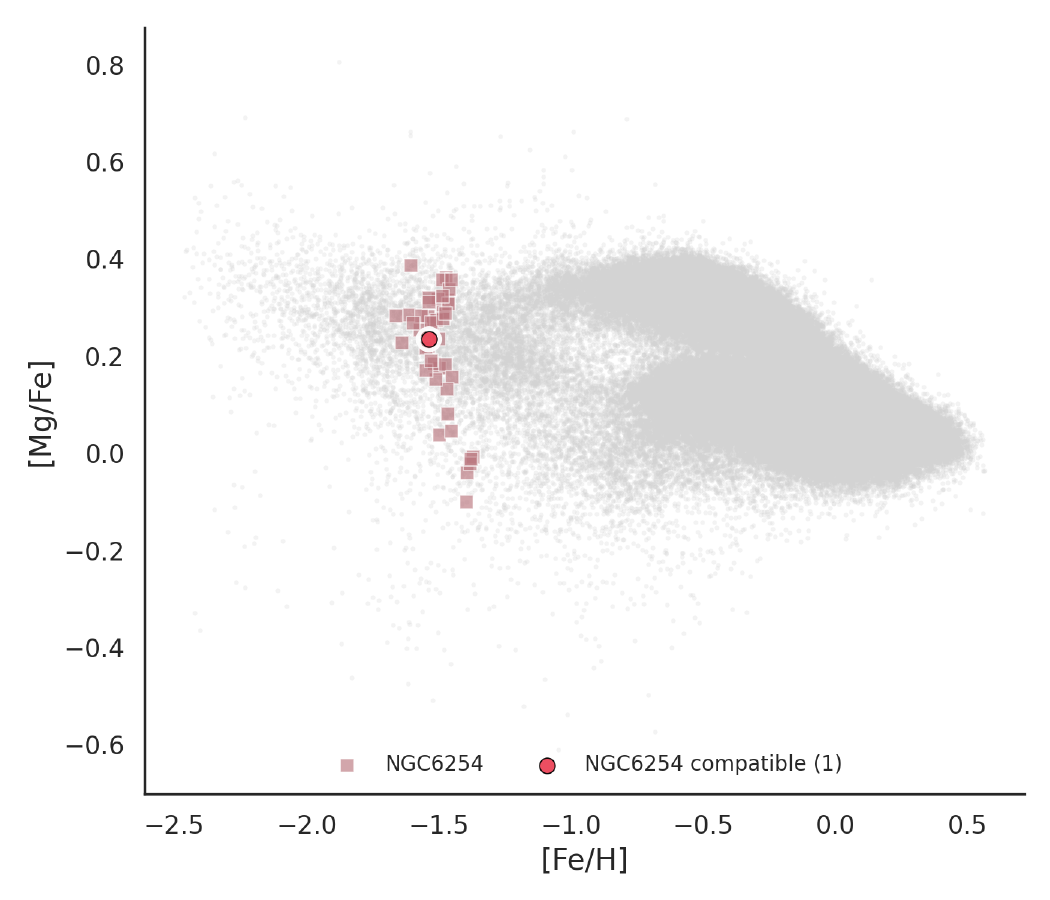}
\includegraphics[clip=true, trim = 3mm 0mm 0mm 3mm, width=0.33\linewidth]{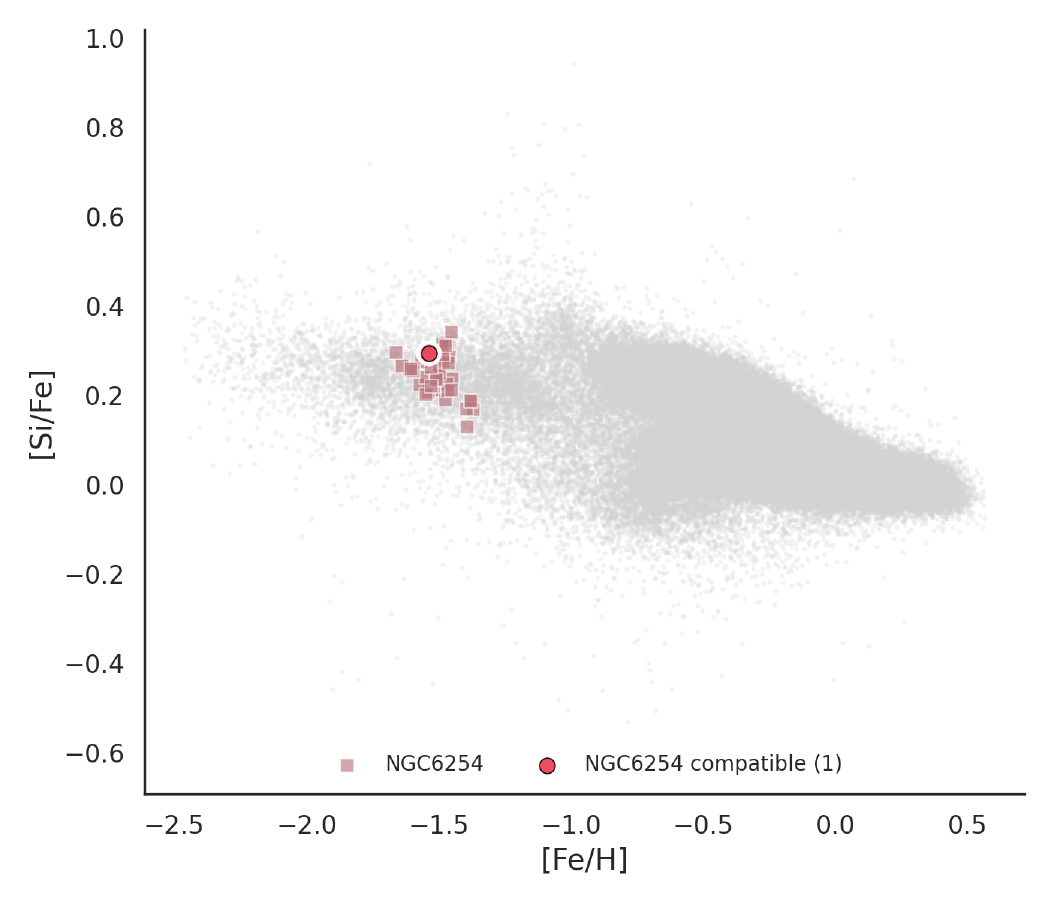}
\includegraphics[clip=true, trim = 3mm 0mm 0mm 3mm, width=0.33\linewidth]{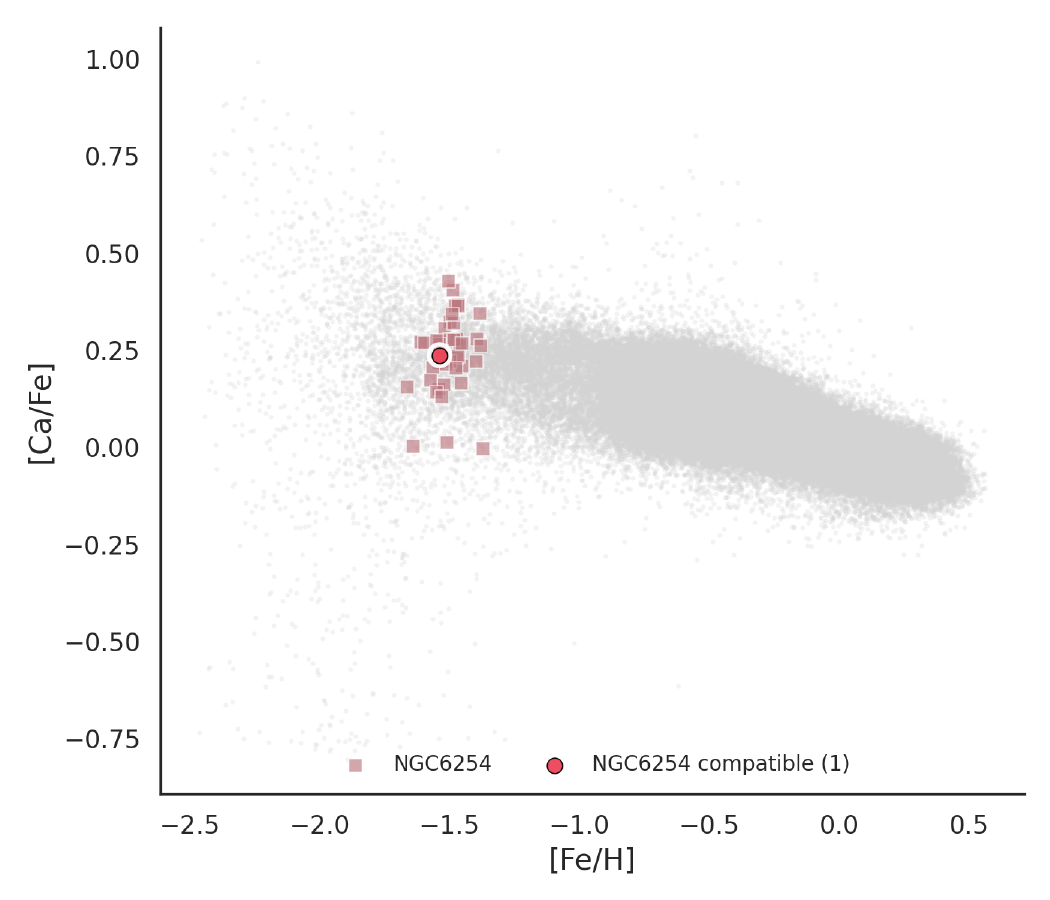}\par
\includegraphics[clip=true, trim = 3mm 0mm 2mm 0mm, width=0.33\linewidth]{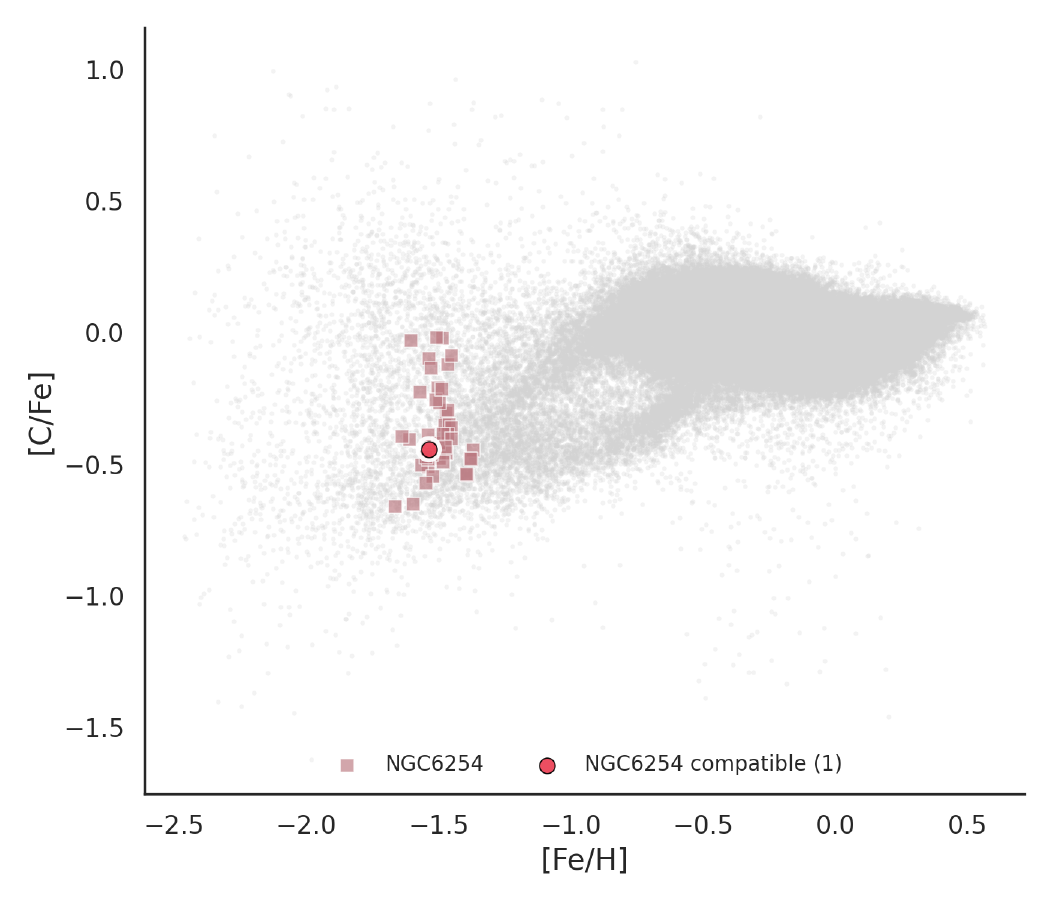}
\includegraphics[clip=true, trim = 2mm 0mm 0mm 1mm, width=0.33\linewidth]{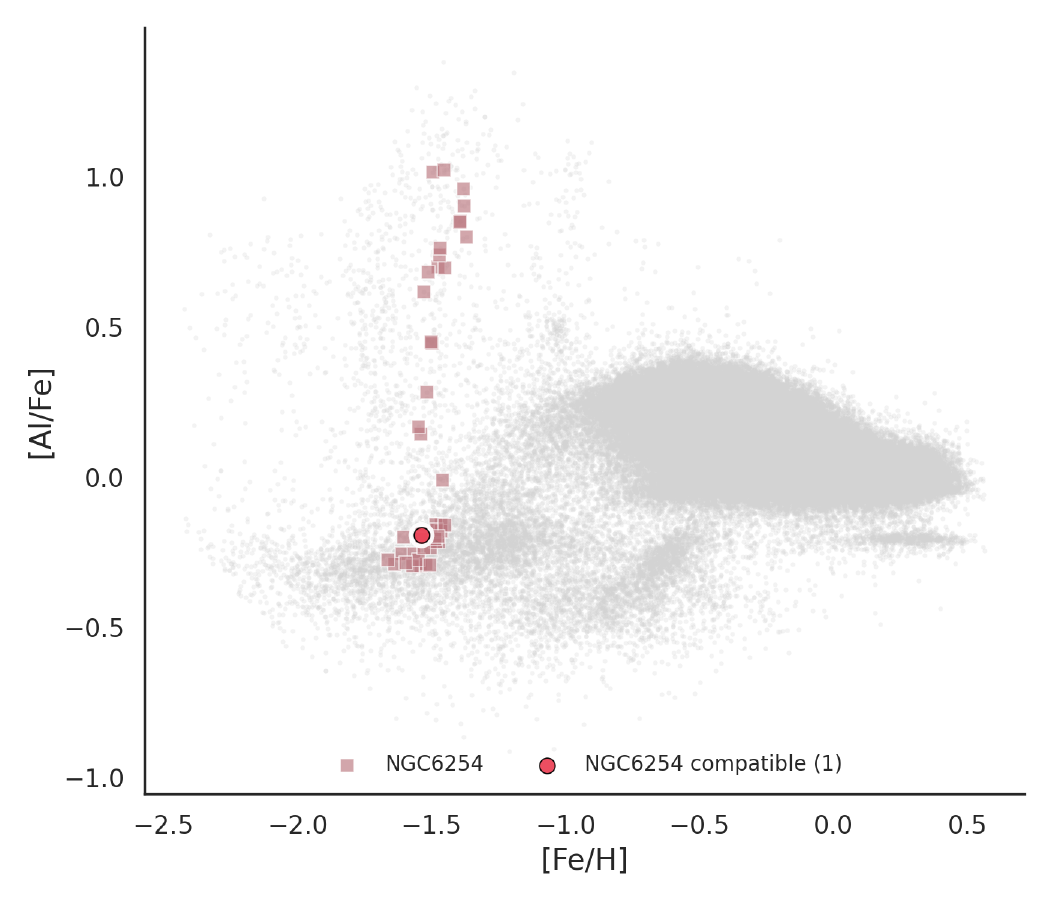}
\includegraphics[clip=true, trim = 2mm 0mm 0mm 1mm, width=0.33\linewidth]{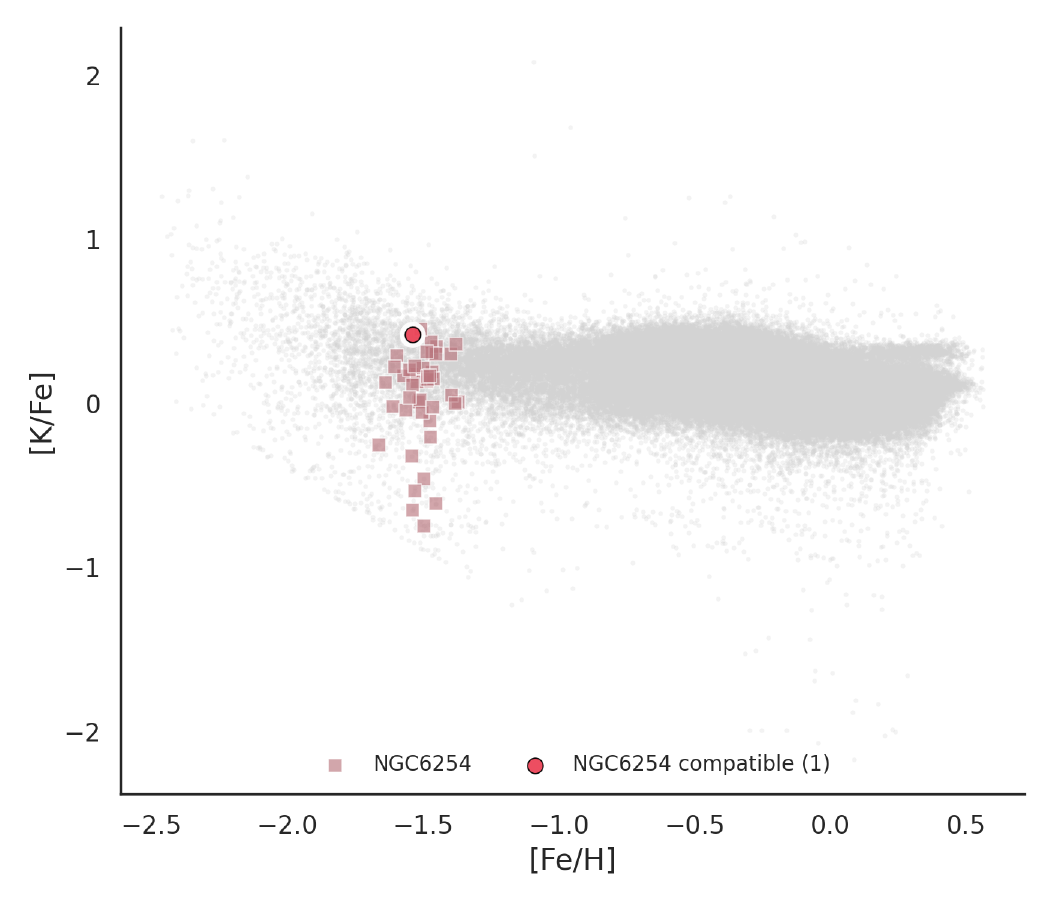}\par
\includegraphics[clip=true, trim = 10mm 80mm 10mm 80mm, width=0.33\linewidth]{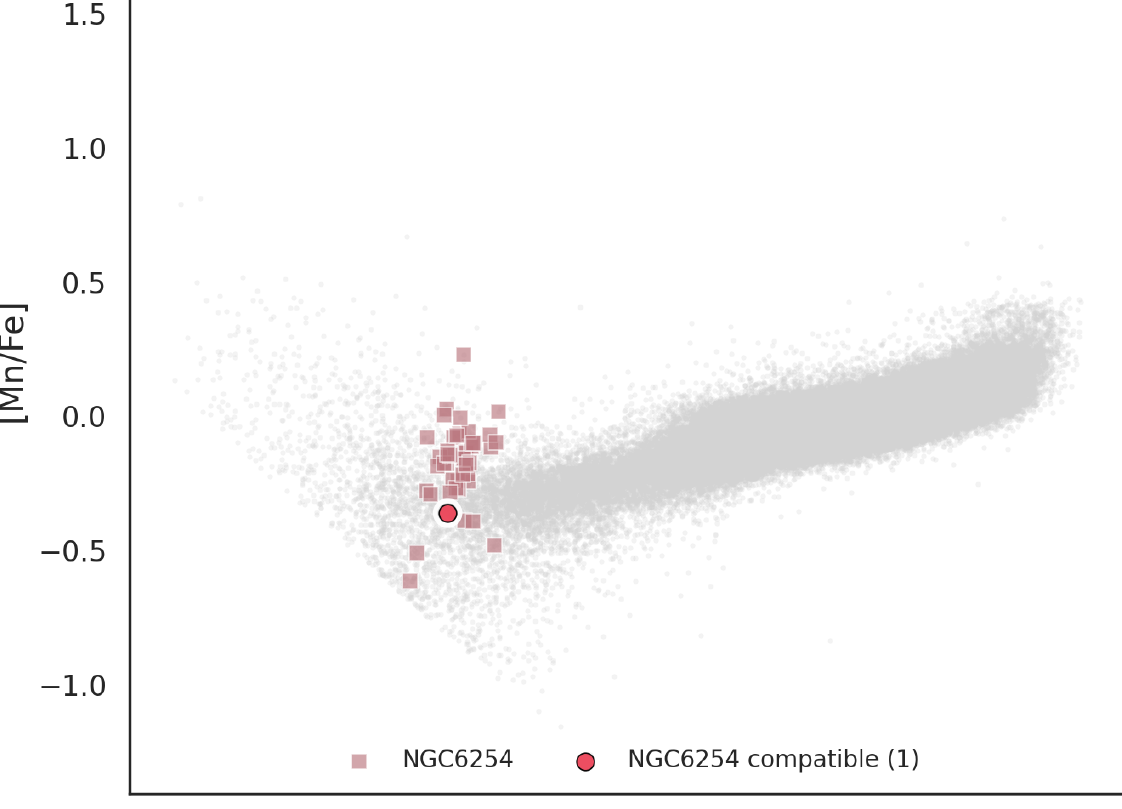}
  \caption{Same as Fig.~\ref{ngc6205_chem} but for NGC~6254.}
              \label{ngc6254_chem}%
    \end{figure*}

\begin{figure*}\centering
\includegraphics[clip=true, trim = 3mm 0mm 0mm 3mm, width=0.33\linewidth]{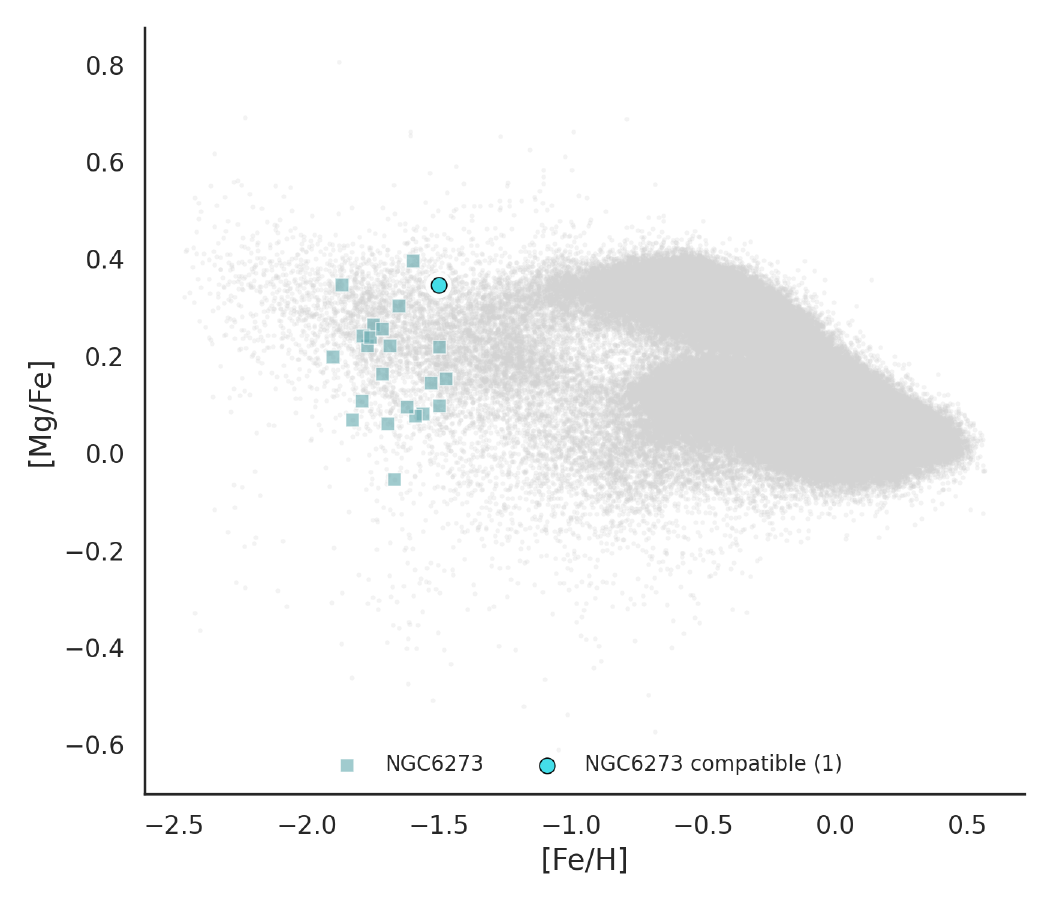}
\includegraphics[clip=true, trim = 3mm 0mm 0mm 3mm, width=0.33\linewidth]{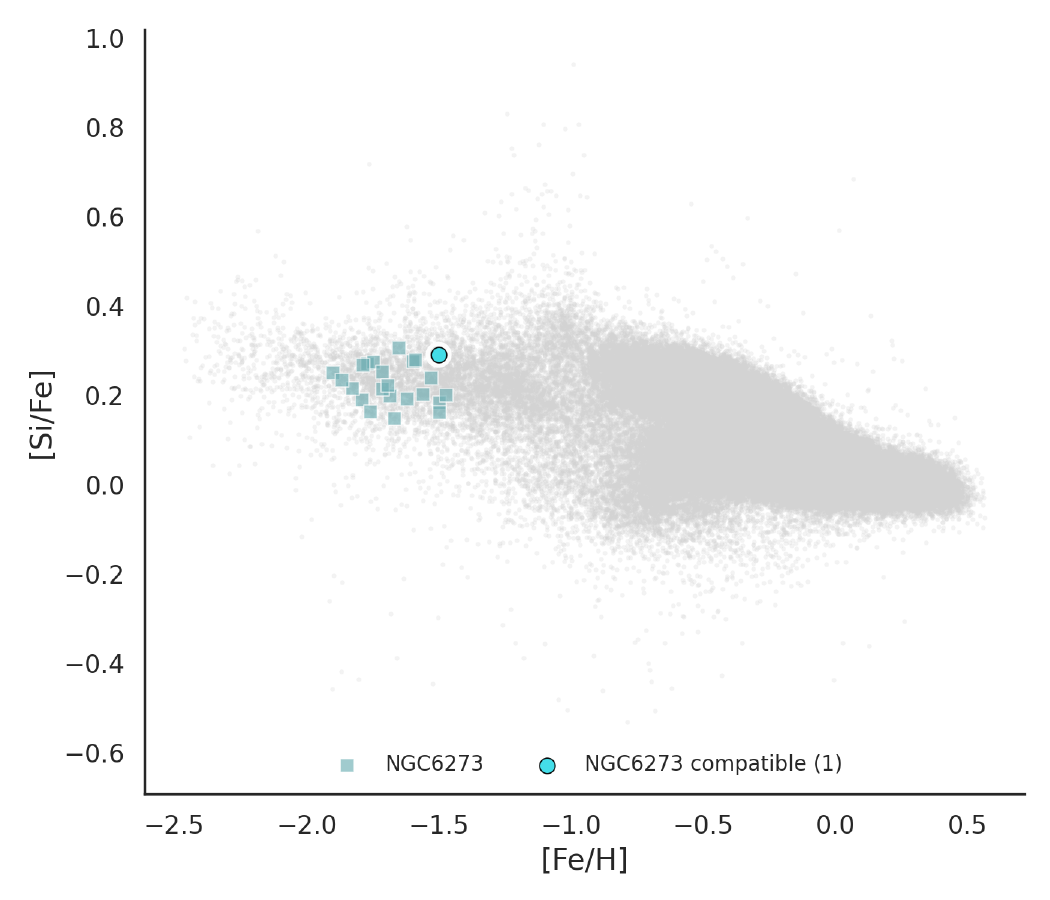}
\includegraphics[clip=true, trim = 3mm 0mm 0mm 3mm, width=0.33\linewidth]{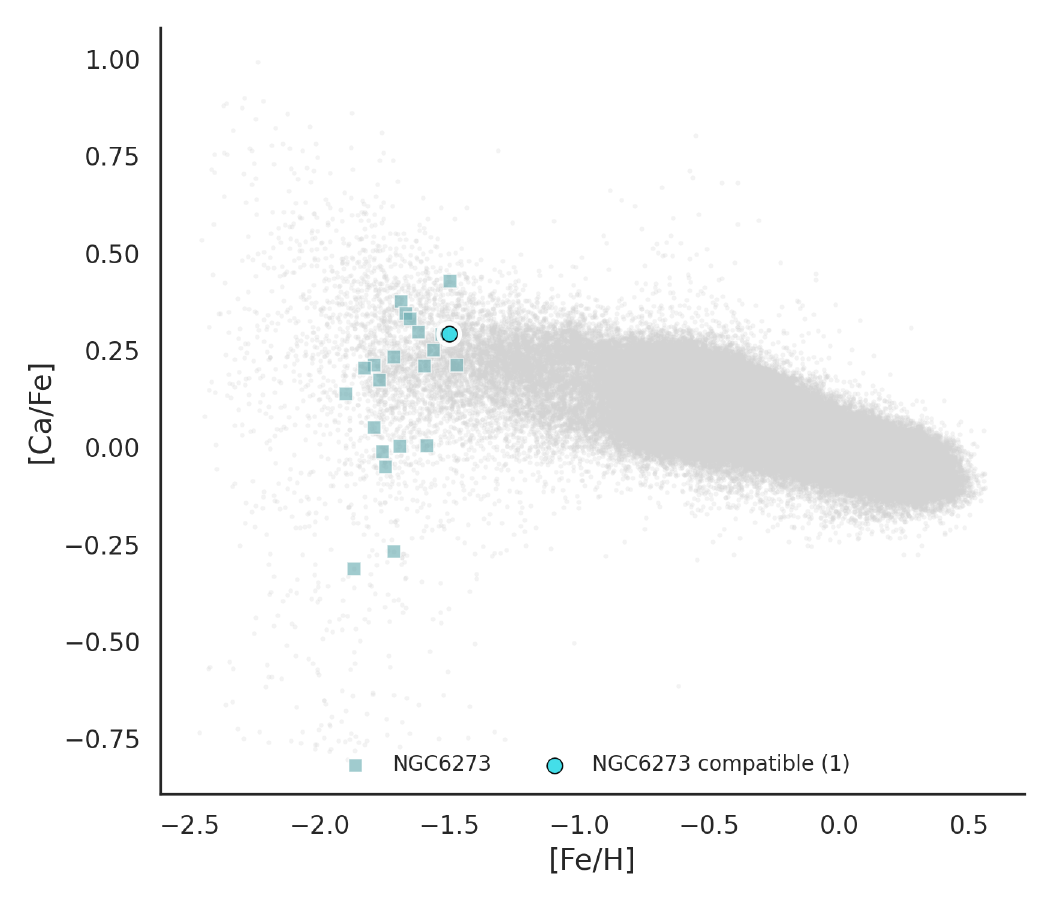}\par
\includegraphics[clip=true, trim = 3mm 0mm 2mm 0mm, width=0.33\linewidth]{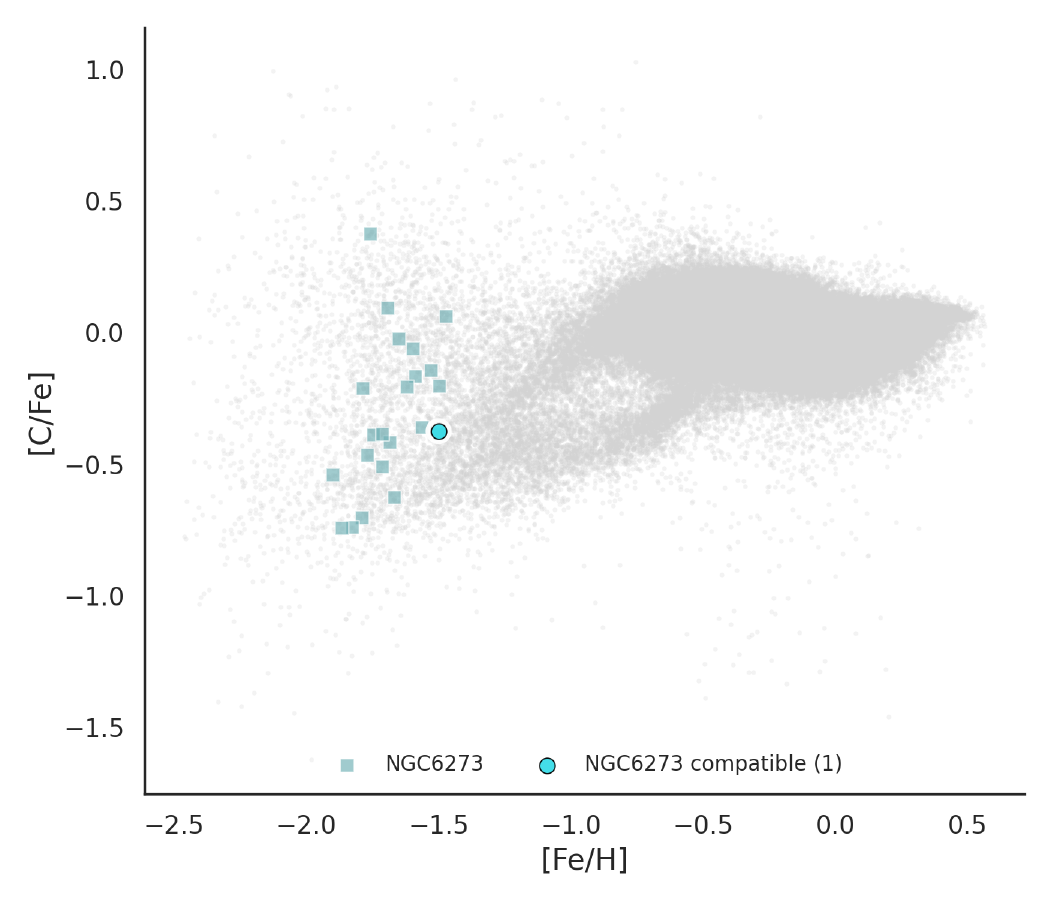}
\includegraphics[clip=true, trim = 2mm 0mm 0mm 1mm, width=0.33\linewidth]{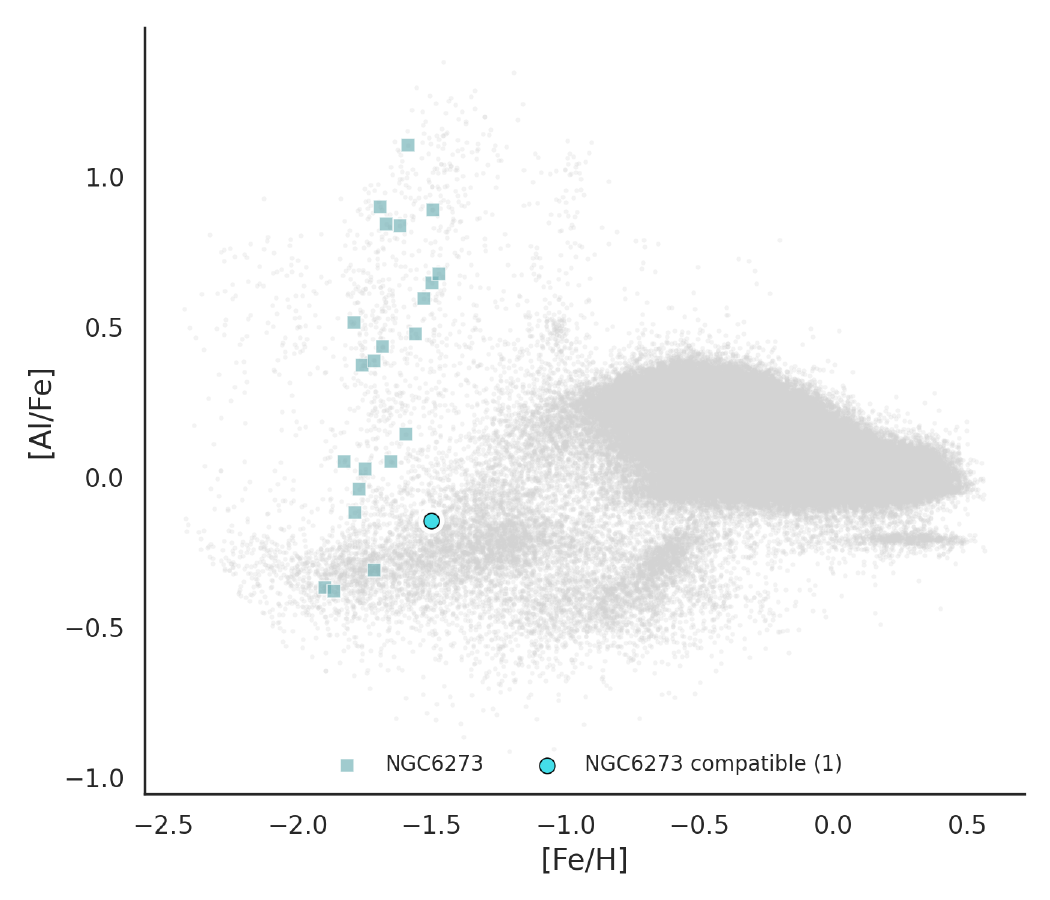}
\includegraphics[clip=true, trim = 2mm 0mm 0mm 1mm, width=0.33\linewidth]{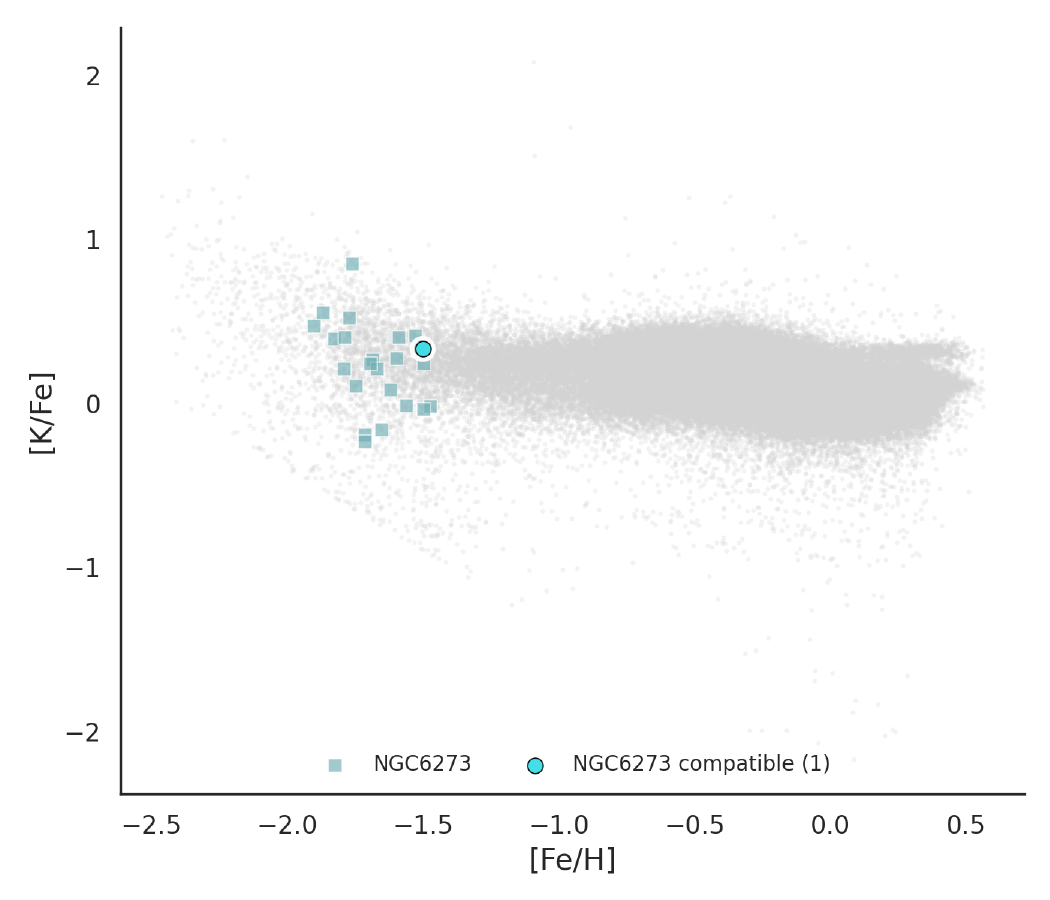}\par
\includegraphics[clip=true, trim = 1mm 0mm 0mm 1mm, width=0.33\linewidth]{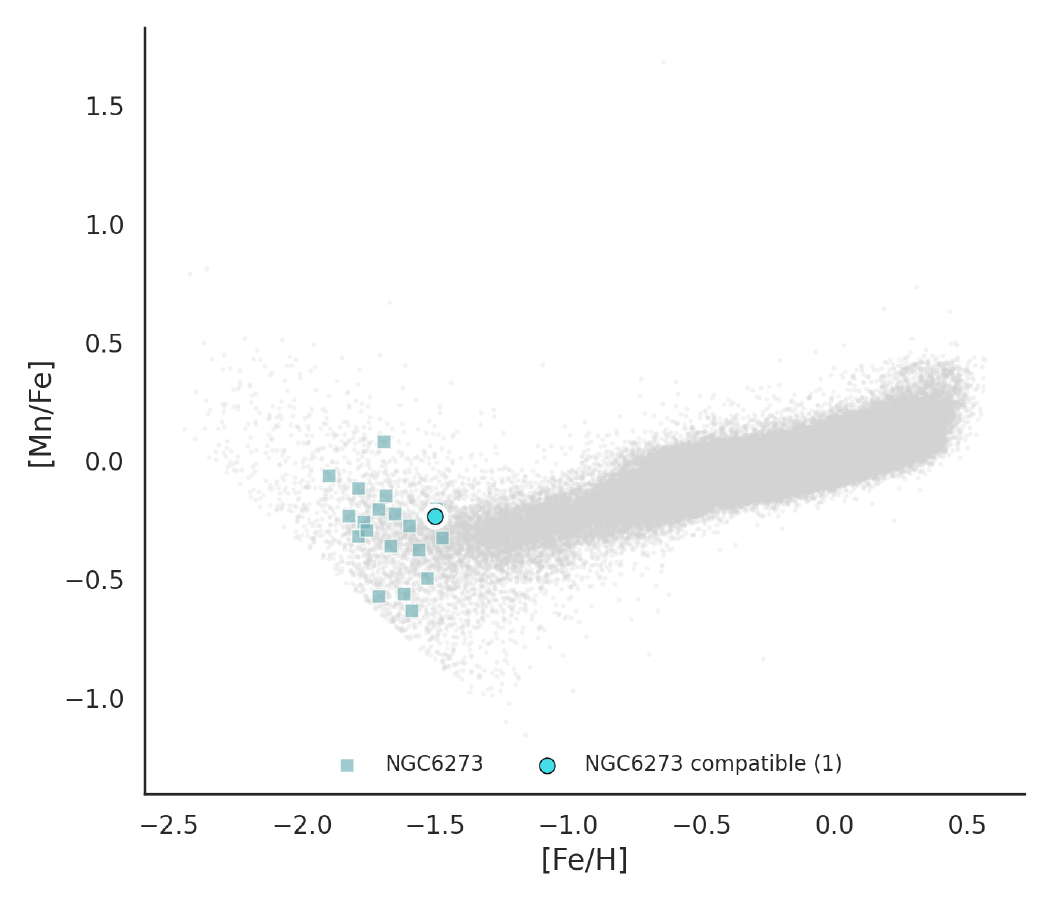}
  \caption{Same as Fig.~\ref{ngc6205_chem} but for NGC~6273.}
              \label{ngc6273_chem}%
    \end{figure*}

\begin{figure*}\centering
\includegraphics[clip=true, trim = 3mm 0mm 0mm 3mm, width=0.33\linewidth]{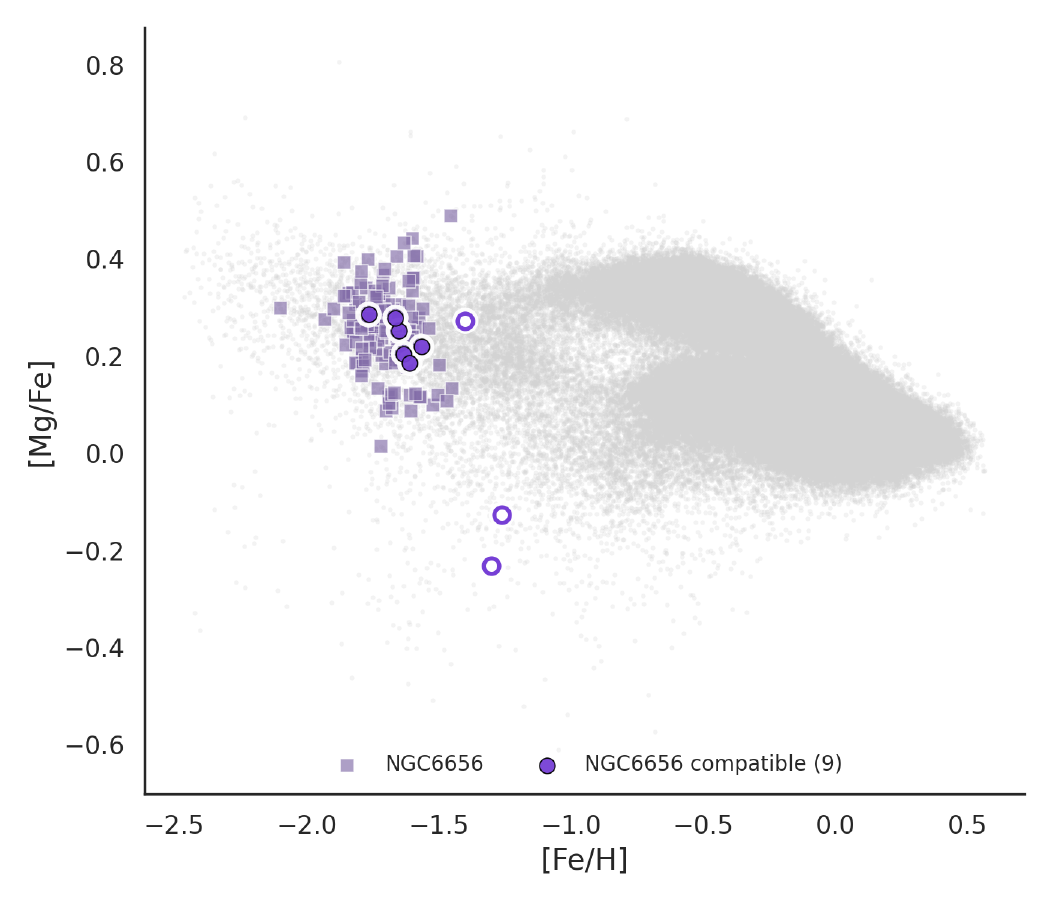}
\includegraphics[clip=true, trim = 3mm 0mm 0mm 3mm, width=0.33\linewidth]{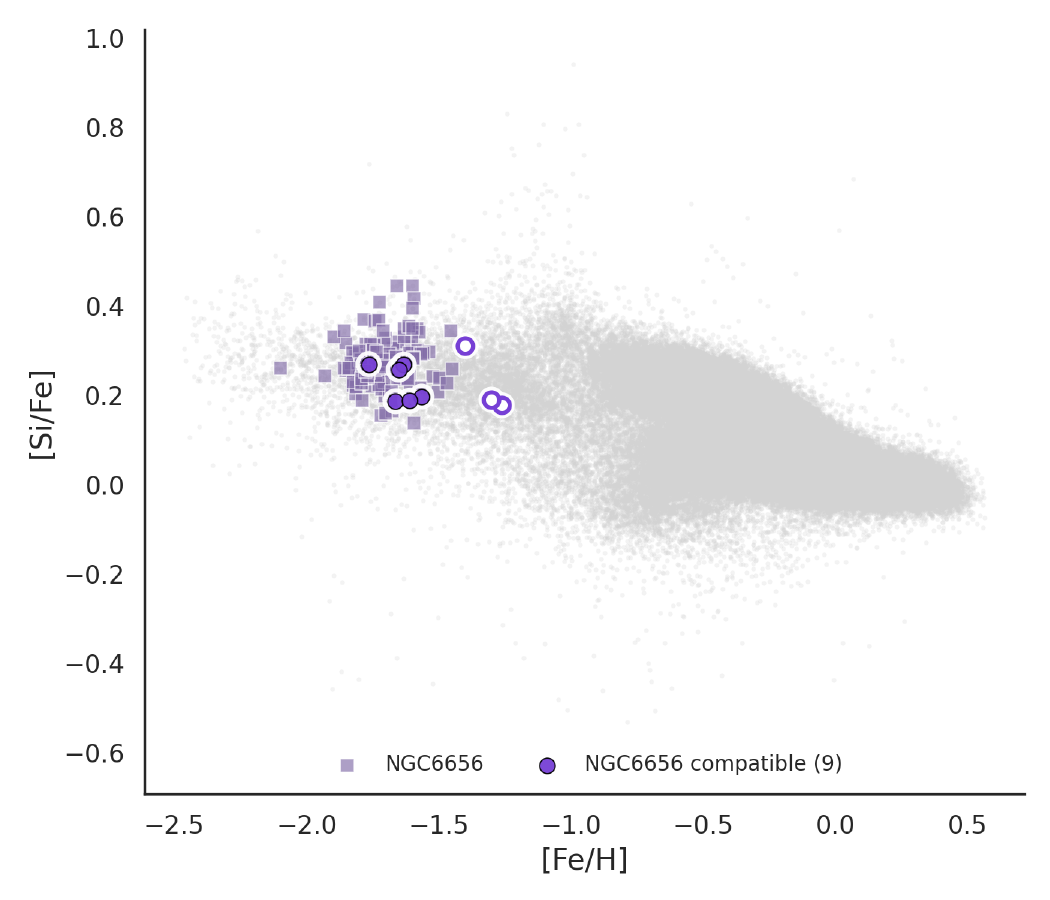}
\includegraphics[clip=true, trim = 3mm 0mm 0mm 3mm, width=0.33\linewidth]{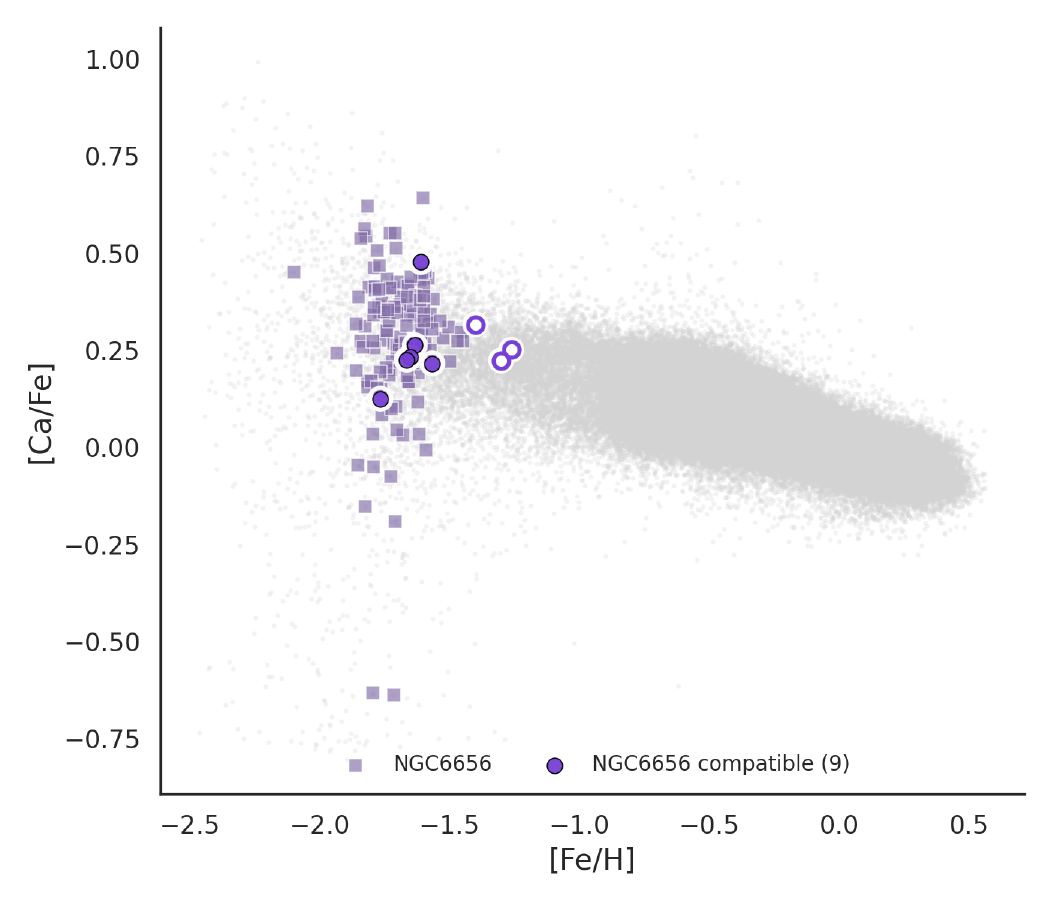}\par
\includegraphics[clip=true, trim = 3mm 0mm 2mm 0mm, width=0.33\linewidth]{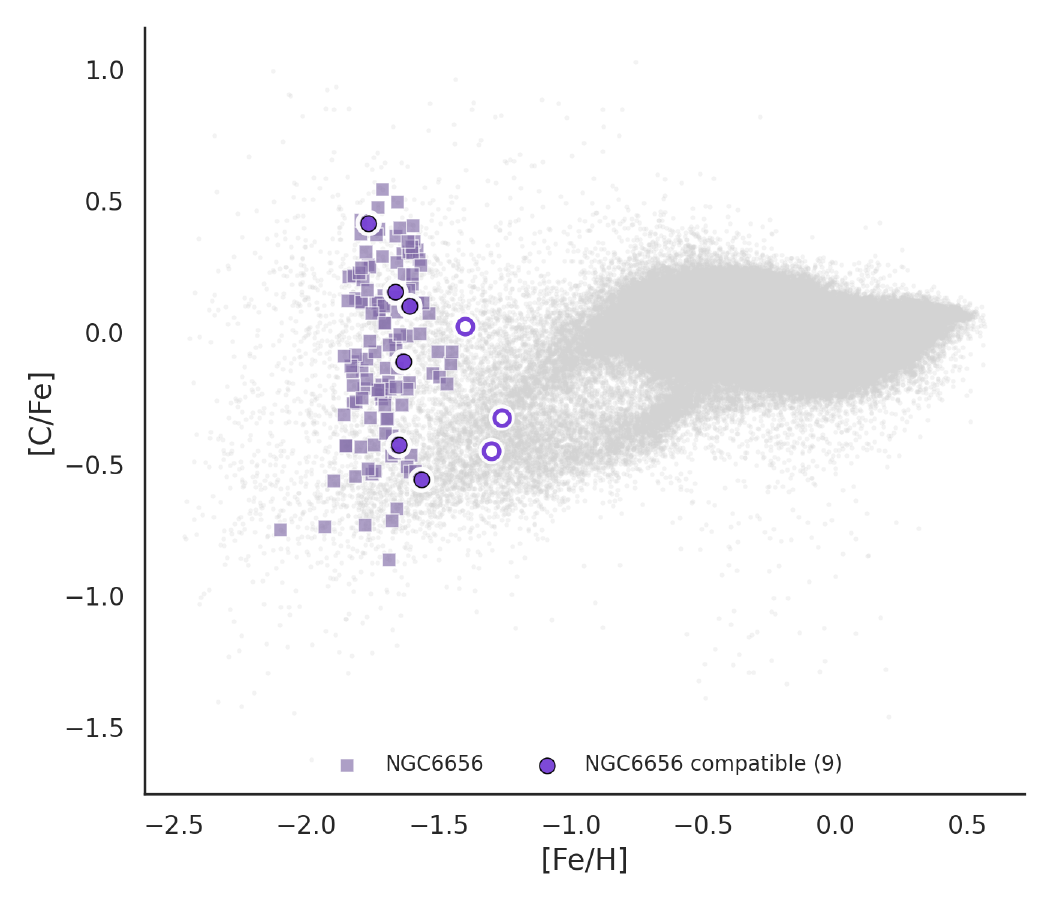}
\includegraphics[clip=true, trim = 2mm 0mm 0mm 1mm, width=0.33\linewidth]{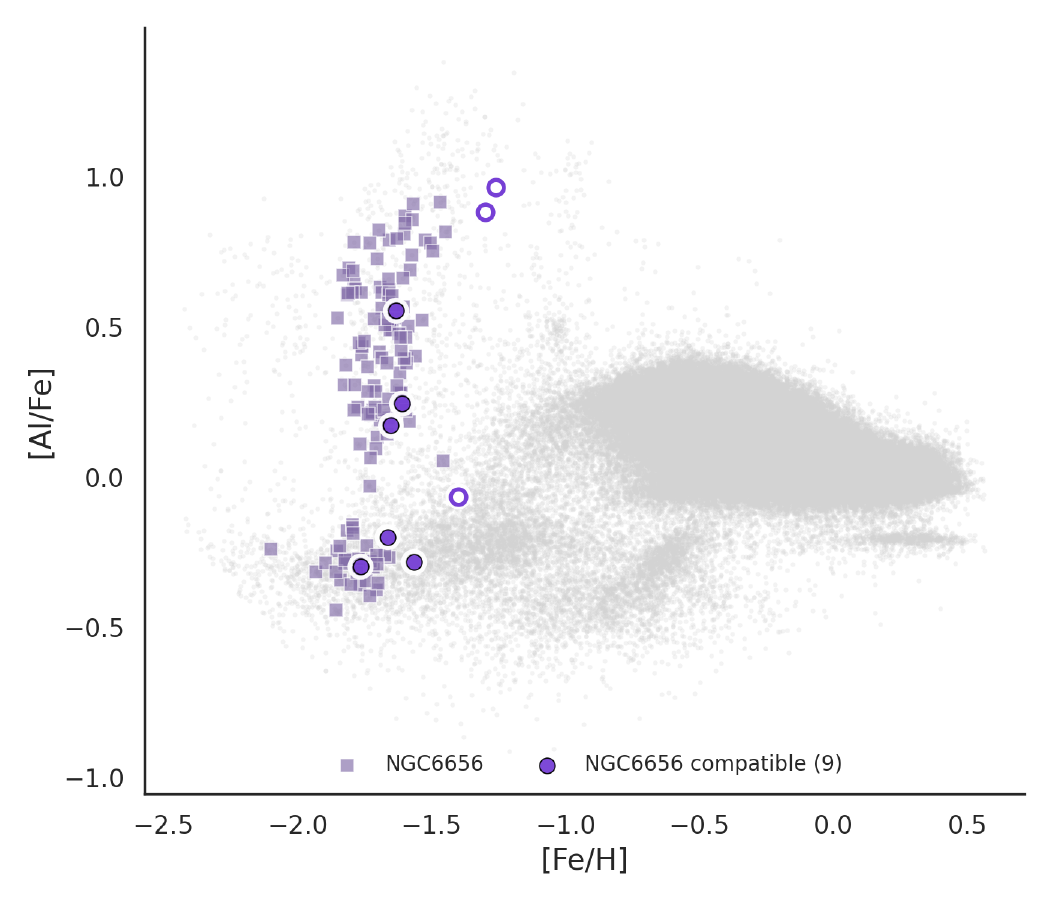}
\includegraphics[clip=true, trim = 2mm 0mm 0mm 1mm, width=0.33\linewidth]{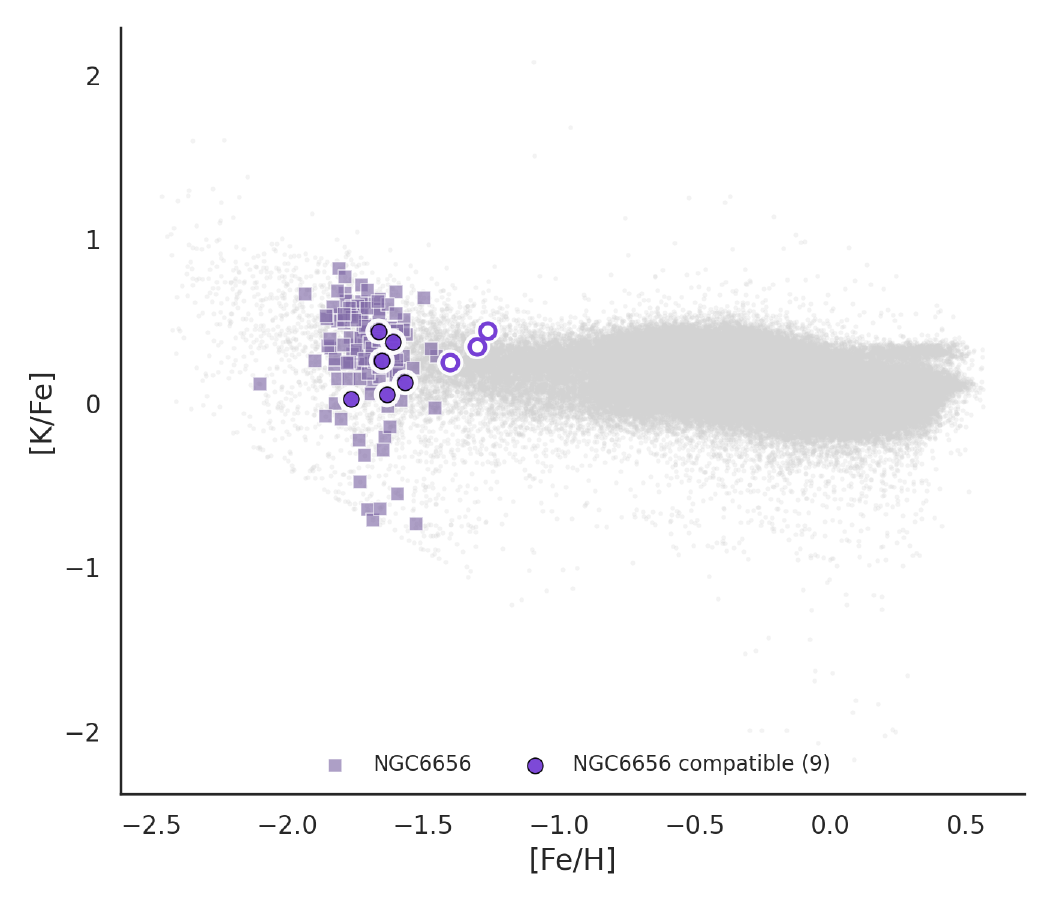}\par
\includegraphics[clip=true, trim = 1mm 0mm 0mm 1mm, width=0.33\linewidth]{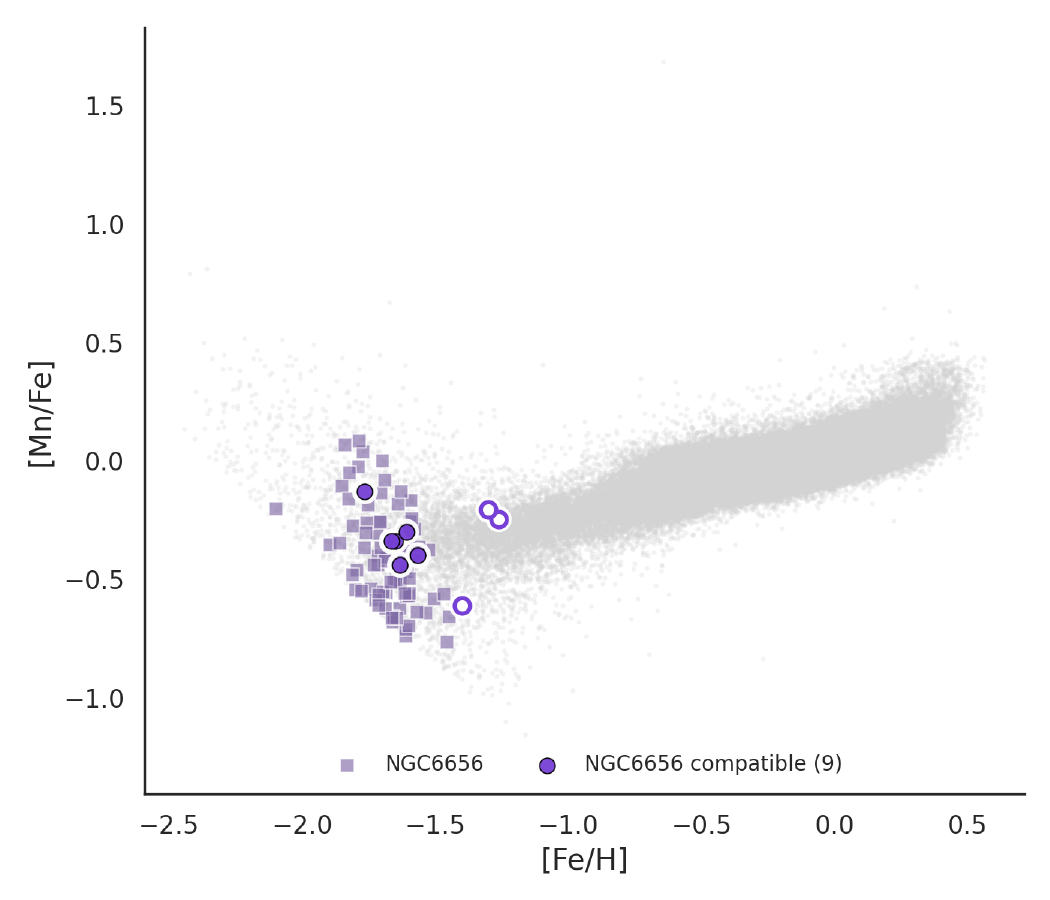}
  \caption{Same as Fig.~\ref{ngc6205_chem} but for NGC~6656.}
              \label{ngc6656_chem}%
    \end{figure*}

\begin{figure*}\centering
\includegraphics[clip=true, trim = 3mm 0mm 0mm 3mm, width=0.33\linewidth]{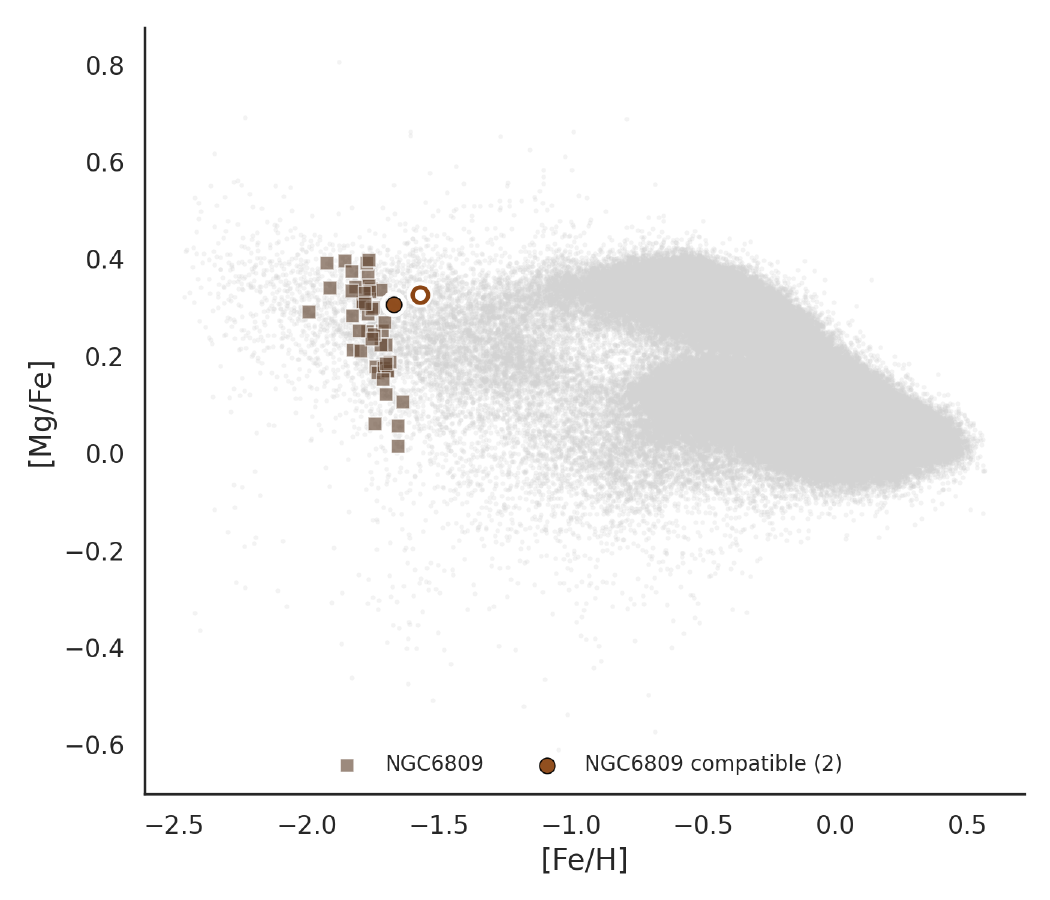}
\includegraphics[clip=true, trim = 3mm 0mm 0mm 3mm, width=0.33\linewidth]{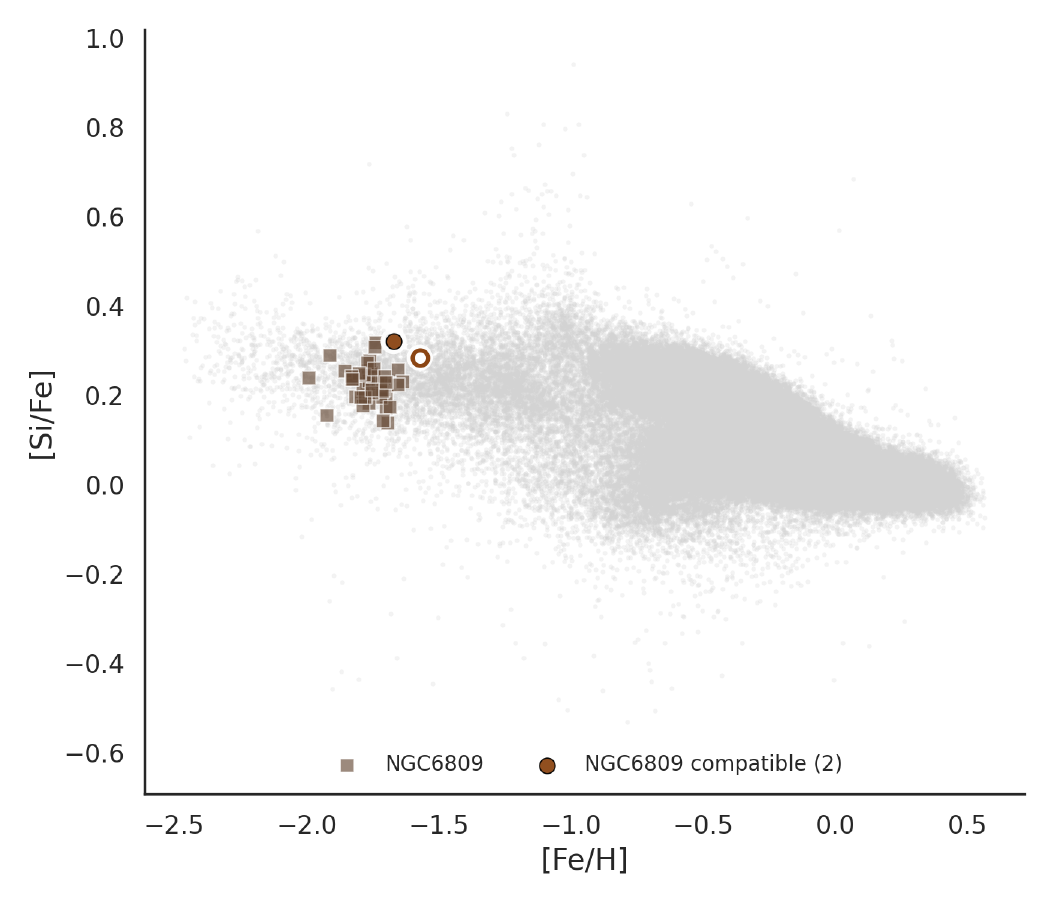}
\includegraphics[clip=true, trim = 3mm 0mm 0mm 3mm, width=0.33\linewidth]{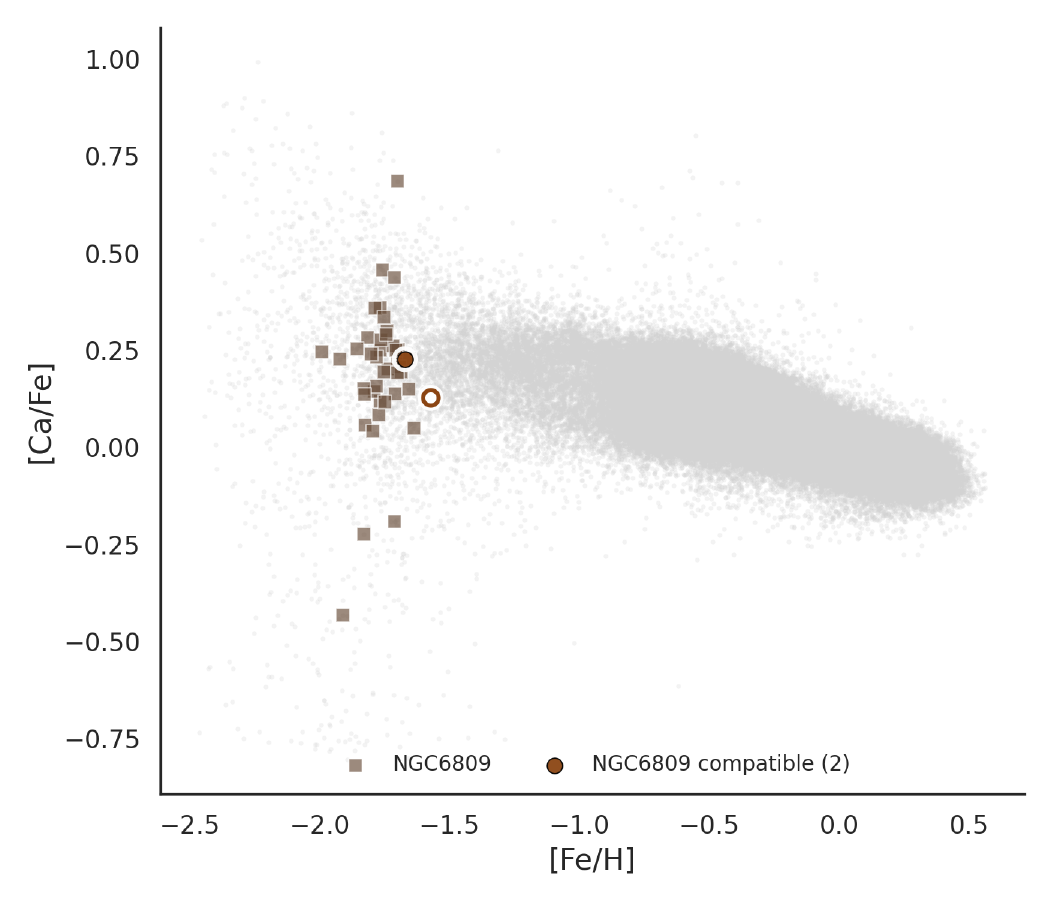}\par
\includegraphics[clip=true, trim = 3mm 0mm 2mm 0mm, width=0.33\linewidth]{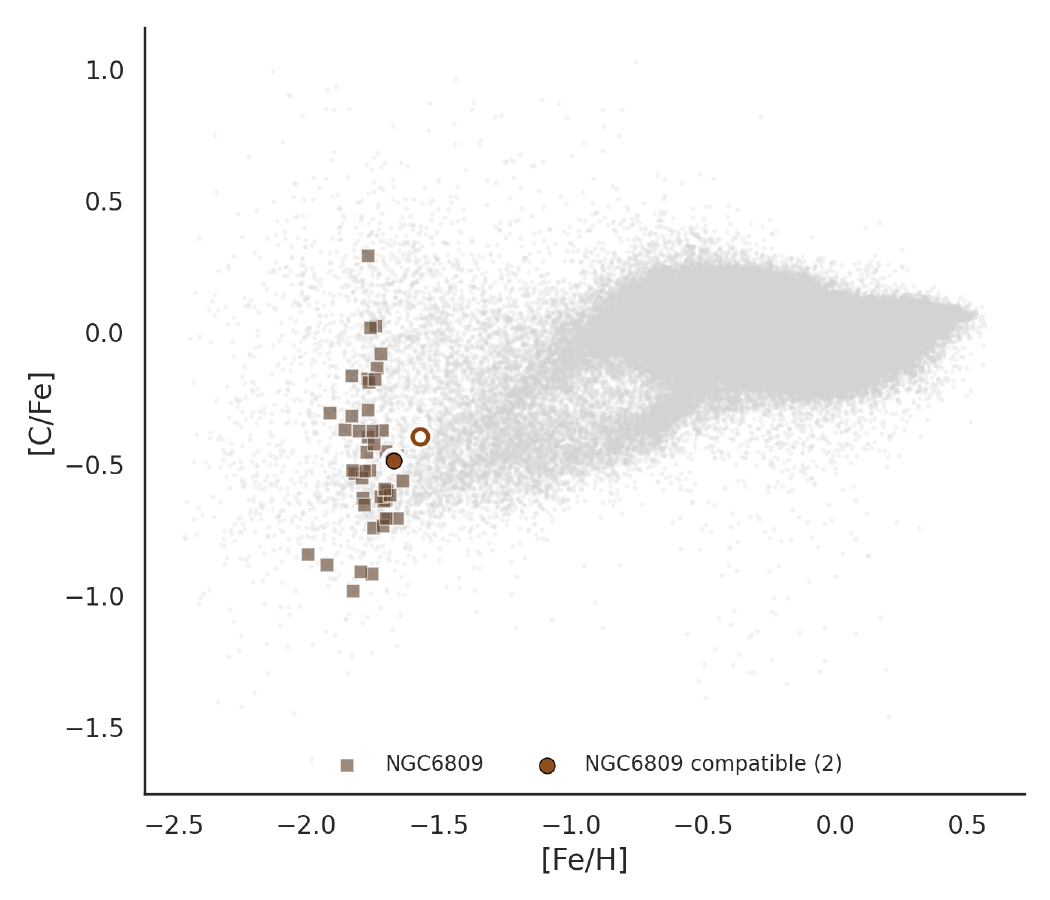}
\includegraphics[clip=true, trim = 2mm 0mm 0mm 1mm, width=0.33\linewidth]{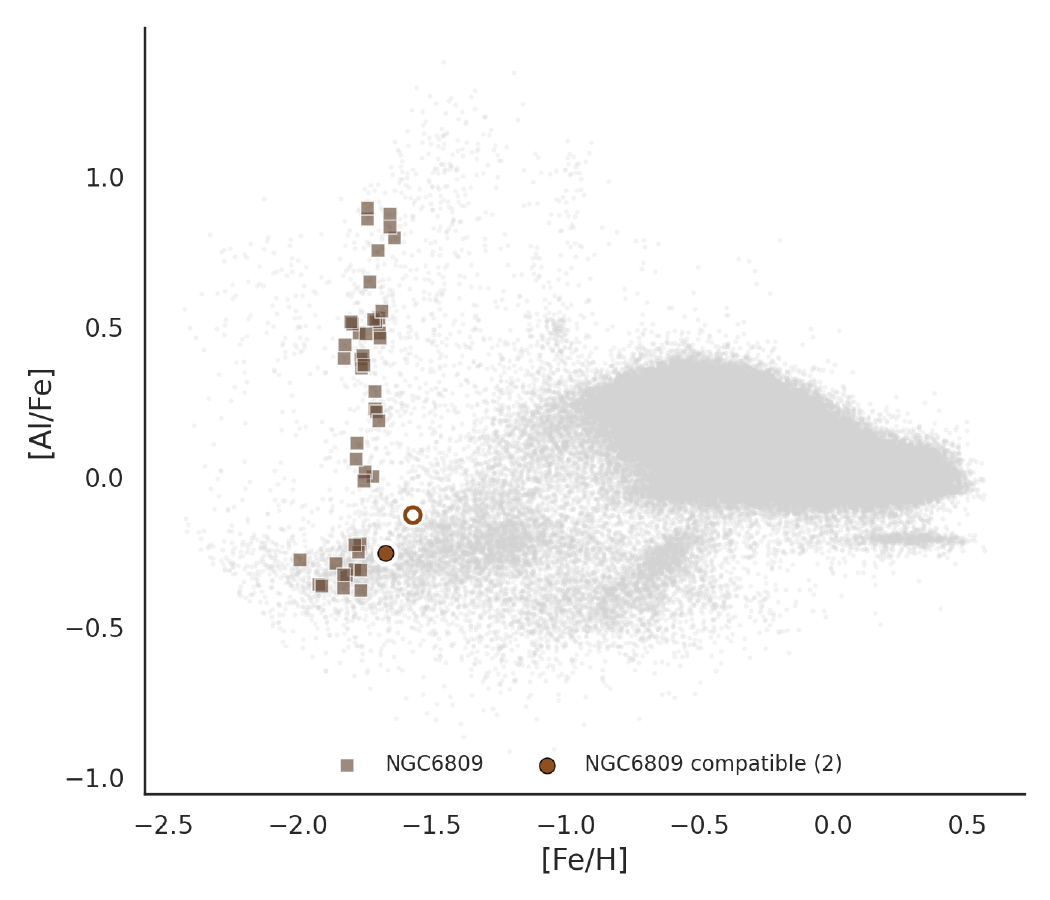}
\includegraphics[clip=true, trim = 2mm 0mm 0mm 1mm, width=0.33\linewidth]{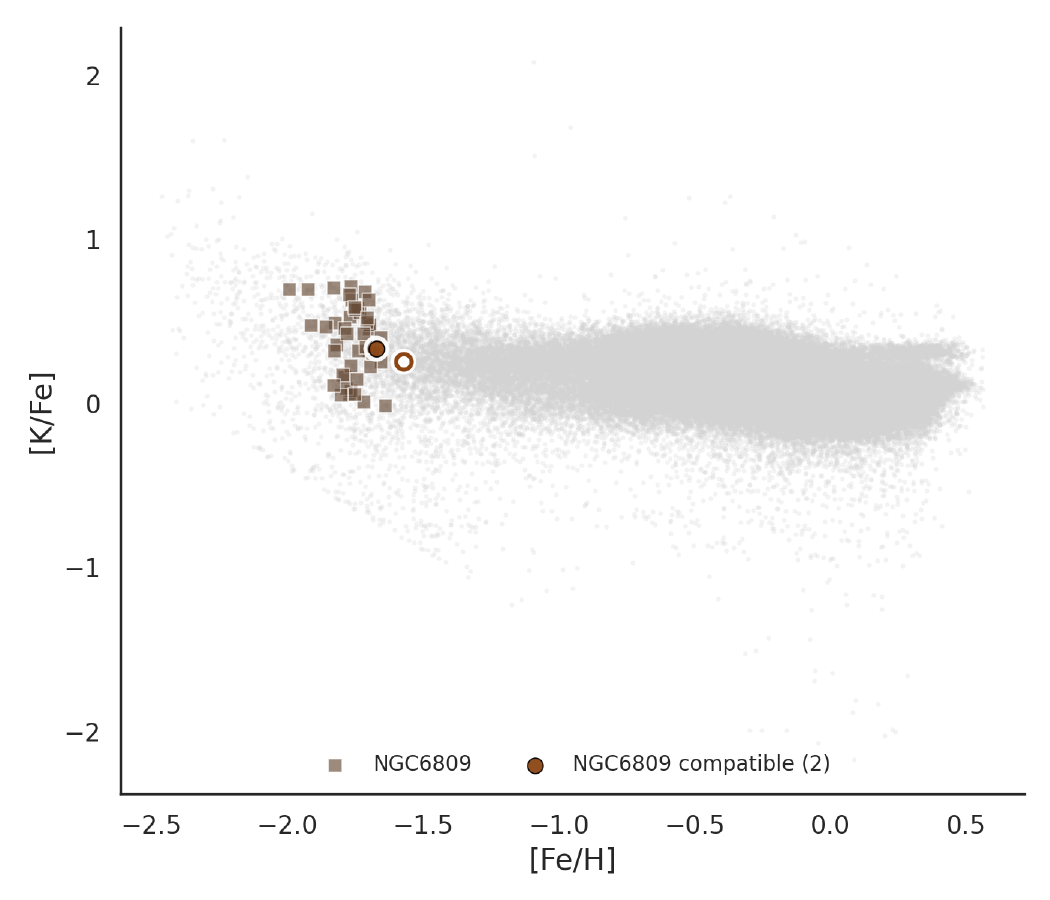}\par
\includegraphics[clip=true, trim = 1mm 0mm 0mm 1mm, width=0.33\linewidth]{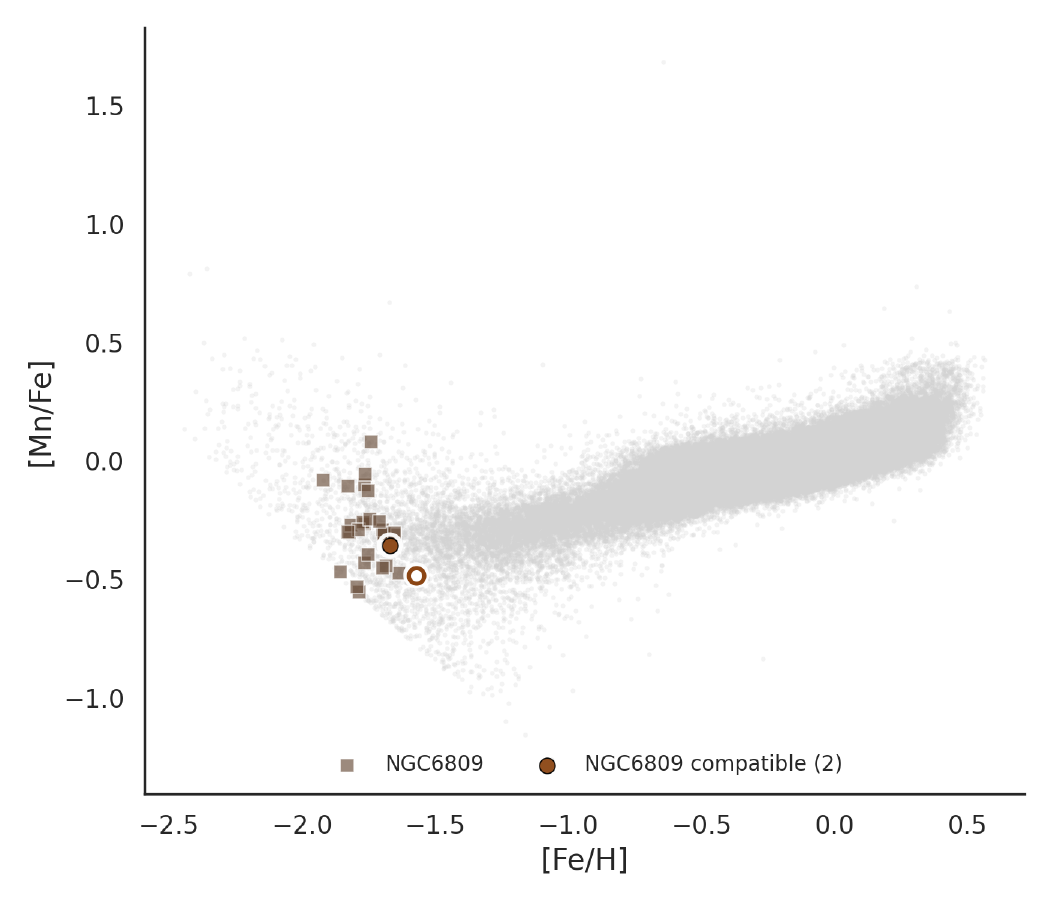}
  \caption{Same as Fig.~\ref{ngc6205_chem} but for NGC~6809.}
              \label{ngc6809_chem}%
    \end{figure*}

For each Nephele's cluster, Figures \ref{ngc5139_chem}, \ref{ngc6205_chem}, \ref{ngc6254_chem}, \ref{ngc6273_chem}, \ref{ngc6656_chem}, and \ref{ngc6809_chem} show the comparison of its chemical abundances with the ones of field stars
that have been found as chemically compatible with $\omega$~Cen and kinematically compatible with
the cluster itself. This comparison highlights the fact that, especially for some clusters such as NGC~6656 and NGC~6809, it is possible that, among the stars selected as possible members of the corresponding stellar stream, the contamination of field stars may not be negligible, and could in fact originate from stars in the main body of the progenitor galaxy Nephele, whose chemical properties are compatible with those of $\omega$~Cen. These stars are referred as "interloper" in Table~\ref{stream_ID}. In this stricter selection we retain four Al-rich stream stars in total (one for $\omega$~Cen and three for NGC~6656), so that Al-rich stars represent $4/16 = 25\%$ of the sample, i.e. a fraction still comparable to or slightly higher than the $17\%$ found in the full Nephele candidate sample. This confirms that Al-rich stars remain well
represented among the GC-stream candidates even under the more conservative cuts.

\begin{table}\centering
\caption{APOGEE identifiers of stream stars retrieved by our GMMKin, together with their associated parent cluster, and an \texttt{Al-rich} flag indicating membership in the Al-rich population, according Eq.~\ref{eq:alrich}.}
\resizebox{\columnwidth}{!}{
\begin{tabular}{llcc}
\toprule
GC NAME & APOGEE ID & Al-rich &comments \\
\midrule
NGC 5139 & 2M10332941-8541290 & &- \\
NGC 5139 & 2M13275828-4810111 & &GC member: \tt{RV\_Prob} = 0.98 \\
NGC 5139 & 2M14555718+2143047 & &- \\
NGC 5139 & 2M16102042+2600410 & &- \\
NGC 5139 & 2M16141203+1944342 & $\checkmark$ &- \\
NGC 5139 & 2M22030484-4736227 & &- \\
\midrule
NGC 6205 & 2M14515461+4628484 & &- \\
\midrule
NGC 6254 & 2M15574647-2334261 & &- \\
\midrule
NGC 6273 & 2M17305742-2319057 & &- \\
\midrule
NGC 6656 & 2M05180778-5930455 & &- \\
NGC 6656 & 2M11462612-1419069 & &interloper \\
NGC 6656 & 2M11592083+2156402 & &- \\
NGC 6656 & 2M15264680+2908444 & &interloper \\
NGC 6656 & 2M17015278+2152062 & $\checkmark$ &interloper \\
NGC 6656 & 2M18360558-2416056 & $\checkmark$ &GC member: \tt{RV\_Prob} = 0.98 \\
NGC 6656 & 2M18362044-2420197 & $\checkmark$ &GC member: \tt{RV\_Prob} = 0.98 \\
NGC 6656 & 2M18373436-2340254 &  &GC member: \tt{RV\_Prob} = 0.98 \\
NGC 6656 & 2M18374802-2400334 & $\checkmark$ &GC member: \tt{RV\_Prob} = 0.97 \\
\midrule
NGC 6809 & 2M15132332+0148026 & &interloper \\
NGC 6809 & 2M21383576-3057135 & &- \\
\bottomrule
\end{tabular}}
\tablefoot{Additional details, i.e. if the star is an interloper and not a robust stream candidate (see Sect.~\ref{neph_GCstreams} for an explanation and discussion on this point) or if the star is considered as a cluster member according to the RV-based membership \citep[see][]{schiavon2023},  are described in the third column.}
\label{stream_ID}
\end{table}

\section{Nephele and Gaia Sausage-Enceladus in chemical spaces} \label{app:nephele_gse}
\begin{figure*}\centering
\includegraphics[clip=true, trim = 3mm 5mm 0mm 3mm, width=0.33\linewidth]{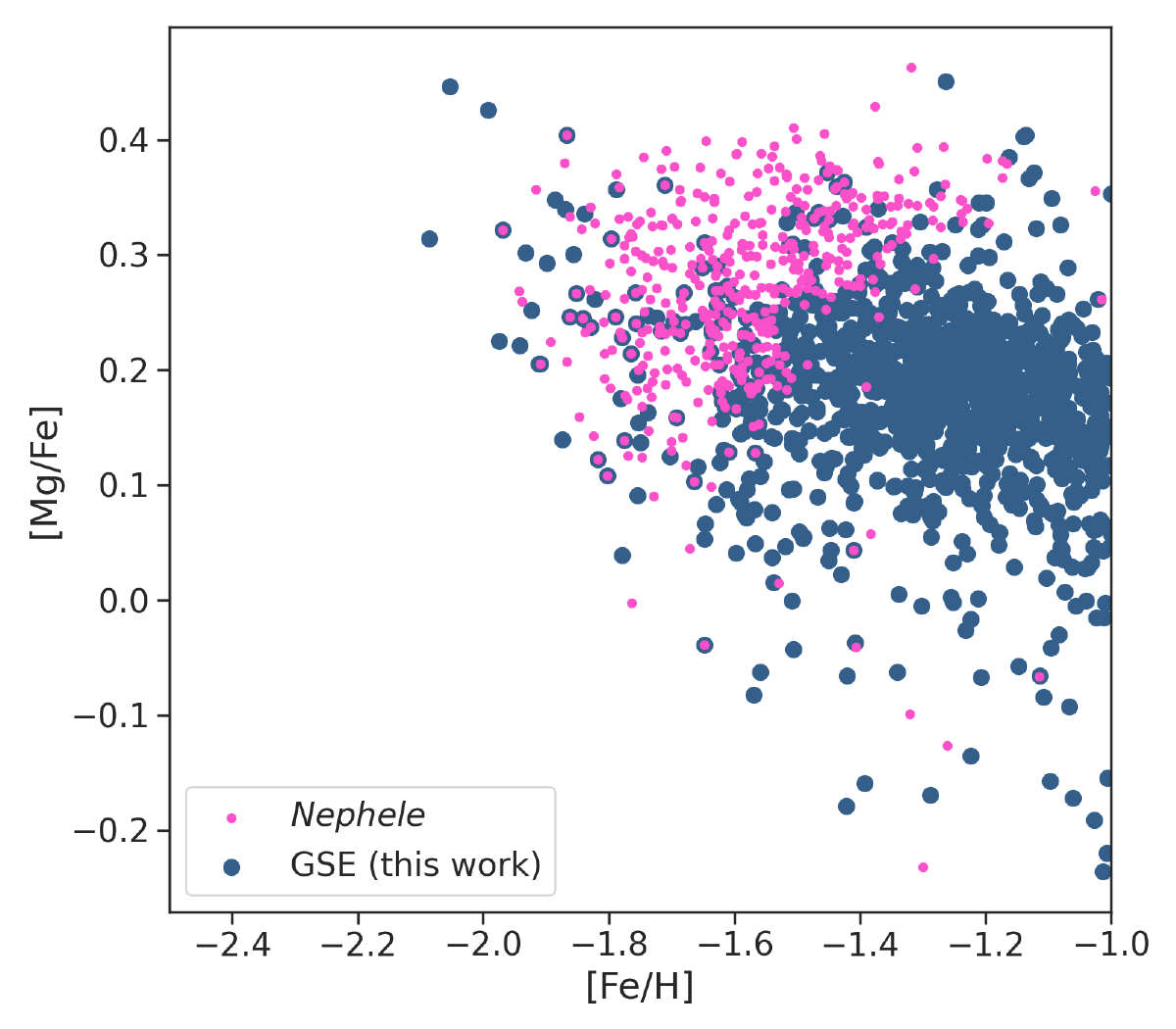}
\includegraphics[clip=true, trim = 3mm 5mm 0mm 3mm, width=0.33\linewidth]{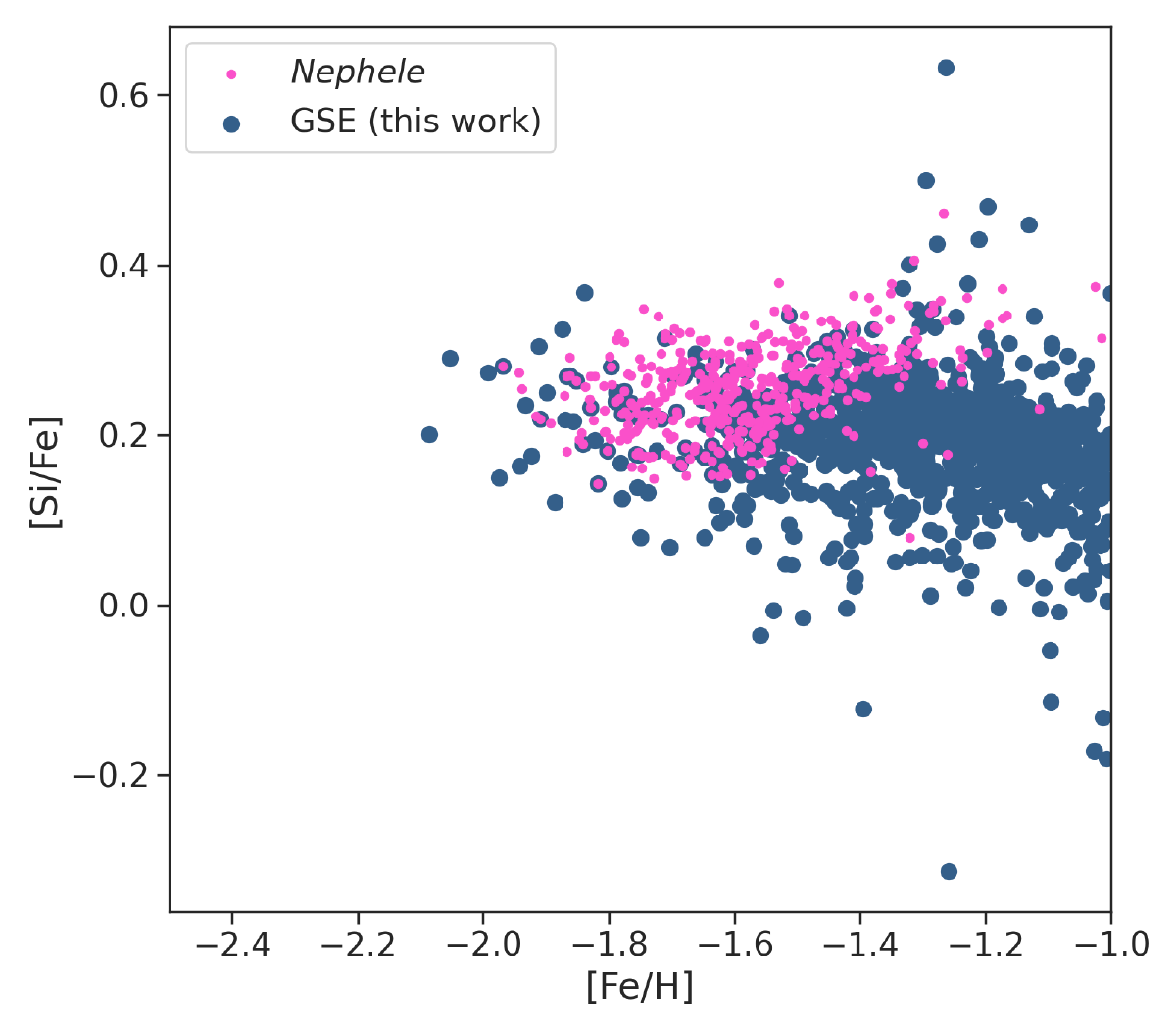}
\includegraphics[clip=true, trim = 3mm 5mm 0mm 3mm, width=0.33\linewidth]{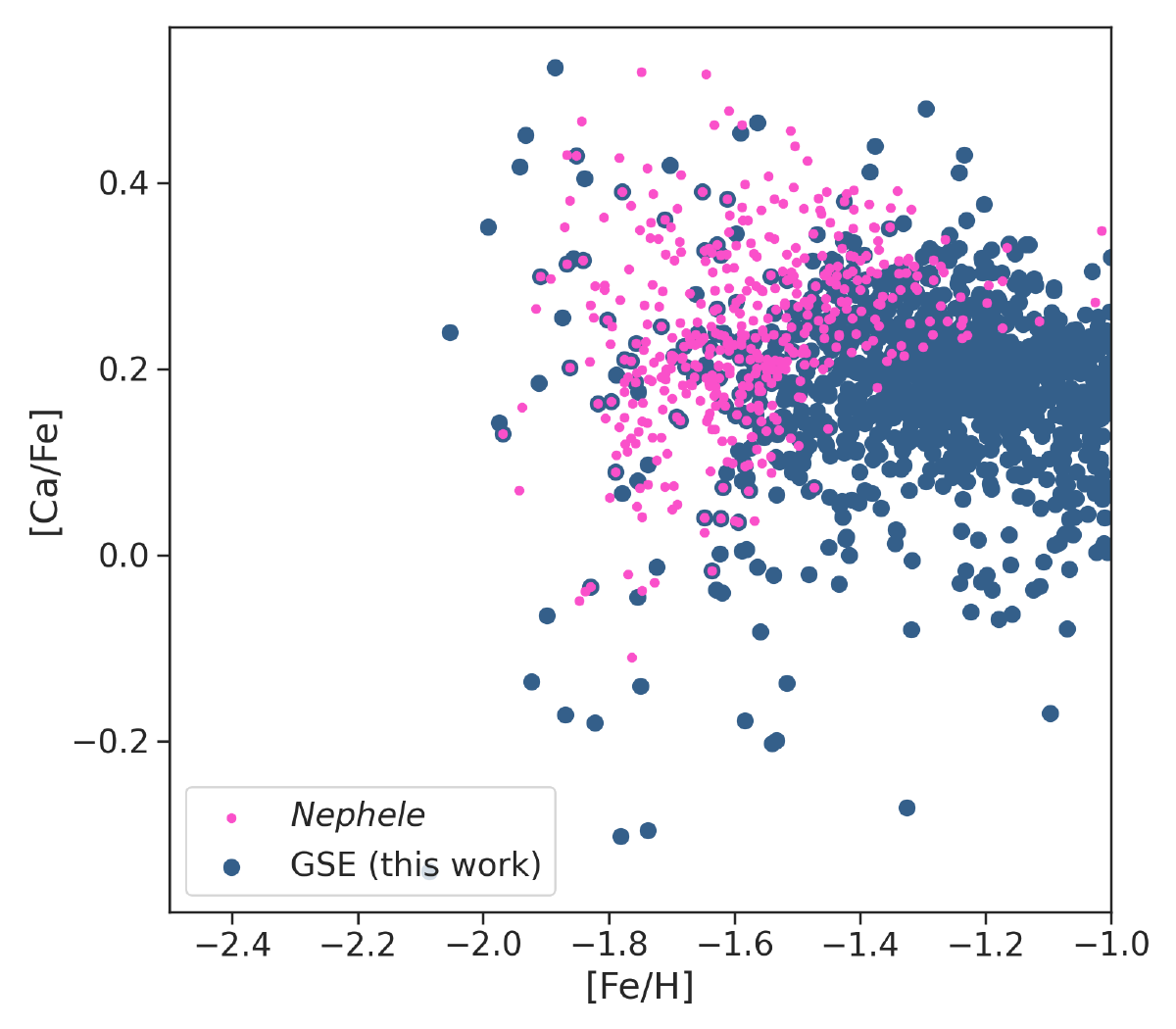}\par
\includegraphics[clip=true, trim = 3mm 5mm 2mm 0mm, width=0.33\linewidth]{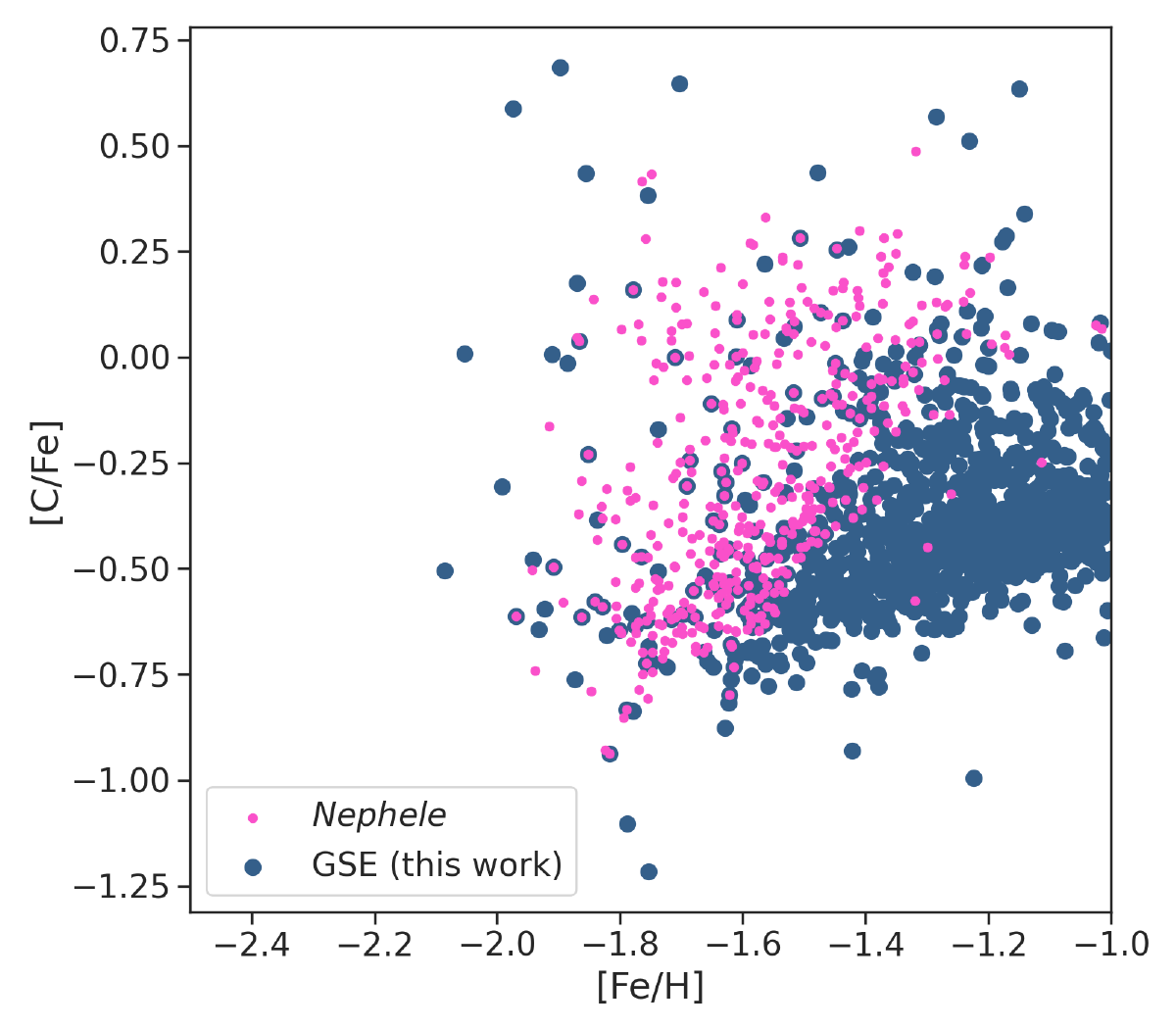}
\includegraphics[clip=true, trim = 2mm 5mm 0mm 1mm, width=0.33\linewidth]{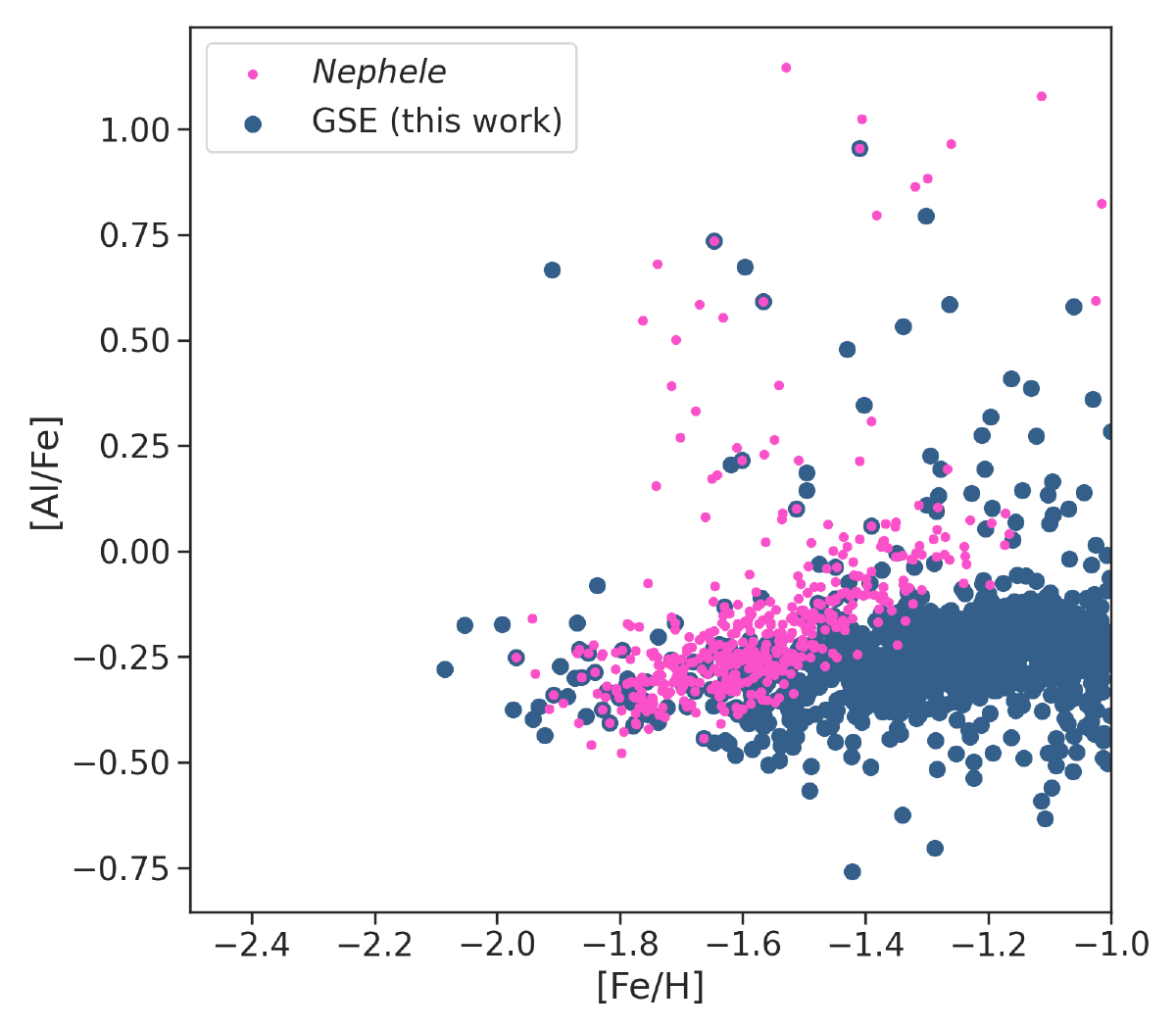}
\includegraphics[clip=true, trim = 2mm 5mm 0mm 1mm, width=0.33\linewidth]{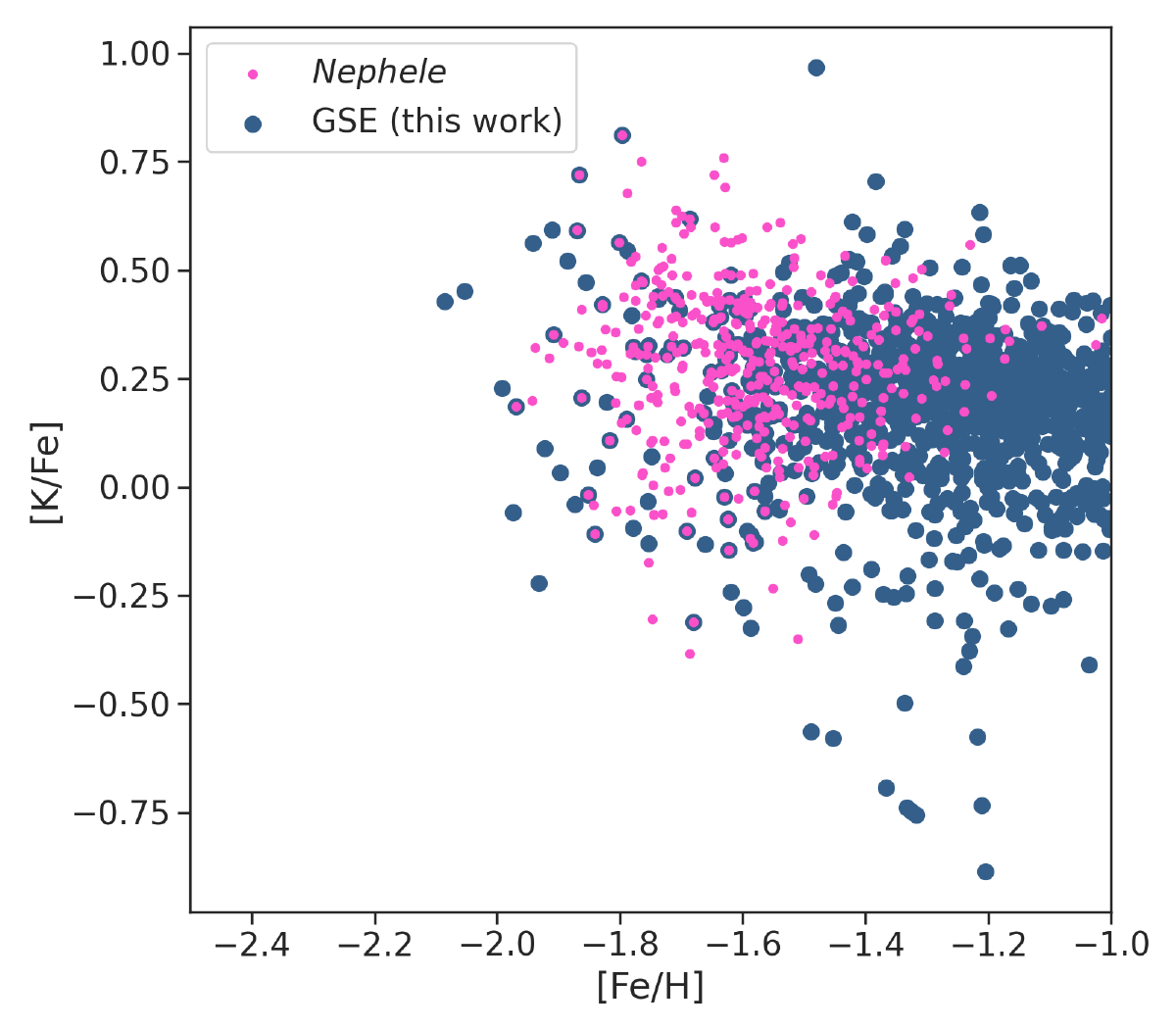}\par
\includegraphics[clip=true, trim = 1mm 0mm 0mm 1mm, width=0.33\linewidth]{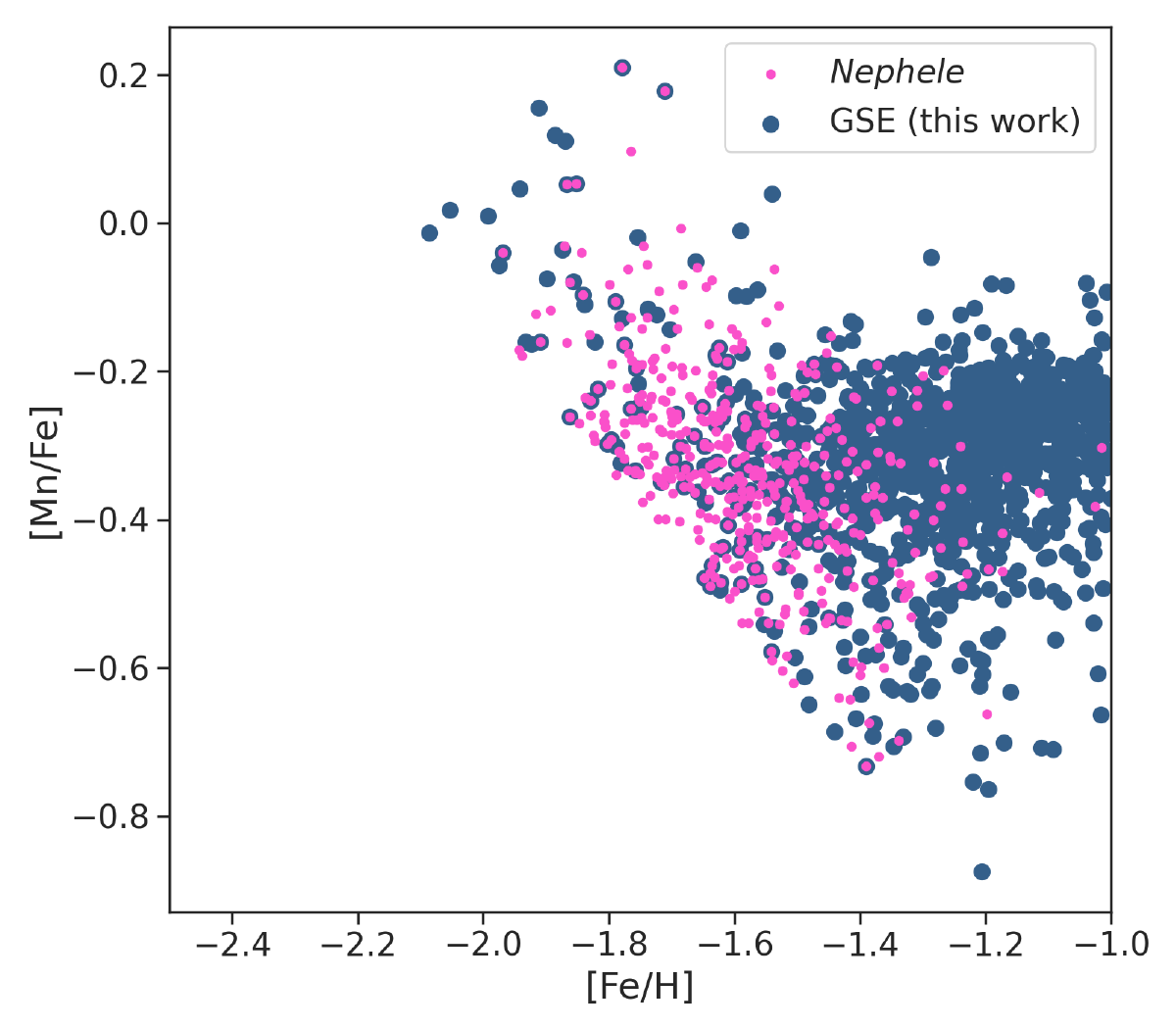}
  \caption{APOGEE abundance trends for Nephele candidates (magenta) and
    the GSE sample defined in this work (blue, see Sect.\ref{neph_gse}). Each panel shows
    [X/Fe] versus [Fe/H] for a different element.}
              \label{nephvsgse}%
    \end{figure*}

In Fig.~\ref{nephvsgse} we compare the [X/Fe]-[Fe/H] trends of our Nephele candidates with those of our GSE sample (see Sect.~\ref{neph_gse}). The two systems do not occupy identical regions of abundance space. Within our APOGEE selection, Nephele stars are somewhat more biased towards lower metallicities and, at fixed [Fe/H], they tend to show slightly higher [Mg/Fe] and [Si/Fe], as well as generally more scattered [Al/Fe] and [K/Fe], than GSE. In this qualitative sense, these ratios most clearly highlight subtle differences between the two samples in APOGEE. However, Fig.~\ref{nephvsgse} also shows substantial overlap: no single [X/Fe] ratio provides a clean one-dimensional separation, especially around $-1.8 \lesssim \mathrm{[Fe/H]} \lesssim -1.5$.

\end{appendix}

\end{document}